\documentclass[preprint]{aastex}

\usepackage{epsfig}
\usepackage{graphicx}
\usepackage{psfrag}
\usepackage{color}
\usepackage{dcolumn}
\usepackage{bm}
\usepackage{longtable}
\usepackage{natbib}
\usepackage[usenames,dvipsnames]{xcolor}
\usepackage[bookmarks=true]{hyperref}
\def\lesssim{\mathrel{\hbox{\rlap{\hbox{\lower4pt\hbox{$\sim$}}}\hbox{$<$}}}}
\def\gtrsim{\mathrel{\hbox{\rlap{\hbox{\lower4pt\hbox{$\sim$}}}\hbox{$>$}}}}

\newcommand{\bea}{\begin{eqnarray}}
\newcommand{\eea}{\end{eqnarray}}

\newcommand{\dF}{{^{^*}\!\!F}}

\newcommand{\bF}{{\bf F}}
\newcommand{\bU}{{\bf U}}

\newcommand{\del}{{\partial}}

\newcommand{\prim}{{{\mathbf{P}}}}

\newcommand{\beq}[1]{\begin{equation} #1 \end{equation}}
\newcommand{\beqa}[1]{\begin{eqnarray} #1 \end{eqnarray}}
\newcommand{\deriv}[2]{\frac{ d #1 }{ d #2 }}
\newcommand{\pderiv}[2]{\frac{ \partial #1 }{ \partial #2 }}

\newcommand{\harm}{\texttt{Harm3d}}
\newcommand{\Rmax}{r_\mathrm{max}}
\newcommand{\Rmin}{r_\mathrm{min}}
\newcommand{\rpmax}{r_{p}}
\newcommand{\rin}{r_\mathrm{in}}
\newcommand{\rout}{r_\mathrm{out}}

\newcommand{\Omegabin}{\Omega_\mathrm{bin}}
\newcommand{\bbh}{\mbox{BBH\ }}

\newcommand{\xx    }[1]{x^{\left(#1\right)}}
\newcommand{\NN    }[1]{N^{\left(#1\right)}}
\newcommand{\QQ    }[1]{Q^{\left(#1\right)}}
\newcommand{\xone  }{\xx{1}}
\newcommand{\xtwo  }{\xx{2}}
\newcommand{\xthree}{\xx{3}}
\newcommand{\None  }{\NN{1}}
\newcommand{\Ntwo  }{\NN{2}}
\newcommand{\Nthree}{\NN{3}}

\def\lambdabar{%
\relax
\bgroup
\def\@tempa{\hbox{\raise.73\ht0
\hbox to0pt{\kern.25\wd0\vrule width.5\wd0
height.1pt depth.1pt\hss}\box0}}%
\mathchoice{\setbox0\hbox{$\displaystyle\lambda$}\@tempa}%
{\setbox0\hbox{$\textstyle\lambda$}\@tempa}%
{\setbox0\hbox{$\scriptstyle\lambda$}\@tempa}%
{\setbox0\hbox{$\scriptscriptstyle\lambda$}\@tempa}%
\egroup
}

\newcommand{\alphascale}{\left(\frac{\alpha}{0.2}\right)}
\newcommand{\hoverrscale}{\left(\frac{H/r}{0.15}\right)}
\newcommand{\runtwo}{\hbox{RunIn\ }}
\newcommand{\runthree}{\hbox{RunSS\ }}
\newcommand{\bbhs}{\mbox{BBHs\ }}

\slugcomment{Submitted to ApJ.}

\begin{document}

\title{Circumbinary MHD Accretion into Inspiraling Binary Black Holes}

\author{Scott C. Noble$^1$,  
  Bruno C. Mundim$^1$, 
  Hiroyuki Nakano$^1$, 
  Julian H. Krolik$^2$, 
  Manuela Campanelli$^1$, 
  Yosef Zlochower$^1$, 
  Nicol\'{a}s Yunes$^3$}

\affil{$^1$ Center for Computational Relativity and Gravitation, 
Rochester Institute of Technology, Rochester, NY 14623}

\affil{$^2$ Physics and Astronomy Department, Johns Hopkins University, Baltimore, MD 21218}

\affil{$^3$ Department of Physics, Montana State University, Bozeman,  MT 59717}

\email{scn@astro.rit.edu}

\begin{abstract}
As two black holes bound to each other in a close binary approach merger
their inspiral time eventually becomes shorter than the characteristic inflow time
of surrounding orbiting matter.
Using an innovative technique in which we represent the changing spacetime in the region
occupied by the orbiting matter with a 2.5PN approximation and the binary orbital evolution
with 3.5PN, we have simulated the MHD evolution of a circumbinary disk surrounding an
equal-mass non-spinning binary.
Prior to the beginning of the inspiral, the structure of the circumbinary disk is predicted
well by extrapolation from Newtonian results.  The binary
opens a low-density gap whose radius is roughly two binary separations, and matter piles up
at the outer edge of this gap as inflow is retarded by torques exerted by the binary;
nonetheless, the accretion rate is diminished relative to its value
at larger radius by only about a factor of 2.  During inspiral, the inner edge of the disk
at first moves inward in coordination
with the shrinking binary, but as the
orbital evolution accelerates, the rate at which the inner edge moves toward smaller radii
falls behind the rate of binary compression.   In this stage, the rate of angular momentum
transfer from the binary to the disk slows substantially, but the net accretion rate
decreases by only 10--20\%.  When the binary separation is tens of gravitational radii, the
rest-mass efficiency of disk radiation is a few percent, suggesting that supermassive binary
black holes in galactic nuclei could be very luminous at this stage of their evolution.
If the luminosity were optically thin, it would be modulated at a frequency that is a beat
between the orbital frequency of the disk's surface density maximum and the binary orbital
frequency.  However, a disk with sufficient surface density to be luminous should also be
optically thick; as a result, the periodic modulation may be suppressed.
\end{abstract}


\section{Introduction}

    There is now excellent evidence that every galaxy with a bulge contains a supermassive
black hole at its center \citep{Gultekin09}.  In addition, the prevailing theory of galaxy
formation posits that today's massive galaxies were assembled from smaller pieces, as
dark-matter haloes of progressively greater size merged \citep{Davis85,Bardeen86}.
If massive black holes were already present in those progenitors, they would bring their
black holes with them into the new combined galaxy, creating an opportunity for the black
holes to merge.  Such an event would be very exciting to detect for many reasons:  It would
reveal the presence of supermassive black holes early in the life of galaxies.  It would
shed important light on the growth of the strong correlations between nuclear black hole
mass and galaxy structure \citep{Gultekin09}.  Most of all, it would provide a concrete example
of one of general relativity's most spectacular predictions and possibly also allow a test
of the validity of general relativity in a truly strong-field regime.

An extremely large amount of energy is very rapidly released in a binary black hole
(BBH) merger event, almost all of it through gravitational radiation [several percent
of the black hole masses in a timescale of $\sim (M_{\rm BBH}/M_\odot)\times 493\,\mu$s].
Gravitational radiation may be strong enough to eject the final remnant from its host
galaxy, with recoil velocities or ``kicks'' up to $\approx 10^3$ km/s predicted by numerical relativity simulations
\citep{Baker:2007gi,Campanelli:2007ew,Campanelli:2007cga,Gonzalez:2007hi,Herrmann:2007ac,
Koppitz:2007ev,Baker:2008md,Healy:2008js,Lousto:2009ka,Lousto:2011kp,Lousto:2012su}.
Unfortunately, a gravitational wave observatory with adequate sensitivity in the appropriate
frequency range is still well in the future, whether it operates by direct detection or
through pulsar timing \citep{Jennrich09,wen11}.
On the other hand, even if only a small part of the energy is deposited in nearby gas, the
associated photon signals might be much more readily seen with instruments operating today.
Because the energy given to the gas comes from work done by gravitational forces, one would
expect, on the basis of the Equivalence Principle, that the total energy added to the gas 
would be proportional to its mass.  
If most of this added energy is dissipated into heat (local irregularities are likely to
drive shocks), the total energy radiated in photons would then be similarly proportional to
the gas mass \citep{Krolik10}.  The question is, therefore: ``How much mass would one expect
in the neighborhood of a black hole merger?"

    Even if a BBH were supplied with mass at a rate characteristic of high luminosity quasars
($\sim 10 M_{\odot}$~yr$^{-1}$), several effects may severely reduce how much gas remains
close to the binary.  Torques exerted by the binary on the inflowing
gas may hold back the inflow, preventing much of it from approaching closer than a few times
the binary separation $a$ \citep{P91,MM08}.  As the binary compresses, whether by interactions
with passing stars and external gas or by gravitational radiation, the gas follows, but is
held off at a distance of at least $\simeq 2a$.  Toward the end of the binary's evolution, 
gravitational radiation losses grow rapidly and dominate the orbital shrinkage.  Ultimately,
the orbit shrinks on a timescale shorter than the characteristic accretion inflow time
and the BBH is expected to decouple from the disk.  After such decoupling, there would not
be enough time for much disk mass to catch up with the black holes before they merge
\citep{MP05}.  Thus, for a given external supply rate, the amount of gas available to be
heated in a merger is determined by a competition between the internal stresses that drive
inflow and a pair of dynamical mechanisms that tend to keep gas at ``arms-length" from
the merging black holes. 

    Until recently, efforts to quantify these effects have relied almost entirely
on the phenomenological Shakura-Sunyaev $\alpha$-disk model to describe internal stresses
\citep{MP05,MM08,LiuSLS10,Tanaka10}, in which the vertically-integrated and time- and
azimuthally-averaged internal stress is supposed to be a factor $\alpha$ times the similarly
integrated and averaged
pressure; the only exceptions were studies focusing on binaries
with large mass ratios between primary and secondary, a limit primarily relevant to planet
formation and to extreme-mass ratio inspiral sources
\citep{WBH03,PapNel03a,PapNel03b,Kocsis:2011dr,Yunes:2011ws}.  Moreover, with the exception
of \cite{Farris11} and \cite{Bode12}, which assumed these stresses were negligible,
all these calculations also assumed Newtonian dynamics.  However, there is strong reason to
think that the actual mechanism of these stresses is magnetohydrodynamic (MHD)
turbulence, stirred by the magnetorotational instability (MRI) \citep{BH98}.  In contrast
to ordinary, isotropic turbulence, orbital shear makes this turbulence highly anisotropic,
so that there is a non-zero correlation between the radial and azimuthal components of the
magnetic field; this correlation creates the stress.  MHD calculations
are therefore required, at the very least, to establish the appropriate scale of the stresses
and the approximate magnitude of $\alpha$.  In addition, although the $\alpha$-model may give a
reasonable description of time-averaged behavior well inside the body of an accretion flow,
it is particularly ill-suited for predicting dynamical behavior on shorter timescales
\citep{Hirose09b} and at disk edges \citep{KHH05,Noble10}.  Because the key issues in how much
gas reaches a merging BBH depend, of course, on the time-dependent behavior of gas near and
within the inner edge of a disk, explicit calculation of the MHD turbulence is also
required for an accurate treatment of the time- and spatial-dependence of the internal stress. 

    In this paper, we present the first simulation of a circumbinary accretion disk around
a binary black hole system during the epoch in which the binary's inspiral time grows
shorter than the inflow time through the disk.  Generically, this period occurs not long
before the binary's final merger.
Our physics treatment includes fully relativistic MHD.  This study differs from that
of~\cite{Shi11}, who presented similar MHD simulations, but concentrated on the Newtonian
regime, when the black holes are very widely separated. Moreover, their Newtonian treatment did
not allow for the black holes to inspiral, a reasonable assumption when the semi-major
axis is hundreds of thousands of gravitational radii, but a terrible assumption in the
late inspiral.  We here focus on BBHs with separations of $\sim (10$--$20)\,r_g$, where 
$r_g \equiv GM/c^2$ and $M$ is the total mass of the binary 
(we adopt geometric units with $G=c=1$ for the remainder of this paper). 
The spacetime associated with the BBH orbital dynamics is described
through a vacuum post-Newtonian (PN) approximation (see the review paper of~\citet{Blanchet:2002av} and references therein),
where we neglect the back reaction of the disk on the BBH dynamics.  Our work also
contrasts with that of \cite{Giacomazzo12}, who employed full numerical relativity to
compute the spacetime in which an initially uniform gas distribution with an externally-imposed
uniform magnetic field evolved during the last 3 orbits before merger.

    As we will describe below, the PN approximation is adequate to describe the spacetime evolution for our needs.
The PN scheme is a method to describe approximately the dynamics of 
physical systems in which motions are slow compared to the speed of 
light and gravitational fields are weak. That is, one solves
the Einstein field equations perturbatively,
expanding in $(v/c)^2 \ll 1$ and $r_g/r = GM/(rc^2) \ll 1$
Here $v$, $M$ and $r$ are the characteristic velocity, mass and size or separation of
the system.  This approximation has been remarkably effective in describing the
perihelion precession of Mercury \cite{Einstein:1915bz}, and the gravitational-wave loss
from binary systems, such as the Hulse-Taylor pulsar, PSR B1913+16
(see e.g.~\cite{Weisberg:2004hi,Will:2011nz}). 
PN theory also plays a key role in the construction of the gravitational-waveform
templates \citep{Sathyaprakash:2009xs} for inspiraling compact objects currently used
in the search for gravitational waves by laser-interferometric observatories. 
PN theory has also been recently interfaced with numerical relativity simulations to
serve as initial data for the modeling of
BBH mergers~\citep{Tichy:2002ec,2003PhRvD..68d4019B,Yunes:2005nn,Yunes:2006iw,
Yunes:2006mx,Kelly:2007uc,Campanelli:2008nk,JohnsonMcDaniel:2009dq,Kelly:2009js,Mundim:2010hu}. 
In all cases, the PN approximation
is developed to sufficiently high perturbative order that the error contained 
in the approximation
is much smaller than that associated with either the data in hand 
(in the case of binary pulsars)
or the data expected (in the case of direct gravitational wave detection). 

Using this PN-approximated description of the spacetime, we first evolve 
the BBH at a fixed initial separation $a_0=20 M$ to allow the accretion disk to 
relax to a quasi-steady state.  To study the effect of orbital shrinkage on
the accretion disk, we then follow the binary inspiral until it reaches a separation
$a = 8M$, beyond which the PN approximation ceases to be sufficiently accurate for our
purposes.  In a separate simulation, we kept 
the binary's separation fixed at $20 M$ and continued to evolve until $\simeq 76000M$
to study the secular dynamics of the quasi-steady state, and, by contrasting with
the first simulation, highlight the special effects induced by the inspiral.

Our findings can be summarized as follows.  The mass at $r \simeq 2.5a$ builds steadily
throughout the quasi-steady state, but much of it eventually concentrates in a distinct
``lump''.  At smaller radii, a gap is cleared as torques and forces exerted by the
binary either sweep matter inward or fling it outward.  Much of the small amount of
mass in this gap is found in a pair of streams emanating from the inner edge of the
disk and curving inward toward each black hole.  These streams carry nearly half of the 
mass accreting through the bulk of the disk to the inner boundary of the problem
volume
at $r=0.75a_0$.

As the binary starts to shrink, the inner edge of the disk at first moves inward 
following the orbital evolution of the binary, but eventually cannot keep up, 
as the orbital shrinkage grows faster. Nevertheless, a significant amount of mass 
still follows the binary's inspiral within the gap region. 
We find that the final accretion rate in the inspiral stage is about 
$70\%$ of the corresponding rate in the steady-state stage.

The luminosity of the disk is proportional to its surface density.  If the accretion
rate fed to the disk were comparable to that of ordinary AGN, the surface density, and
therefore the luminosity, of such a circumbinary disk could approach AGN level.
Most strikingly, the luminosity should be modulated periodically at a frequency
determined by the binary orbital frequency and the binary mass ratio---$1.46$ times
the binary orbital frequency in the case of equal masses.  However, the amplitude
of modulation may be reduced by the large optical depth of the disk if the surface
density is large enough to generate a sizable luminosity.

This paper is organized as follows. In Section~\ref{sec:spacetime} we describe the
construction of the dynamical BBH spacetime. In Section~\ref{sec:simulation-details}, we
report the details of the MHD simulations of the disk. In Section~\ref{sec:results} and 
Section~\ref{sec:discussion}, we present the results from these simulations and interpret
them in the light of previous work and from the point of view of potential observational
signatures.  Finally, in Section~\ref{sec:conclusions}, we summarize our principal conclusions.

\section{Binary Black-Hole Spacetime}
\label{sec:spacetime}

While solutions of the Einstein equations for single black holes were
discovered as early as 1916~\citep{Schwarzschild16a}, no exact closed-form 
solution to the two-body problem exists and one generally needs to solve 
the Einstein equation numerically. With the breakthroughs in numerical relativity
\citep{Pretorius:2005gq, Campanelli:2005dd,Baker:2005vv,
Scheel:2006gg}, it is now possible to perform stable and accurate 
full numerical simulations of \bbhs in vacuum for a
wide variety of mass ratios and spins parameters. 
However, because the Einstein
equations can be thought of as modified wave equations, with wave
speeds of $c$, the Courant condition\footnote{The Courant condition is a stability
 condition relating the timestep $dt$ to the spatial resolution $h$. The timestep is
limited by $dt < h / v$, where $v$ is the fastest propagation speed in the system of
interest.  } greatly limits the timestep size,
making full numerical simulations impractical when the characteristic
MHD speeds are significantly smaller than $c$. On the other hand, if
an approximate, but accurate, spacetime is given, the Courant
condition is set by the MHD speeds, allowing for a much larger
timestep. 
 Fortunately, analytic 
perturbative techniques have been successfully developed to tackle 
the spacetime problem both in the regime where the black holes are not too
close, as well as in the {\it close limit} regime where the spacetime can
be treated as a perturbed single black hole.
In this paper, we use the PN approach to model the spacetime 
of an inspiraling binary system prior to merger, neglecting the
effect of the disk on the evolution of the black holes, from orbital separations
of $20M$ down to $\sim 8M$, roughly where the standard PN approximation becomes inaccurate for our purposes.

Using this PN-approximated solution, we then solve for the
relativistic MHD evolution of the circumbinary accretion disk. We stop
the PN evolution at $r = 8 M$, but one can in principle continue the
simulations beyond this regime. To do so, 
one could use a snapshot of the PN metric and MHD data at that radius as initial conditions to
then carry out a fully non-linear GRMHD evolution, using numerical
relativity techniques to solve the coupled GRMHD Einstein system of
equations (which will be the subject of an upcoming paper).

We perform two  simulations: (i) \runthree keeps the semi-major axis
of the binary artificially fixed at $20M$; (ii) \runtwo starts from a snapshot 
of \runthree at $t=40000 \, M$ (or after $\simeq 70$ orbits) and 
then lets the black holes inspiral at the PN-theory prescribed rate down to a separation
of $\sim 8M$. 
\runthree is used to study the secular evolution of the 
accretion disk at fixed binary separation, while \runtwo is used to investigate how the diminishing 
separation alters this secular evolution. 
We describe our PN approach to model the spacetime metric below.

\subsection{The post-Newtonian Approximation}

The PN approximation is based on a perturbative
expansion of all fields, assuming slow motion $v/c \ll 1$ and weak fields 
$GM/(r c^2) \ll 1$\footnote{Note that we have explicitly re-introduced $c$ and $G$ in this 
section in order to discuss the PN approximation.}.  
These assumptions allow us to search for solutions that can be expressed as a divergent asymptotic series
about a flat Minkowski background spacetime.
These perturbations obey differential equations determined
by the PN-expanded Einstein field equations. One then solves such equations perturbatively
and iteratively to construct an approximate solution. 

The two body problem in the slow-motion/weak-field limit is better
understood by classifying the spacetime into different regions, where
different assumptions hold and different approximations can be used
(see e.g.~\cite{Thorne:1980ru,Alvi:1999cw,Alvi:2003pn,Yunes:2005nn,Yunes:2006iw,Yunes:2006mx,
JohnsonMcDaniel:2009dq} for a review).  Here we concentrate on the
{\emph{near zone}}, which is the region sufficiently far from the
horizons  that the weak-field approximation of PN is valid, but less than a
reduced gravitational wave wavelength $\lambdabar$ away from the center of
mass of the system, so that retardation effects can be treated
perturbatively. We note that in the {\emph{far zone}}, i.e.\ the
radiation zone where retardation effects can no longer be treated
perturbatively, a multipolar post-Minkowskian expansion can be used rather than a
PN one.  Very close to each black hole (i.e.\ in the {\emph{inner
zone}}), 
perturbed
Schwarzschild solutions are used (which can be extended to include
spin by using perturbed Kerr solutions).
In this paper, the binary black hole metric will be approximated with
only the near zone solution.

The PN approximation, of course, has its limits: binary systems eventually become so closely
separated that a slow-motion/weak-field description is inappropriate. For example, consider  
the simple case of a test particle spiraling into a non-spinning (Schwarzschild) black hole
in a quasi-circular orbit. Eventually, the particle will reach
the innermost stable circular orbit, at which point its orbital velocity $v/c \sim 0.41$. Clearly, such a
velocity is not much less than unity, and thus, the PN approximation need not be an accurate
description of the relativistic orbital dynamics. Similarly, when
comparable-mass \bbhs
inspiral, they eventually reach a separation at which the PN approximation is a bad predictor 
of the dynamics, since the small-velocity/weak-field assumptions are violated. 

The determination of the formal region of validity of the PN approximation is crucial, but it can
only be assessed when one possesses a more accurate, perhaps numerical, description of the orbital 
dynamics. This is indeed the case when considering extreme mass-ratio inspirals (EMRIs), 
consisting of a stellar-mass compact object spiraling into a supermassive black hole. 
When considering EMRIs, one can model the spacetime through black hole perturbation theory, 
i.e., by decomposing the full metric as that of the supermassive black hole plus a perturbation 
induced by the stellar-mass compact object, 
without assuming slow motions (see e.g.~Section 4 of~\cite{Hughes:2009iq} for a recent review,
\cite{Mino:1997bx,Sasaki:2003xr,Barack:2009ux,Poisson:2011nh} for related topics). 
To leading-order, the orbital dynamics are then described by geodesics of the small object in the
spacetime of the supermassive black hole.
The orbital motions are slowly perturbed by the radiation reaction due to the emission of gravitational waves.

The comparison of black hole perturbation theory 
and PN theory predictions has allowed for the construction of 
different measures to estimate the PN region of validity.
When considering the reduction in signal-to-noise ratio induced by
filtering an ``exact" black-hole perturbation theory gravitational wave with a 
4PN\footnote{This was extended to 5.5PN order in the erratum and addendum of \cite{Poisson:1995vs}.}
order filter, 
\cite{Poisson:1995vs} found that PN theory is sufficiently accurate provided $v \lesssim 0.2$.
When considering the 5.5 and 4PN order predictions for the loss of the binary's binding energy for non-spinning
and spinning background black holes respectively, relative to an ``exact" black hole perturbation theory prediction, \cite{Yunes:2008tw} and 
\cite{Zhang:2011vh} found that the former is accurate provided $v \lesssim 0.29$, which
corresponds to an orbital separation of $a \gtrsim 11 M$. The difference
between these estimates is due to the different measures used and the different order of the PN
approximation employed\footnote{It is not surprising that beyond 3PN order the region of validity
of the PN approximation shrinks. This is a property of divergent asymptotic series, whose behavior
in the context of PN theory was analyzed by \cite{Yunes:2008tw} and 
\cite{Zhang:2011vh}.}.

The region of validity of the PN approximation for comparable-mass
binaries has not been as well studied. This is because black hole
perturbation theory is not applicable here, and one must rely on full
numerical relativistic simulations. Currently, state-of-the-art
simulations can only model the last few tens of orbits prior to
merger, while the determination of the formal region of validity would
require knowledge of at least the last thousand orbits. Nonetheless,
there exist analytical arguments suggesting that the region of
validity in the comparable-mass case is larger than in the EMRI case
(i.e., the PN expansion is valid for even larger
velocities)~\citep{Simone:1996db,Blanchet:2002xy,Mora:2003wt}.
Moreover, the NINJA (Numerical INJection Analysis)-2
project~\citep{Ajith:2012tt}, a collaboration between numerical
relativists and gravitational wave data analysts, has established that
certain 3PN order gravitational waveforms are sufficiently accurate
for use as templates provided $v \lesssim 0.33$ ($a \gtrsim 8 M$).

\subsection{Near-Zone PN Evolution}

The PN order of a given term is determined by the exponent of the perturbation parameter
contained in that term. 
In the near zone, the PN expansion of a \bbh spacetime metric is a series expansion
in the orbital velocity $v/c \ll 1$ and the field strength $G M_{A}/(r_{A} c^{2})$, 
where $M_{A}$ and $r_{A}$ are the masses of the $A^\mathrm{th}$ particle
and the distances from the $A^\mathrm{th}$ particle to a field point respectively. 
Here, we may consider $1/c$ as the PN parameter which goes to zero
in the Newtonian limit $c \to \infty$.
Notice that by the virial theorem $v^2/c^2 = {\cal{O}}[GM/(a \, c^2)]$.
A term proportional to $(1/c)^n$ beyond the Newtonian (leading-order) expression
is said to be of $(n/2)^\mathrm{th}$ PN order.

The near zone metric will be described here by a resummed PN expression. One begins with the
2.5PN expansion of the metric for non-spinning point particles in a quasi-circular orbit in harmonic coordinates, given for
example in \cite{Blanchet:1998vx}.
Such a metric, however, describes black holes as point-particles, which is 
why one then applies a ``background resummation",
as in \cite{Yunes:2006iw,Yunes:2006mx,JohnsonMcDaniel:2009dq}. 
This resummation is intended to improve the strong-field behavior of the metric close to
each point-particle, i.e., it recovers the horizon of each individual black hole.
The metric can then be formally written as
\beq{
g_{\mu\nu}(t,\,\vec{x}) = 
g_{\mu\nu}[\vec{x};\,\vec{y}_{A}(t),\,\vec{v}_{A}(t)] \,, 
}
where $\vec{x}$ is a spatial vector from the binary's center of mass to a field point, 
while $\vec{y}_{A}(t)$ and $\vec{v}_{A}(t)$ are the particle's spatial location
and 3-velocity with $A=(1,\,2)$. This metric depends on the mass of each individual 
black hole, but also on the binary orbital evolution $\{\vec{y}_{A}(t),\,\vec{v}_{A}(t)\}$ 
that must be prescribed separately.  We will use Greek letters (e.g., $\mu,\,\nu,\,\lambda,\,\kappa$) 
to represent  spacetime indices $\left[0,\,1,\,2,\,3\right]$, and Roman letters 
(e.g., $i,\,j,\,k,\,l$) to represent spatial 
indices $\left[1,\,2,\,3\right]$.  

The orbital evolution can also be prescribed within the PN approximation. 
We simplify the analysis by considering only quasi-circular orbits. This
simplification is justified because gravitational wave emission tends to circularize
binaries very efficiently, as demonstrated in the weak~\citep{Peters:1964zz}
and strong field regimes~\citep{Sperhake:2007gu,Hinder:2007qu,Hinder:2008kv}.
We will here use a 3.5PN expansion for the orbital phase evolution $\phi(t)$, 
as given for example by Equation~(234) of \cite{Blanchet:2002av}, which depends
on the quantity $\Theta =  \nu ( t_c -t ) / (5 M) $, where $t$ is time, $t_c$ is the time
of coalescence, $M$ is the total mass and $\nu = M_1 M_2/M^2$ is the symmetric mass ratio.
The PN orbital frequency is calculated from $\Omega_{\rm bin}=d\phi/dt$ in this paper.

Although waveforms can be fully characterized by harmonics of the orbital phase, the latter depend on 
the orbital trajectories. One may model these in harmonic coordinates via
\beqa{
y_{1}^{i}(t) &=& \frac{M_{2}}{M} a(t) \left[\cos{\phi(t)},\,\sin{\phi(t)},\, 0 \right]\,,
\\
y_{2}^{i}(t) &=& -\frac{M_{1}}{M} a(t) \left[\cos{\phi(t)},\,\sin{\phi(t)},\, 0 \right]\,,
}
where  $a(t)=|\vec{y}_{1}(t)-\vec{y}_{2}(t)|$ is the orbital separation as a function of time.
This separation can be calculated via the balance law (see e.g.,
\citep{Blanchet:2002av}), 
which states that the local rate of change of the binary's orbital binding energy is exactly balanced 
by the gravitational wave luminosity carried out to future null infinity and into black hole horizons, namely 
\beq{
\frac{dE_{\rm Orb}}{dt} = - \cal{L} \,.
}
The rate of change of the orbital separation is then given by
\beq{
\frac{da}{dt} = - \left(\frac{dE_{\rm Orb}}{da}\right)^{-1}  {\cal L}  \,. 
}
Assuming the initial condition $a(t=0)=a_0$, we then find
\beq{
t = t_c - 
\int_0^{a} da' \left(\frac{dE_{\rm Orb}}{da'}\right) \; {\cal L}^{-1}
\,,
\label{eq:t_r_12}
}
where the integrand is expanded in a Taylor PN series
and the coalescence time $t_c$ is defined by 
\beq{ 
t_c = \int_0^{a_0} da \left(\frac{dE_{\rm Orb}}{da}\right) \; {\cal L}^{-1} \,.
\label{eq:t_c}
}
In this paper, we only use the part of ${\cal{L}}$ that is carried to future null infinity and, although Equations~(\ref{eq:t_r_12},\ref{eq:t_c})
are typically Taylor expanded to evaluate $t(a)$, here
they are inverted through Newton-Raphson minimization to yield $a(t)$.

Given the above analysis, we can now calculate the time of coalescence and the number of orbits
in each of the simulations carried out. As already discussed, \runthree is artificially kept at a fixed
semi-major axis, so $a(t)=a_0$ for all times, and thus, formally $t_{c} = \infty$. On the other hand, 
\runtwo keeps $a(t)$ fixed to $a(t) = a_0=20M$ for $t < t_\mathrm{shrink} \equiv 40000 M$, 
after which it is allowed to evolve according to the PN equations of motion. The time of coalescence
for this run can be computed by inverting Equation~(\ref{eq:t_r_12}) to obtain $t_c \sim 14000M$, 
although to leading order it is approximately described by \citep{Peters:1964zz}
\beq{
t_c \sim \frac{5}{256 \nu}\left(\frac{a_0}{M}\right)^4 M 
= \frac{5}{256}\left(\frac{a_0}{M}\right)^4 \frac{(1+q)^2}{q} M
\,,
}
where $q = M_2/M_1$ is the binary mass ratio. A \bbh clearly takes longer to merge 
for systems that start at larger initial separations and that possess extreme mass ratios. Obviously,
$t_{c}$ is always defined as the length of time to coalesce, when the binary is allowed to inspiral.
The total simulation time of \runtwo is then $t_{c} +  t_\mathrm{shrink} \sim 54000 M$. 

\begin{figure}[htb]
\centerline{\includegraphics[width=12cm]{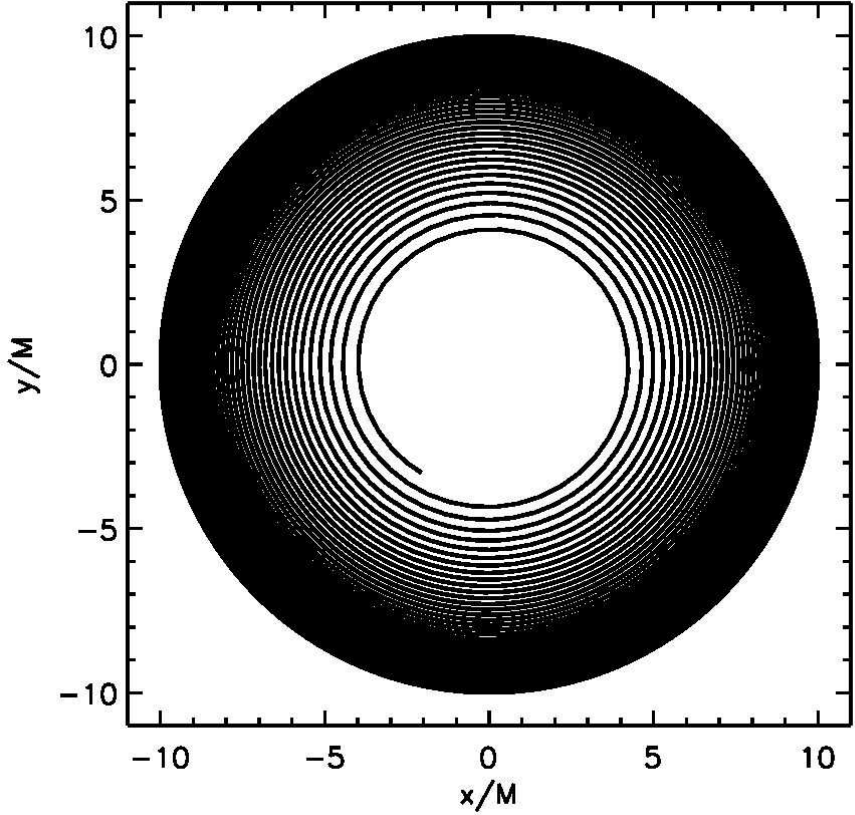}}
\caption{The orbital motion of one of the black holes in the binary from \runtwo.  Its trajectory starts 
from an initial separation of $a(0)=20M$ and stops at $a(t_f)\simeq 8M$. 
Its black hole companion is located at a parity-symmetric point 
across the origin (track not drawn in the figure for the sake of clarity).}\label{fig:traj}
\end{figure}

\begin{figure}[htb]
\centerline{
\includegraphics[width=12cm]{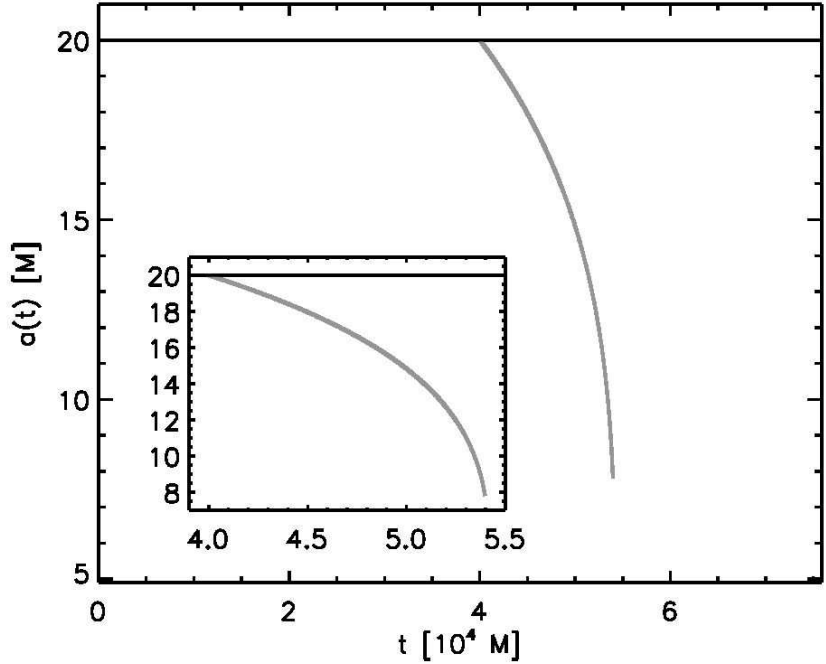}
}
\caption{The evolution of the orbital separation with respect to time, $a(t)$.
We turn on the gravitational radiation reaction at $t=40000M$.}\label{fig:asep}
\end{figure}

\begin{figure}[htb]
\centerline{
\includegraphics[width=12cm]{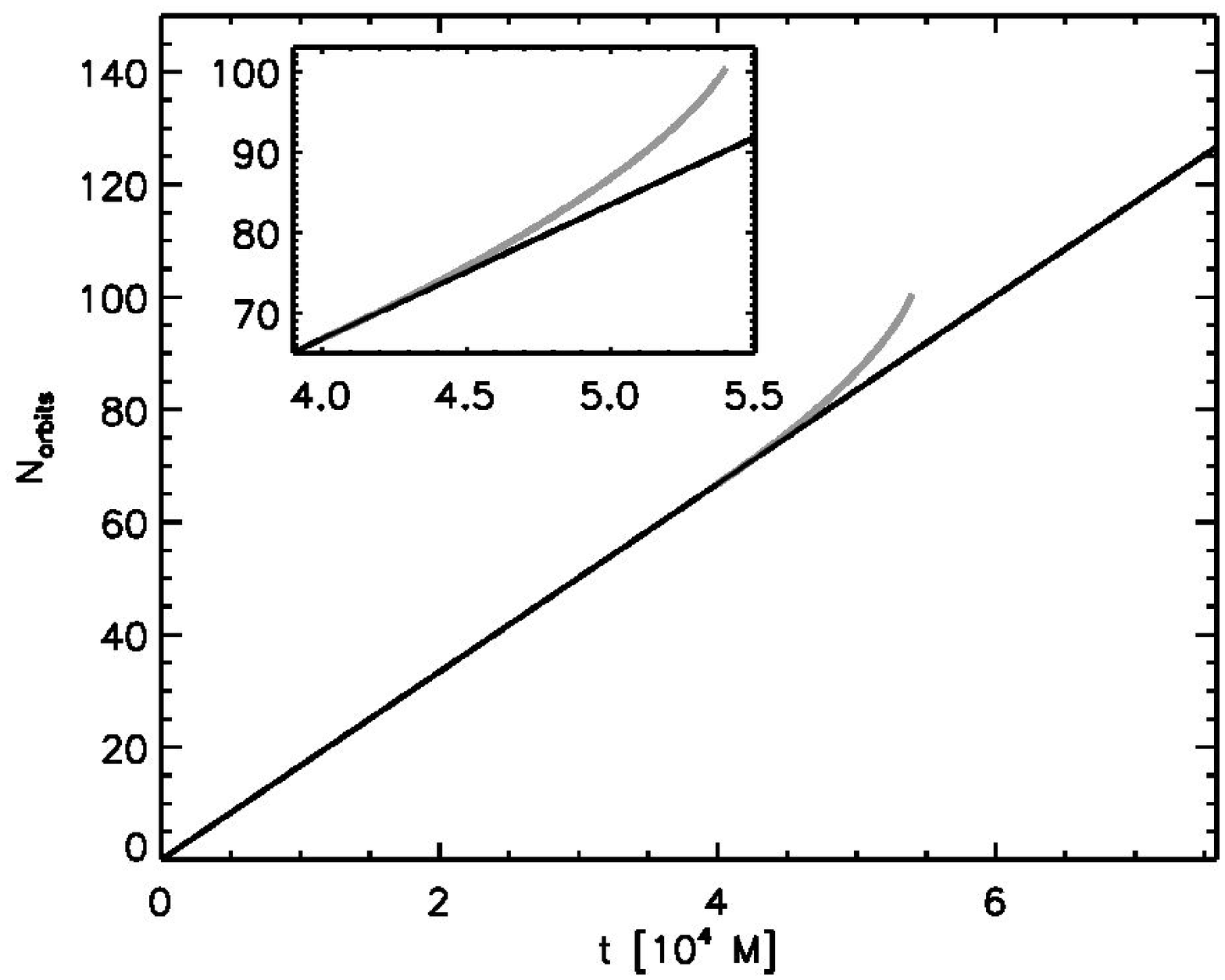}
}
\caption{The number of orbits as a function of time. (Black) \runthree. (Grey) \runtwo, in 
which the first part from $t=0$ to $40000M$ shows the circular orbit without the radiation reaction.}\label{fig:norbits}
\end{figure}

We have plotted a few diagnostics to get a sense of the evolution of the binary system
in \runtwo. Figure~\ref{fig:traj} shows the orbital evolution of the binary 
in the $x$-$y$ plane, after it is allowed to inspiral. Figure~\ref{fig:asep} plots the orbital separation
as a function of time. Observe that initially the semi-major axis is artificially kept fixed, while after
$t>t_\mathrm{shrink}$ it is allowed to decrease due to gravitational radiation reaction. 
Figure~\ref{fig:norbits} plots the number of orbits $N_{\rm orbits}$ traced by the binary system as a function of time, 
which is given by 
\beqa{
N_{\rm orbits} =
\left \{
\begin{array}{l}
{\displaystyle \frac{1}{2\pi}} \,\Omega_{\rm bin,0}\,t 
\ ({\rm if} \ t < t_{\rm shrink}) \,,
\\[2ex]
{\displaystyle \frac{1}{2\pi}} \,
\left[
\Omega_{\rm bin,0}\,t_{\rm shrink} + \phi(t-t_{\rm shrink})
\right]
\ ({\rm if} \ t \geq t_{\rm shrink}) \,.
\end{array}
\right.
}
We have here defined $\Omega_{\rm bin,0}=\Omega_{\rm bin}(r=a_0)$ to be the (constant) PN orbital frequency at a fixed semi-major axis,
and we have set $\phi(0)=0$ for the PN phase evolution. The number of orbits is obviously a piece-wise function since
when $t < t_\mathrm{shrink}$, $N_{\rm orbits}$ increases linearly, as the binary is artificially kept at fixed $a_{0}$, while when
$t > t_\mathrm{shrink}$, $N_{\rm orbits}$ can be approximated to leading order by
\beq{
N_{\rm orbits} \sim \frac{1}{64 \pi \nu}\left(\frac{a_0}{M}\right)^{5/2}
= \frac{1}{64 \pi}\left(\frac{a_0}{M}\right)^{5/2} \frac{(1+q)^2}{q}
\,.
}
Therefore, for \runtwo there are approximately $100$ total orbits, while for \runthree there are approximately $127$ orbits.

\section{Simulation Details}
\label{sec:simulation-details}

Like accretion disks around single black holes, circumbinary accretion flows are well described by the 
ideal MHD equations of motion (EOM) in the curved spacetime of only the black hole or holes.  We therefore 
neglect the matter's contribution to spacetime curvature and the accumulation of mass and momentum by the 
black holes from  gas accretion.  
Many codes have been written to simulate the single black hole case 
(e.g., \cite{1999ApJ...522..727K,dVH03,GMT03,Noble06,2005ApJ...635..723A,2005MNRAS.359..801K,2007MNRAS.379..469T,Noble09}), 
while only the equations of electrodynamics \citep{Pal09}, force-free MHD \citep{2010Sci...329..927P,2010PhRvD..82d4045P} 
and nonmagnetized hydrodynamics (e.g., \cite{2010ApJ...715.1117B,2010PhRvD..81h4008F,Farris11,Bode12}) have been 
solved in the relativistic circumbinary setting.  Unfortunately, these latter simulations employ methods, like 
block-structured adaptive mesh refinement (AMR) in Cartesian coordinates, that typically lead to poor conservation of 
fluid angular 
momentum and excessive dissipation at refinement boundaries.  These two effects alter the disk's angular momentum transport mechanism
and thermodynamics in a nontrivial way.   Furthermore, they require the solution of the Einstein equations, 
which---in turn---imposes a significant computational burden.   In order to avoid these problems, we take an alternate route and 
solve the MHD EOM using a code designed for single black hole systems:  \harm\ \citep{Noble09}.   
Fortunately, \harm\ was written to be almost independent of coordinate system or choice of spacetime, so modifying it to handle 
non-axisymmetric, time-dependent spacetimes was straightforward.  In fact, the only differences between the algorithm described
in \cite{Noble09} and here are that the metric (and its affine connection or gravitational source terms) 
needs to be updated every sub-step of the second-order Runge-Kutta time-integration procedure\footnote{Note that many other 
technical changes were made that do not affect the algorithm, but do affect the runtime efficiency and design of the code.}.
Below, we describe the equations solved, initial data setup and other details of the disk evolution. 

\subsection{MHD Evolution}
\label{sec:mhd-evolution}

Since we assume that the gas does not self-gravitate and alter the spacetime dynamics, we need only solve the GRMHD equations
on a specified background spacetime, $g_{\mu \nu}(x^\lambda)$, where $\left\{x^\lambda\right\}$ represents a set of general spacetime coordinates.  
The EOM originate from the local conservation 
of baryon number density, the local conservation of energy, and the induction equations from 
Maxwell's equations (please see \cite{Noble09} for more details).  They take the form of a set of conservation laws:
\beq{
\del_t \bU\left(\prim\right) = 
-\del_i \bF^i\left(\prim\right) + \mathbf{S}\left(\prim\right) \, 
\label{conservative-eq}
}
where $\bU$ is a vector of ``conserved'' variables, $\bF^i$ are the fluxes, 
and $\mathbf{S}$ is a vector of source terms.  Explicitly, these 
are 
\beq{
\bU\left(\prim\right) = \sqrt{-g} \left[ \rho u^t ,\, {T^t}_t 
+ \rho u^t ,\, {T^t}_j ,\, B^k  \right]^T
\label{cons-U}
}
\beq{
\bF^i\left(\prim\right) = \sqrt{-g} \left[ \rho u^i ,\, {T^i}_t + \rho u^i ,\, {T^i}_j ,\, 
\left(b^i u^k - b^k u^i\right) \right]^T
\label{cons-flux}
}
\beq{
\mathbf{S}\left(\prim\right) = \sqrt{-g} 
\left[ 0 ,\, 
{T^\kappa}_\lambda {\Gamma^\lambda}_{t \kappa} - \mathcal{F}_t ,\, 
{T^\kappa}_\lambda {\Gamma^\lambda}_{j \kappa} - \mathcal{F}_j ,\, 
0 \right]^T \, 
\label{cons-source}
}
where $g$ is the determinant of the metric, ${\Gamma^\lambda}_{\mu \kappa}$ is
the metric's affine connection, $B^i = \dF^{it}/\sqrt{4\pi}$ is our magnetic field (proportional to the field measured by observers traveling 
normal to the spacelike hypersurface), $\dF^{\mu \nu}$ is the Maxwell tensor, $u^\mu$ is the fluid's $4$-velocity, 
$b^\mu =  \frac{1}{u^t} \left({\delta^\mu}_{\nu} + u^\mu u_\nu\right) B^\nu$ is the magnetic $4$-vector or the magnetic field projected 
into the fluid's co-moving frame, and $W = u^t / \sqrt{-g^{tt}}$ is the fluid's Lorentz function.  
The MHD stress-energy tensor, $T_{\mu \nu}$,  is 
defined as 
\beq{
T_{\mu \nu} = \left( \rho h + 2 p_{m} \right) u_\mu u_\nu   + \left( p + p_{m}\right) g_{\mu \nu} - b_\mu b_\nu \label{mhd-stress-tensor}
}
where $p_m = b^\mu b_\mu / 2$ is the magnetic pressure, $p$ is the gas pressure, $\rho$ is the rest-mass density, $h = 1 + \epsilon + p/\rho$ 
is the specific enthalpy, and $\epsilon$ is the specific internal energy,
We evolve the quantity $( \rho u^t + {T^t}_t )$ instead of ${T^t}_t$ 
in order to reduce the magnitude of the internal energy's 
numerical error \citep{GMT03}.  
Note that the terms proportional to ${\Gamma^{\lambda}}_{t\kappa}$ and ${\Gamma^{\lambda}}_{\phi \kappa}$ in the source
no longer vanish as the metric is now dependent on time and azimuthal coordinate, $\phi$.  Also, note
that we add a negative source term $\left(-\mathcal{F}_\mu\right)$ to the local energy conservation equation to model 
energy/momentum loss from radiative cooling; please see Section~\ref{sec:thermodynamics} for more details. 

The MHD evolution is facilitated by calculating and using so-called primitive variables: 
the rest-mass density ($\rho$), the internal energy density ($u = \rho \epsilon$), the velocities relative to 
the observer moving normal to the spacelike hypersurface, $\tilde{u}^i = u^i - u^t g^{ti} / g^{tt}$.  The magnetic field $B^i$ is considered
both a primitive and a conserved variable.  We employ piecewise parabolic reconstruction of the primitive variables for calculating 
the local Lax-Friedrichs flux at each cell interface \citep{GMT03}.   We use a 3-dimensional version of the FluxCT 
to impose the solenoidal constraint, $\partial_i \sqrt{-g} B^i = 0$ \citep{2000JCoPh.161..605T}.  
The EMFs (electromotive forces) are calculated midway along each cell edge 
using piecewise parabolic interpolation of the fluxes from the induction equation. 
A second-order accurate Runge-Kutta method is used to integrate the EOM using the method of lines once
the numerical fluxes are found.  The primitive variables are found from the conserved variables using the ``2D'' scheme of \cite{Noble06}. 
A conservation equation for the entropy density is evolved and used to replace the total energy equation of the 2D method whenever the plasma 
becomes too magnetically dominated, or---specifically---when $\rho \epsilon < 0.02 p_{m}$; this procedure helps us avoid 
numerical instabilities and negative pressures from developing. Please see \cite{Noble09} for more details. 

The MHD evolution is performed in the same way as in single black hole cases except that the metric is evaluated\footnote{We remind the reader that the 
metric is known in closed form, requiring only direct evaluation except for the Newton-Raphson iteration to find the current time's binary separation.} 
at the present sub-step's time
before the MHD fields are updated.  The metric is required in many facets of the update 
procedure.  For example, it is used in calculating the $4$-velocity from the primitive velocities, source terms and geometric factors in the EOM, 
and for deriving the primitive variables from the conserved variables.  The affine connection is calculated via finite differencing the PN metric to evaluate 
\beq{
{\Gamma^\mu}_{\nu \kappa} = \frac{1}{2} g^{\mu \sigma} \left( \partial_\nu g_{\kappa \sigma} + \partial_\kappa g_{\nu \sigma}  - \partial_\sigma g_{\nu \kappa} \right)
\quad . \label{connection}
}
The spatial finite differences use fourth-order centered stencils away from the physical boundaries, and backward/forward stencils adjacent to the physical boundaries.  
Since the metric is evaluated and stored at the cell centers and faces, but the connection is only evaluated at the centers, fourth-order stencils require only 
three cells' worth of data to compute.  The time derivatives are second-order accurate, but use a time spacing $10^{-3}$ times that used in the MHD integration.  This means 
that additional evaluations of the metric are made at advanced and retarded times at the cell centers to  calculate the time derivatives for each connection evaluation.  
We have verified that the truncation error from the time derivatives is smaller than that from the spatial derivatives.  Also, the connection's spatial finite differencing is 
one order more accurate than that of the MHD procedure, implying it is not the primary source of error in the calculation.  Please see 
Appendix~\ref{sec:resolution-requirements} for a discussion on our resolution tests. 

\subsection{Initial Conditions}
\label{sec:initial-conditions}

In this project, we avoid evolving the gas in the neighborhood of the black holes, choosing instead to focus 
on establishing reasonable prior conditions for the gas that ultimately feeds the BBH. We therefore
excise a spherical domain, which includes the binary, from our calculation.  

It is common to begin with initial conditions devoid of 
large transient artifacts.  This is often done by starting from a torus of material in equilibrium (via pressure and rotational support) 
about the central gravitating source (e.g., a black hole) \citep{1985ApJ...288....1C,dVHK03}.  Unfortunately, such tori
will not be near equilibrium in our spacetime as it is $\left(t,\phi\right)$-dependent.  Plus, the equations describing their structure assume that the metric 
has the same form (i.e. share the same zero-valued elements) as the Kerr metric in Boyer-Lindquist coordinates.  
We resolve these issues in the following way.  First, since we hold the binary at fixed separation for several orbits, 
the spacetime initially has a helical Killing symmetry, with Killing vector 
$\mathcal{K}^a = \Omega_\mathrm{bin} \left(\partial_\phi \right)^a + \left(\partial_t \right)^a$.  
In other words, the spacetime is invariant in a frame rotating with the binary, 
while the separation is held constant.   Since the torus will lie a few $a_0$ away from the binary, its dynamical response time---comparable to its orbital
period---will be longer than the binary period, implying that a torus near equilibrium in this helically-symmetric spacetime will also be 
near equilibrium in its time average.  Due to its helical symmetry, its time average is also its azimuthal average.  We therefore start with a
torus in equilibrium in a $\left(t,\phi\right)$-independent spacetime, $\hat{g}_{\mu \nu}$, found by averaging over $\phi$:
\beq{
\hat{g}_{\mu \nu} = \frac{ \int g_{\mu \nu} \sqrt{g_{\phi \phi}} \, d\phi }{ \int \sqrt{g_{\phi \phi}} \, d\phi } 
\quad . \label{phi-average-metric}
}
We have verified that the same components that are zero-valued in Boyer-Lindquist coordinates are 
consistent with zero to within our PN-order accuracy in the $\hat{g}_{\mu \nu}$ metric.  This means we can employ a similar 
torus solution method as described in \cite{1985ApJ...288....1C}.  A description of our modifications to the procedure---including the generalization 
to our $\phi$-averaged spacetime---is provided in Appendix~\ref{sec:hydr-torus-solut}.  
Note that we now ensure that the equilibrium solution is found iteratively to greater precision, instead of the 
approximate method described in \cite{dVHK03} which has been used in prior work of the authors in single black hole disk evolutions \citep{Noble09,Noble10} and by 
others studying the hydrodynamic circumbinary  case \citep{Farris11}.   We find that our procedure produces initial tori that are much closer to equilibrium than 
the approximate scheme.  Please see Appendix~\ref{sec:hydr-torus-solut} for more details. 
 
Previous studies have shown that a gap develops near $2.5a$ for equal mass binaries \citep{MM08,Shi11}.  We aim to study how this gap develops, so we choose
to start material outside this radius.  We therefore set up a disk with inner edge located at $\rin = 3 a_0$ and pressure maximum located at $\rpmax = 5 a_0$; 
from prior experience, $\rpmax$ is approximately the radius at which the disk transitions from accreting to decreting since matter must shed its angular 
momentum to fluid elements further out in order to accrete.  These outer elements gain angular momentum and form a time-averaged decretion flow away from the 
central potential.   We will therefore focus on $r < \rpmax = 5 a_0$ in our analyses.   The initial disk extends to $\rout \simeq 12 a_0$, is isentropic 
with $p/\rho^\Gamma = 0.01$, and is tuned to 
have an aspect ratio of $H/r=0.1$ at $r=\rpmax$, where $H$ is the density scale height defined as the 
first moment of the rest-mass density with respect to distance from the midplane:
\beq{
H \equiv \frac{\langle \rho \sqrt{g_{\theta\theta}} \, |\theta - \pi/2| \rangle}{\langle \rho \rangle} \quad ,
\label{scaleheight}
}
and where the $\langle X \rangle$ denotes the average over origin-centered spheres: 
\beq{
\langle X \rangle  \equiv  \frac{\int X \, \sqrt{-g} \, d\theta d\phi}{\int \sqrt{-g} \, d\theta d\phi}  \quad . 
\label{shell-average-def}
}
More information about the initial torus and its solution method is
given in Appendix~\ref{sec:hydr-torus-solut}.   In the disk, we add random, 
cell-scale noise to the the internal energy, $u$, in order to hasten the development of turbulence; the 
random noise is evenly distributed over the range $\pm 5\times10^{-3}$. 

Once the torus is in place on the grid, 
a surrounding nonmagnetized atmosphere is added as our numerical scheme requires us to 
maintain positive values of $\rho$ and $p$.  The atmosphere
is initially static, $u^i=0$, and in approximate pressure equilibrium: 
$\rho_\mathrm{atm} = 1 \times 10^{-7} \rho_\mathrm{max} \left(r/M\right)^{-3/2}$, 
$u_\mathrm{atm} = 3.3\times 10^{-6} \, u_\mathrm{max} \left(r/M\right)^{-5/2}$, 
where $\rho_\mathrm{max}$ and $u_\mathrm{max}$ are---respectively---the initial maxima of $\rho$ and $u$. 
We note that when either $\rho$ or $u$ are 
found to go below, respectively, $\rho_\mathrm{atm}$ or $u_\mathrm{atm}$, they are set to those atmosphere values without any 
modification to the magnetic field or fluid velocity; this happens very rarely once the disk's turbulence saturates.  

The magnetic field is initialized as a set of dipolar loops that follow density contours in the disk's interior.  We set the 
azimuthal component of the vector potential and differentiate it to yield $B^i$; $B^\phi(t=0) = 0$ in our configuration.  
The vector potential component is 
\beq{
  A^\phi = A^\phi_0 \, \max\left[ \left(\rho - \frac{1}{4}\rho_\mathrm{max} \right) , 0 \right] \quad . \label{initial-vector-potential}
} 
The magnitude of the field, $A^\phi_0$, is set such that the ratio of the disk's 
total internal energy to its total magnetic energy is $100$. 

\subsection{Grid, Boundary Conditions, and Parameters}
\label{sec:grid-bound-cond}

The domain on which the 
MHD EOM are solved is a uniformly discretized space of spatial coordinates $\left\{\xx{i}\right\}$ that are isomorphic to 
spherical coordinates $\left\{r,\theta,\phi\right\}$:
\beq{
r(\xone) = M e^{\xone}  \quad , \label{r-to-x1} 
}
\beq{
\theta(\xtwo) = 
\frac{\pi}{2} \left[  1  + \left(1-\xi \right) \left(2 \xtwo - 1 \right) 
+ \left( \xi - \frac{2 \theta_c}{\pi} \right) \left( 2 \xtwo - 1 \right)^n \right] 
\quad ,  \label{theta-x2-2-def}
}
and $\phi = \xthree$.  We set $n=9$, $\xi=0.87$, and $\theta_c=0.2$.  
The logarithmic radial coordinates are such that the radial cell extents are smaller at smaller radii in order to 
resolve smaller scale features of the accretion flow there.  The $\xtwo \leftrightarrow \theta$ mapping concentrates more cells near the plane of the disk and the binary's orbit, 
the equator of our coordinate system.   Let each grid cell in our numerical domain be labeled by three spatial indices that each cover $[0,\NN{n}-1]$, where 
$\left\{\NN{n}\right\}$ are the number of cell divisions along each dimension.  A cell with indices $\left(i,j,k\right)$ is 
located at $\left(\xone_{i}, \xtwo_{j}, \xthree_{k} \right)$, where 
$\xx{n}_{j} = \xx{n}_{b} + \left(j + \frac{1}{2} \right) \Delta \xx{n}$.   The grid we used is completely specified by: 
$\xone_{b} = \ln\left(\Rmin/M\right)$, $\Delta \xone = \ln\left(\Rmax/\Rmin\right)/\None$, $\Rmin=15M$, $\Rmax=260M$, $\None = 300$, 
$\xtwo_{b} = 0$, $\Delta \xtwo = 1/\Ntwo$, $\Ntwo = 160$, 
$\xthree_{b} = 0$, $\Delta \xthree = 2\pi/\Nthree$, $\Nthree = 400$. 

We chose our resolution and grid extent based upon a number of criteria.  First, our $\theta$ and $\phi$ resolutions were set 
in order to adequately resolve the MRI based on guidelines of \cite{hgk11} and \cite{2011arXiv1106.4019S}.  The radial resolution was chosen 
to resolve the spiral density waves---generated by the binary's time-varying tidal field---by several radial zones.   
We find that our grid adequately resolves the MRI, as measured
by the criteria of \cite{hgk11} and \cite{2011arXiv1106.4019S}, 
throughout the domain of interest (i.e. $r < 5a_0$) $\forall t$. 
Please see Appendix~\ref{sec:resolution-requirements} for a quantitative 
description of these resolution criteria and for a demonstration of how well we resolve the MRI. 

The radial extent of the grid was inspired by \cite{Shi11} and the limits of our near-zone PN metric.  
Since the near-zone PN metric that we use is only valid at distances more than $10 M_i$ from the black hole with mass 
$M_i$ \citep{Yunes:2006iw,Yunes:2005nn,JohnsonMcDaniel:2009dq}, then---in the equal mass case considered here---we have
$\Rmin \ge 10 \frac{M}{2} + \max(a(t))/2 = 5M + a_0/2 = 15M$.  
\cite{Shi11} found that an inner radial boundary located at $\Rmin \lesssim 1.1 a$ was sufficiently 
far away from the gap and deep in the potential as to not significantly alter the development and evolution of the surface density peak at the edge of 
the gap.  These two constraints justify our choice of $\Rmin = 15M = 0.75 a_0$ and suggest that our inner boundary condition 
may begin affecting the gap's evolution when 
$a(t) \lesssim \Rmin/1.1 \simeq 13.6M \simeq 0.68 a_0$, which occurs after approximately $t=51235M$ in \runtwo.    We set  $\Rmax=260M=13a_0$ to 
encompass the initial torus.  

All cells are advanced in time with the same time increment ($\Delta x^0 = \Delta t$), which itself changes in 
time; $\Delta x^0$ is set to $0.45 \Delta t_\mathrm{min}$, where $\Delta t_\mathrm{min}$ is the shortest 
cell crossing time of any MHD wave over the entire domain. 

Boundary conditions were imposed through assignment of primitive variables and $B^i$
in ghost zones. 
Outflow boundary conditions are imposed at $r=\Rmin$ and $r=\Rmax$ which amounts to extrapolating the
primitive variables at $0^\mathrm{th}$-order into the ghost zones.  Additionally, 
$u^r$ is set to zero---and $\tilde{u}^i$ recalculated---whenever
it points into the domain at $r=\Rmax$.  We note that we attempted to 
implement a similar condition on $u^r$ at $r=\Rmin$, but found it to be unstable 
during the earliest part of the simulation.  Even though it was successfully used in \cite{MM08} and \cite{Shi11}, 
we found that this condition was inconsistent with the tendency of negative radial pressure
gradients developing ahead of each black hole as it moved around its orbit.  This pressure gradient moves
small amounts of material onto the grid, elevating the density just above the floor ahead of the black holes.  
Even without the special condition on $u^r$ at $\Rmin$ and just using $0^\mathrm{th}$-order extrapolation of the primitive 
variables there, we observe insignificant 
amounts ($\ll 1\%$ of the total) of positive mass flux into the domain there and 
a nearly flat $\dot{M}(r,t)$ profile over  $\Rmin < r  < 2 a$ with no noticeable artifacts near $\Rmin$ (e.g., Figure~\ref{fig:accrate_radius}). 

\subsection{Thermodynamics}
\label{sec:thermodynamics}

Depending on internal properties of the gas (e.g., density, accretion rate), the disk may or may not be 
optically thin, geometrically thin, or have a constant aspect ratio $H/r$.  As these disk characteristics are
sensitive to the assumed initial conditions and thermodynamics of the system, we therefore must select which kind of disk to 
model \emph{a priori} and verify the consistency of our assumptions \emph{a posteriori}.  We chose to model a disk 
with intermediate thickness ($H/r=0.1$) in order to both address the fact that binary's torque will likely heat the gas 
efficiently and our expectation that the disk will be dense, optically thick and radiating efficiently. 

We chose the ideal-gas ``$\Gamma$-law'' equation of state to close the MHD EOM:
$P = \left(\Gamma - 1\right) \rho \epsilon$.  We set $\Gamma = 5/3$, which reasonably well describes the behavior of
a plasma whose specific thermal energy is smaller than an electron's rest-mass energy (i.e. it is not relativistically hot). 
The gas is cooled to a target 
entropy,
the initial entropy of the disk, at a rate equal to 
\beq{
\mathcal{L}_c = \frac{\rho \epsilon}{T_\mathrm{cool}} \left( \frac{\Delta S}{S_0} + \left|\frac{\Delta S}{S_0}\right| \right)  \quad , 
\label{cooling-function}
}
where $\Delta S \equiv S - S_{0}$ and $T_\mathrm{cool} =  2 \pi \left(r/M\right)^{3/2}$ is the cooling time, which we approximate as 
the Newtonian period of a circular equatorial 
orbit at radius $r$.  
Our procedure is similar to those used by \cite{Noble09} and \cite{2010MNRAS.408..752P}. 
The term in the parentheses acts as a switch ensuring that $\mathcal{L}_c \ge 0$ always, and is zero
when the local entropy, $S=p/\rho^\Gamma$, is below the target entropy, $S_0 = 0.01$, which is the constant value used 
in the initial data's torus.  
Hence, the cooling function should release any heat generated through dissipation 
since the initial state.    We do not cool unbound material---i.e. fluid elements that satisfy $\left( \rho h + 2p_{m} \right)u_t < - \rho$---since
we do not want to include cooling that results from application of density or pressure floors.
Since $\mathcal{L}_c$ is the cooling rate in the 
local fluid frame, its implementation in the EOM must be expressed in the coordinate frame:
\beq{
\mathcal{F}_\mu = \mathcal{L}_c u_\mu \quad . \label{cooling-flux}
}

Another advantage of the cooling function is that it provides us with a proxy for bolometric emissivity that is consistent with 
the disk's thermodynamics---unlike \textit{a posteriori} estimates of synchrotron and/or bremsstrahlung luminosity that have 
typically been made in numerical relativity simulations (e.g., \cite{2010ApJ...715.1117B,2010PhRvD..81h4008F,Farris11}). 
We will use $\mathcal{L}_c$ to make predictions of the total luminosity from circumbinary disks.  These predictions are made 
by integrating $\mathcal{L}_c$ over the domain in the coordinate frame;  we expect to verify their accuracy using 
full GR ray-tracing in future work. 


\section{Results}
\label{sec:results}

\subsection{Approximate Steady State}

\begin{figure}[htb]
\centerline{
\begin{tabular}{cc}
 \includegraphics[width=8cm]{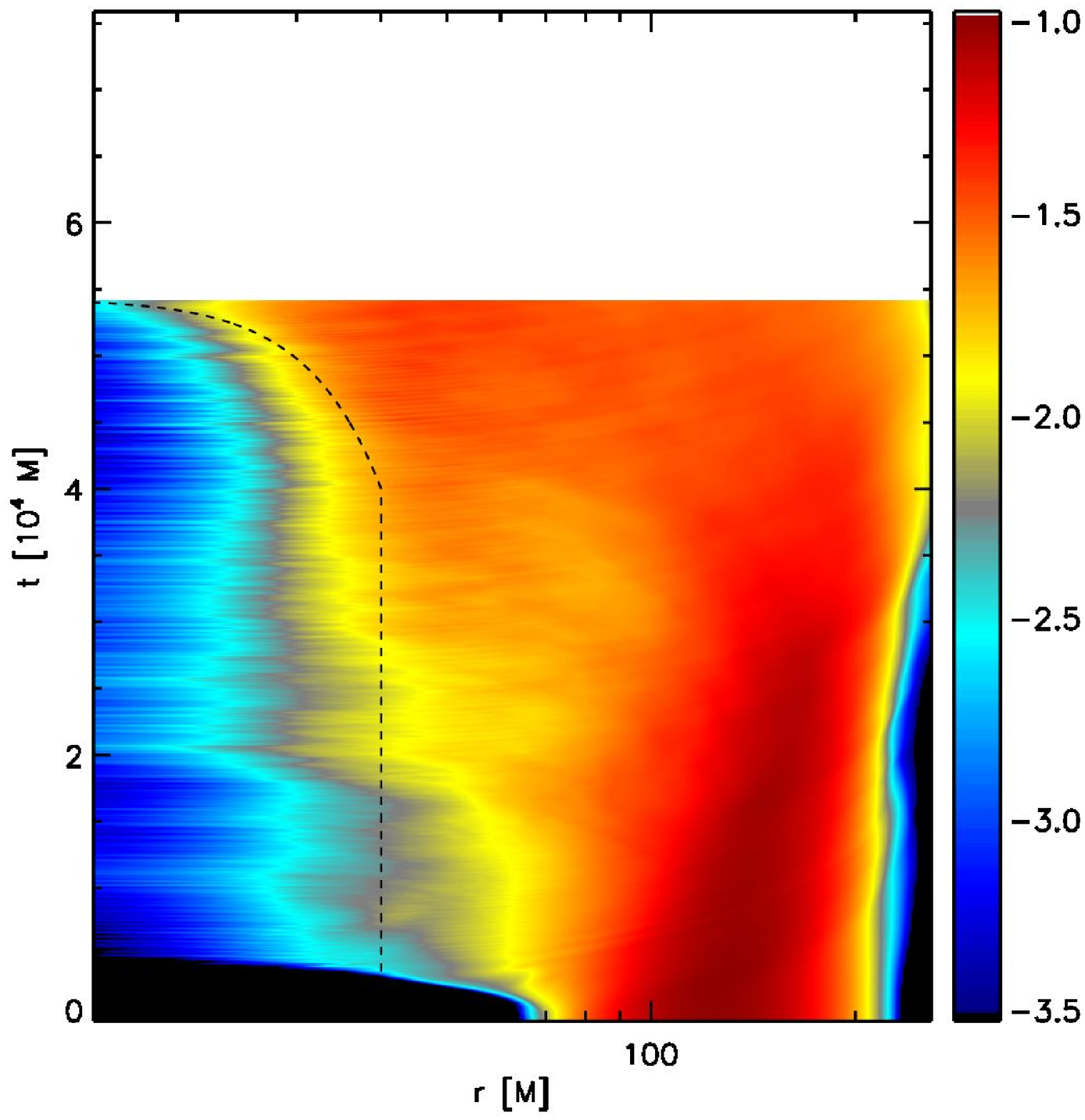}
& \includegraphics[width=8cm]{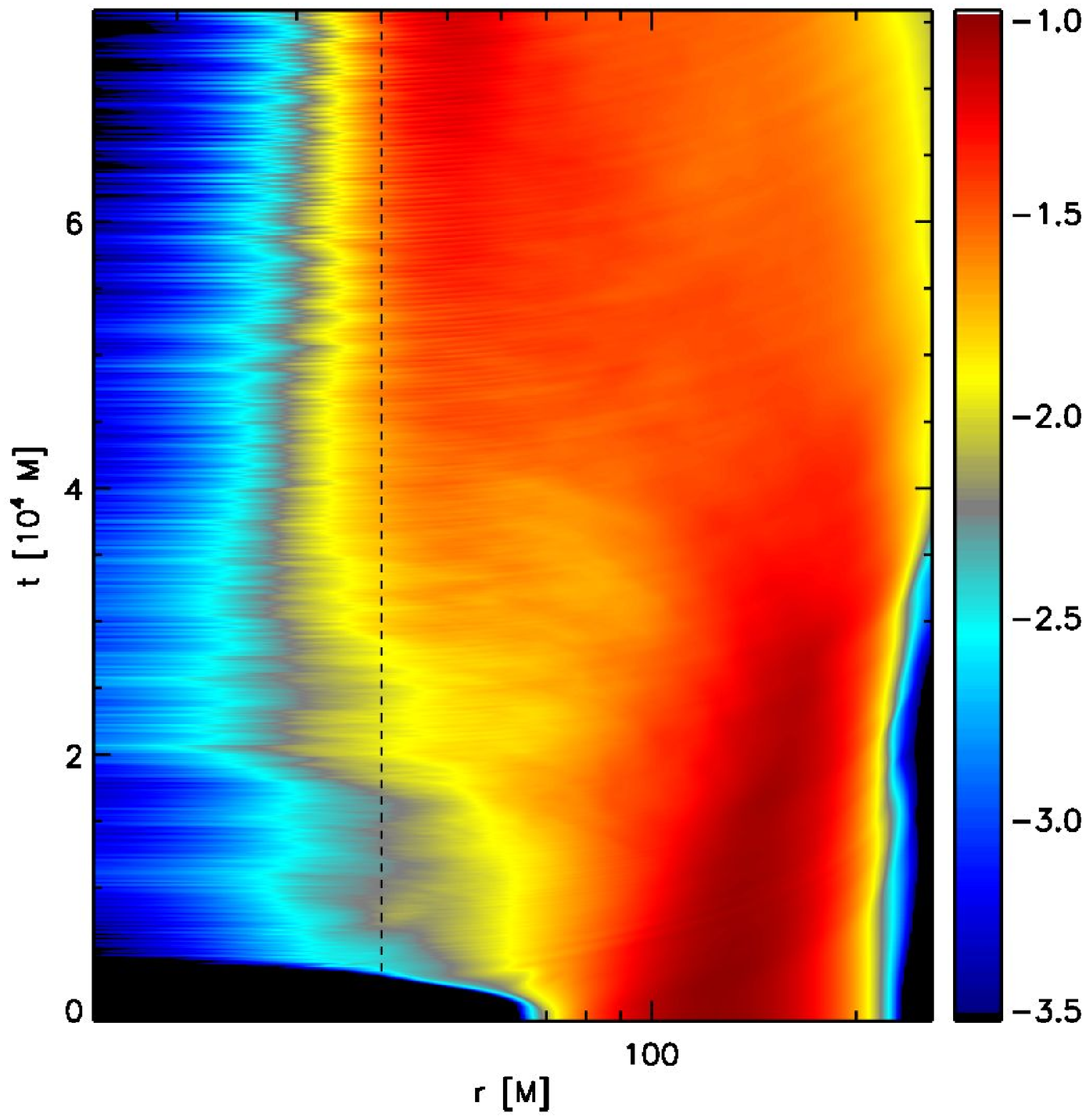}
\end{tabular}
}
\caption{Color contours of $\log \Sigma(r)$ as a function of time.  The scale is shown
in the color bar.  The black dashed curve shows $2a(t)$.  (Left) \runtwo . (Right) \runthree .}
\label{fig:surfdens_r_t}
\end{figure}

    At the beginning of both simulations, orbital shear transforms part of the radial component
of the magnetic field to toroidal, creating a laminar Maxwell stress.  Meanwhile, in the same
region, the magnetorotational instability grows, its amplitude exponentially growing on the local
dynamical timescale, $\simeq 500M$ at the initial inner edge of the disk, $r = 60M$.  The
turbulence in the inner disk reaches nonlinear saturation at $t \simeq 10000M$.  Under the
combined influence of the initial laminar and later turbulent Maxwell stress, matter flows
inward (see Figure~\ref{fig:surfdens_r_t}).

\subsubsection{Surface density}

    Soon after $t \simeq 10000M$, the inward flow begins to pile up at $r \simeq 50M$, between
two and three times the binary separation (the dashed line in both panels of
Figure~\ref{fig:surfdens_r_t} marks the location of $2a(t)$ in order to guide the eye).
We define the surface density $\Sigma$ as
\begin{equation}
\Sigma(r,\phi) \equiv \int \, d\theta \sqrt{-g} \rho / \sqrt{g_{\phi\phi}(\theta=\pi/2)};
\label{sigma}
\end{equation}
when we quote it as $\Sigma(r)$, that denotes an azimuthal average of equation~(\ref{sigma}).  
In later discussion,
we will sometimes normalize the surface density to $\Sigma_0$, the maximum surface density
in the initial condition; in code-units $\Sigma_0 =0.0956$.  In \runthree, $\Sigma(r \sim 2a)$
grows steadily for the duration of the simulation, but after $t \simeq 20000M$,
the logarithmic rate of growth (i.e., $d\ln\Sigma(r)/dt$) gradually becomes slower and slower.
Because a number of azimuthally-averaged properties like $\Sigma(r)$ all become steadier
after $t=40000M$, we call the period from then until the end of \runthree the
``quasi-steady epoch".  For the same reason, we began the binary orbital evolution of \runtwo
at that time.

\begin{figure}[htb]
\centerline{
\begin{tabular}{cc}
 \includegraphics[width=8cm]{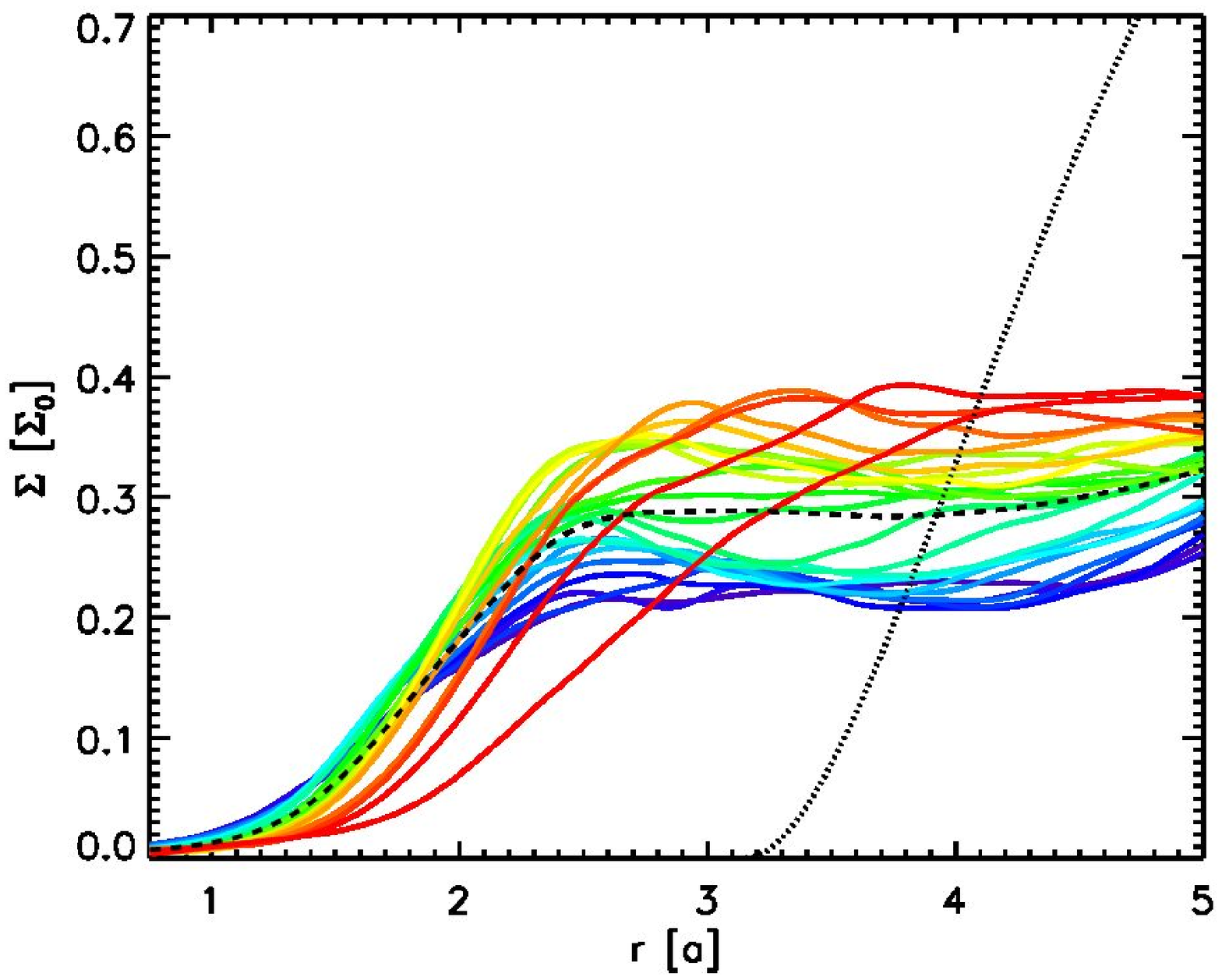}
&\includegraphics[width=8cm]{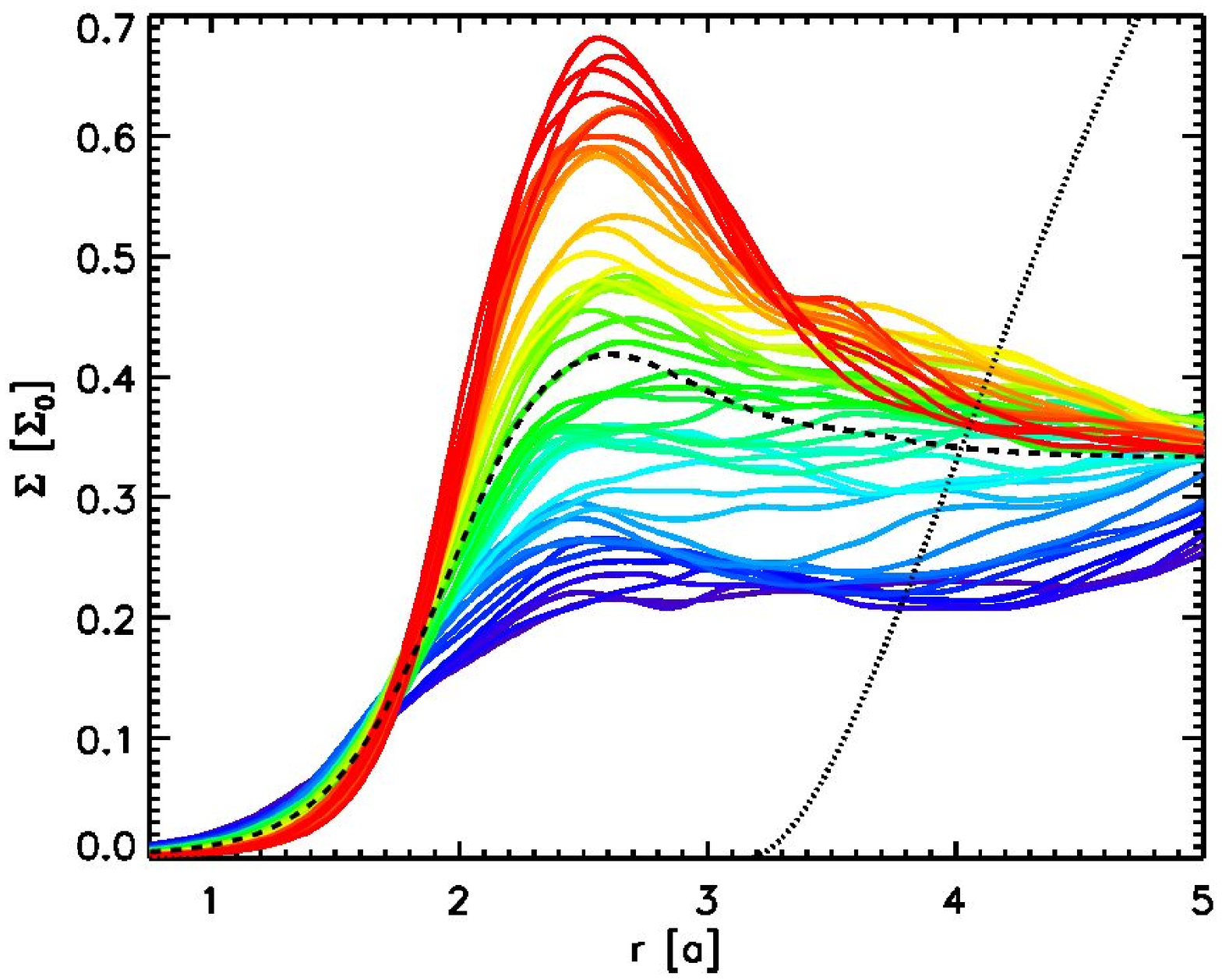}
\end{tabular}
}
\caption{$\Sigma(r/a)$ every $1000M$ in time from $t = 30000M$ to the end of the
simulation.  Time increases from violet color to red.  The dotted curve shows the
initial condition.  The dashed curve shows the average of the colored curves.
(Left) \runtwo, where the time span extends to $53000M$.  Note
that in this simulation $a$ decreases after $t=40000M$, so that a fixed value of
$r/a$ corresponds to a progressively smaller radial coordinate after that time.
(Right) \runthree, where the time span extends to $76000M$.  The binary separation
is fixed throughout this simulation.}
\label{fig:surfdens_r}
\end{figure}

Once this quasi-steady state is reached, $\Sigma(r)$ rises sharply from the inner boundary at
$r=16M$ to $r\simeq 50M$, initially $\propto r^{2.5}$, but at late times in \runthree ,
$\propto \exp(3r/a)$ (Figure~\ref{fig:surfdens_r}).  At first, the azimuthally-averaged
surface density profile forms a relatively flat plateau at
radii greater than $50M \simeq 2.5a$, but by $t=30000M$, a distinct local maximum appears
at $r \simeq 50M$ and persists for the remainder of the simulation.  This maximum is
noticeably asymmetric in the sense that $|d\Sigma/dr|$
is always considerably smaller in the disk body (i.e., $r > 2.5a$) than in the gap region
inside $r=2a$ (Figure~\ref{fig:surfdens_r}).  This behavior resembles closely
what has previously been seen  in the Newtonian regime (e.g., \cite{MM08,Shi11}).

By construction, the behavior of \runtwo is identical to that of \runthree up to $t=40000M$,
when the binary inspiral was begun.  In fact, at large radius, the behavior of the surface
density profile in \runtwo continues to be very similar to that of \runthree even after the
binary begins to shrink.  Near the surface density peak and at smaller radii, however, things
change.  In \runtwo, the location of the peak moves inward as the binary becomes smaller, and
the slope of the disk's inner edge becomes noticeably shallower as the inspiral accelerates.
Comparing the curve of the dashed line in the \runtwo panel of Figure~\ref{fig:surfdens_r_t}
to the curving edge of the colors denoting higher surface density, one can see that the location
of the disk's inner edge follows the evolution of the binary until shortly before the end of
the simulation.

\begin{figure}[htb]
\centerline{
\begin{tabular}{cc}
\includegraphics[width=8cm]{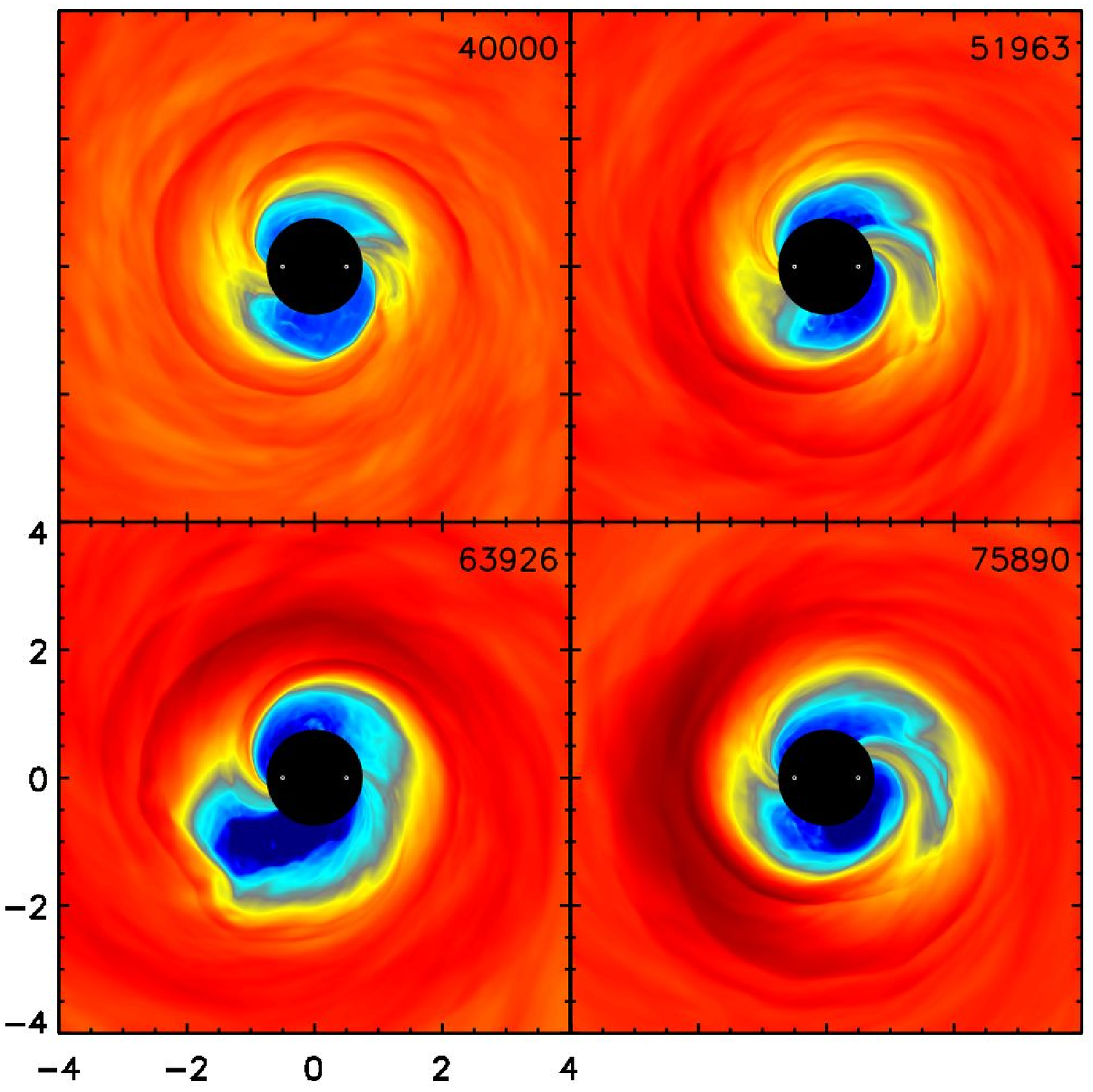}
&\includegraphics[width=8cm]{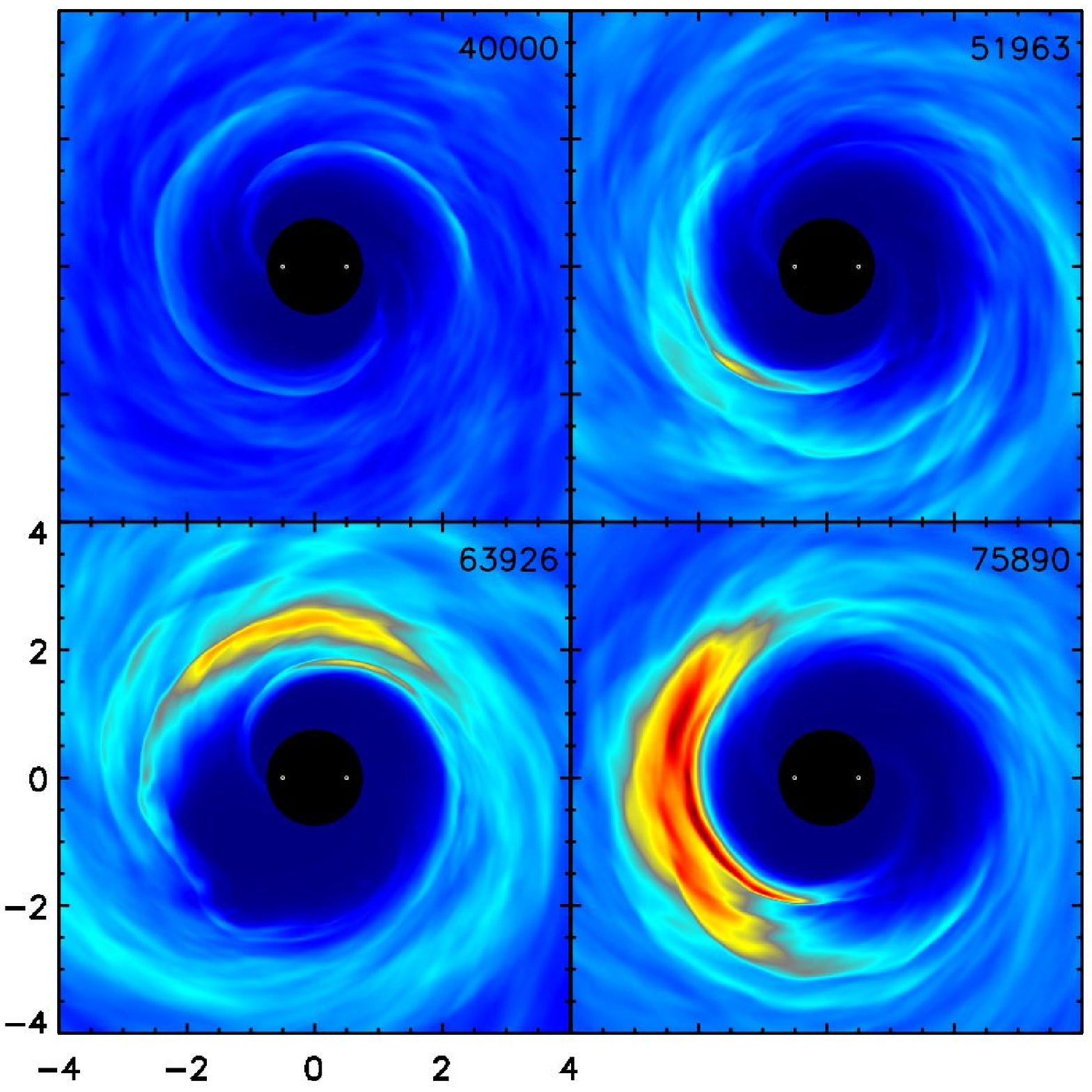}\\
\includegraphics[width=7cm]{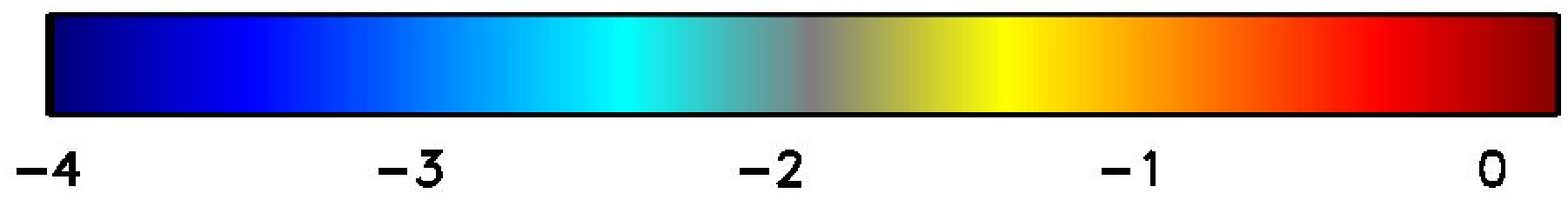}
&\includegraphics[width=7cm]{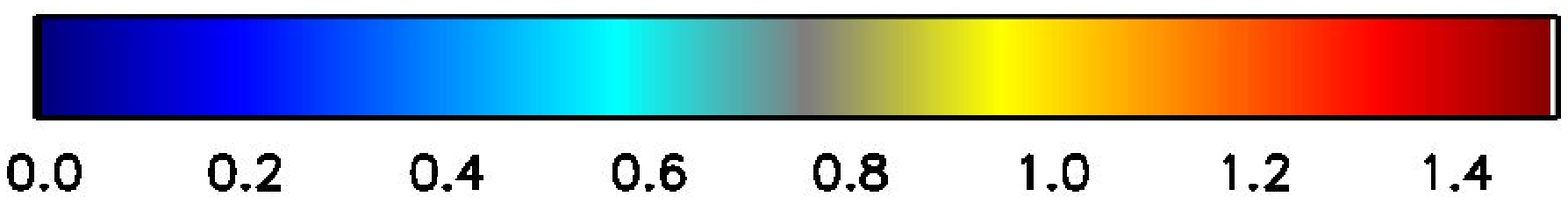}
\end{tabular}
}
\caption{Color contours of surface density in units of $\Sigma_0$ as a function of radius and azimuthal
angle in \runthree at four different times in two different scales: (Left) Logarithmic
color scale emphasizing the streams from the disk toward the binary members. (Right) Linear
color scale emphasizing the growth of asymmetry in the inner disk.  In both panels, the times
shown are $t=40000M$ (upper-left), $t=51963M$ (upper-right),
$t=63926M$ (lower-left), and $t=75890M$ (lower-right).}\label{fig:surfdens_equatorial}
\end{figure}

     However, speaking in terms of azimuthally-averaged surface density obscures an important
aspect of circumbinary disks: near and inside their inner edges, their structure is generically
far from axisymmetric.  In Figure~\ref{fig:surfdens_equatorial}, we show $\Sigma(r,\phi)$ at
$t = 40000M$.  As mentioned previously, at radii smaller than $\simeq 2a \simeq 40M$, there is
relatively little matter.  The reason this gap forms is that, unlike a time-steady,
axisymmetric potential, the time-dependent quadrupolar potential of the binary does not
conserve either the energy or the angular momentum of test-particles.  Consequently, closed
orbits do not exist, and torques driven by the binary can rapidly expel some matter to the
outside, while matter on other trajectories can be forced inward \citep{Shi11}.  As a result,
even though the rate at which matter enters the gap is comparable to the outer-disk accretion
rate, at any given time, relatively little matter can be found in the region within $\simeq 2a$,
and the matter that is present follows trajectories with little resemblance to stationary
circular orbits.  Instead, a pair of streams leave the inner edge of the disk and curve inward
toward each member of the binary.  Part of their flow gains enough angular momentum to return to
the disk, but part crosses the inner simulation boundary, traveling toward the domain of the
binary.

In addition, several tens of binary orbits after matter begins to pile up at $r \simeq 2.5a$,
a distinct ``lump" (as \cite{Shi11} called it) forms in the region of the surface density peak.
The density contrast between this lump and adjacent regions grows steadily in time.  Thus,
despite the relatively slow variation
of azimuthally-averaged disk properties during the period we call ``quasi-steady", the ``lump"
continues to evolve secularly.  Although \runthree was not continued long enough to
see this effect, \cite{Shi11} found that the eccentricity of the lump's orbit also grows slowly.

\subsubsection{Accretion rate, internal stresses, and angular momentum budget}

\begin{figure}[htb]
\centerline{
\includegraphics[width=8cm]{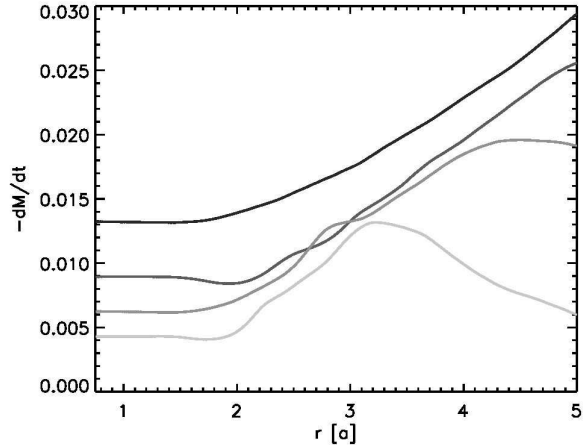}
}
\caption{Time-averaged accretion rate during four equally spaced segments from $t=30000M$ (black) till 
the end of \runthree.  The curves becomes progressively lighter in shade as time advances. }
\label{fig:accrate_radius}
\end{figure}

     It is also useful to characterize the global dynamics of circumbinary disks in terms of
the radial dependence of the net mass flow, i.e., the accretion rate as a function of radius.
We show in Figure~\ref{fig:accrate_radius} how this quantity slowly evolved during the quasi-steady
epoch of \runthree by dividing the time from $30000M$ until the end of the simulation
at $76000M$ into four segments and averaging over each one separately.  The accretion rate is constant
as a function of radius only inside the gap region, at most times increasing gradually outside
$r \simeq 2a$.  During the first part of this period, the accretion rate rises steadily to radii
beyond $5a$, but after $t \simeq 50000M$, the accretion rate in the outer disk gradually falls.
At the end of the simulation, $\dot M(r)$ is actually about a factor of 2 greater at $r \simeq 3a$
than anywhere else.  Averaging over the entire quasi-steady epoch, the rate at which mass passes
through the inner boundary is a bit less than half the accretion rate at $r=5a$.
Although the first analytic theories of circumbinary disks
\citep{P91} assumed that {\it no} accretion would pass the inner edge of such a disk,
Newtonian simulations, both purely hydrodynamic \citep{MM08} and MHD \citep{Shi11}, have
generally seen leakage fractions of a few tens of percent; our fraction is thus only
somewhat greater than previously found.

\begin{figure}[htb]
\centerline{
\begin{tabular}{c}
\includegraphics[width=12cm]{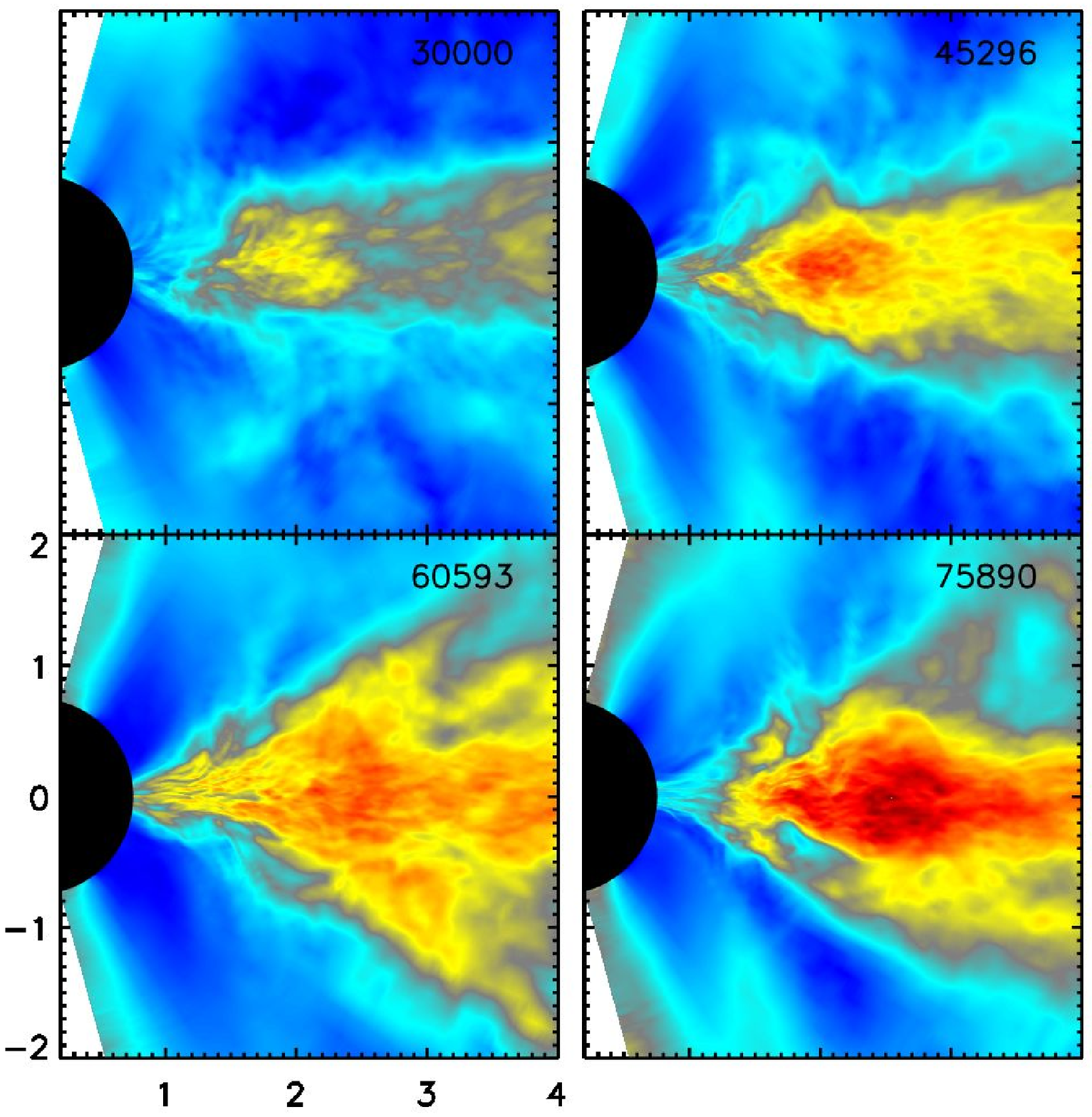}\\
\includegraphics[width=10cm]{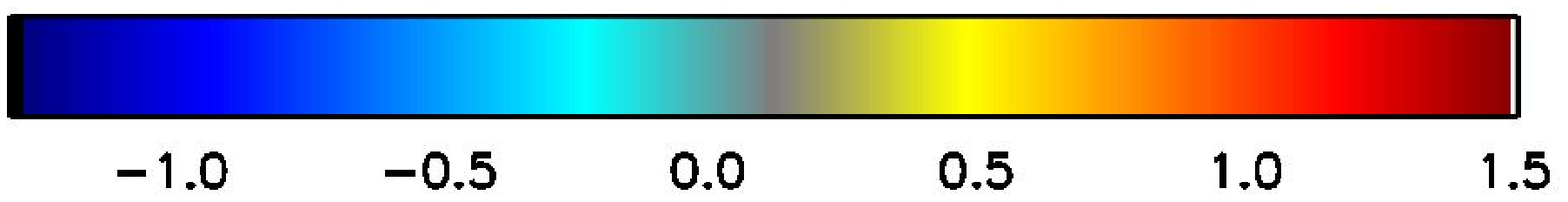}
\end{tabular}
}
\caption{Log$_{10}$ of the azimuthally-averaged plasma $\beta$ parameter at four times during \runthree.}
\label{fig:plasma_beta}
\end{figure}

     As mentioned earlier, Maxwell stresses due to correlations induced in MHD turbulence by
orbital shear dominate angular momentum transport within accretion disks.  Because the ratio
of Maxwell stress to magnetic pressure, $2\langle B^{(r)} B_{(\phi)} \rangle/\langle B^2\rangle$,
is fixed \citep{hgk11} at $\simeq 0.3$--0.4 in a point-mass potential (here the notation
$X^{(\mu)}$ denotes the magnitude of the $\mu$-component of four-vector {\bf X} projected into
the fluid frame), the stress is linearly proportional to the magnetic pressure.  A useful measure
of the strength of magnetic effects is therefore the plasma
$\beta \equiv \langle p \rangle /\langle B^2\rangle$.
In most previous accretion disk simulations, this quantity is $\sim 100$ in the midplane and
drops to $\sim O(1)$ a few scale-heights out of the plane.  We show its dependence on position in
the poloidal plane at several times during the quasi-steady epoch of \runthree in
Figure~\ref{fig:plasma_beta}; to be more precise, we show
the ratio of the time- and azimuthally-averaged gas pressure to the similarly averaged magnetic
pressure.  As that figure illustrates, the level of magnetization is rather larger than
usual (i.e., $\beta$ is smaller than usual), but gradually diminishes over time.   At
$t=30000M$, $\beta \simeq 1$ in the midplane at $r \sim 3$--$5a$ and $\simeq 3$ in the region
of the surface density peak ($r \sim 2$--$3a$); by the end of the simulation, it is
$\gtrsim 10$ in the disk body for the whole range $2a < r < 5a$ and reaches as much as
$\simeq 30$ in the lump.

Ever since the work of \cite{SS73}, it has been popular to measure the vertically-integrated,
azimuthally-averaged, and time-averaged internal disk stress in units of the similarly integrated
and averaged pressure.  In order to avoid unphysical pressures found in the unbound regions,
we computed the stresses and pressures only in bound material.  Outside $r \sim 4a$, where the
disk resembles an ordinary accretion
disk, we find that the Maxwell stress alone has magnitude $\simeq 0.3$--0.5 in these units.
This is roughly 3--5 times larger than the stress levels found in general relativistic
simulations of MHD flows in the Kerr metric \citep{KHH05}.  In the gap region, the ratio of
Maxwell stress to pressure rises about a factor of 2, while the Reynolds stress in the gap rises
dramatically (as also found by \cite{Shi11}).  These large Reynolds stresses are entirely due
to the strong binary torque, which pushes part of the inflowing streams back out to the disk
with additional angular momentum.

\begin{figure}[htb]
\centerline{
\includegraphics[width=12cm]{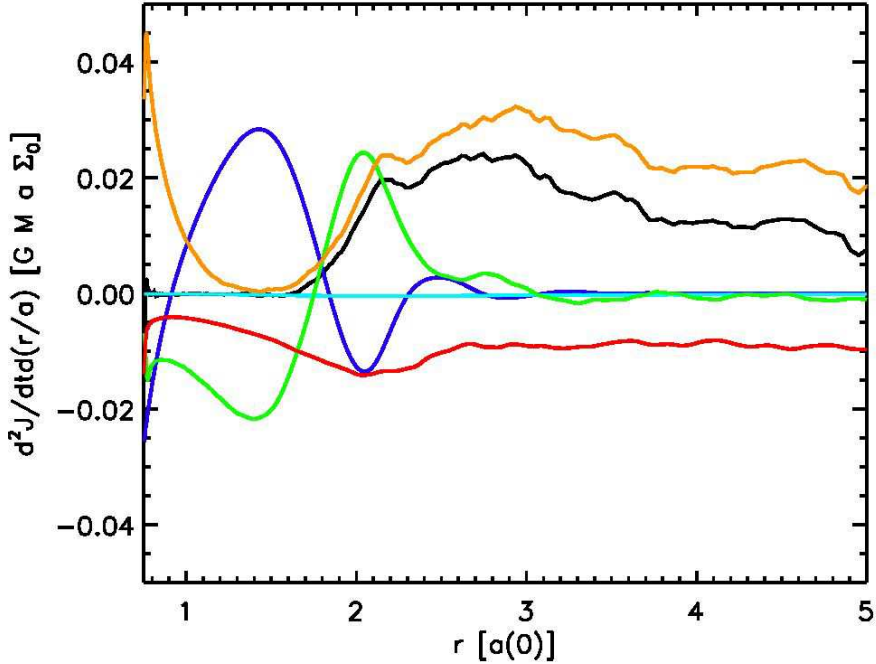}
}
\caption{Radial derivatives of the angular momentum flux due to shell-integrated Maxwell stress
in the coordinate frame (red), the angular momentum flux due to shell-integrated Reynolds stress
in the coordinate frame (green), and advected angular momentum (gold).  Torque densities per
unit radius due to the actual binary potential and radiation losses are shown by blue and
cyan curves, respectively.  The net rate of change of angular momentum $\partial_r \partial_t J$
(solid black).  All quantities are time-averaged over the quasi-steady epoch in \runthree .}
\label{fig:djdtdr}
\end{figure}

An overview of angular momentum flow in the system can be gleaned from Figure~\ref{fig:djdtdr},
in which we show the radial derivatives of the time-averaged angular momentum fluxes integrated on
shells, i.e., the time-averaged torques due to the several mechanisms acting.  Several important
points stand out in this figure.  The first is that the binary torques are delivered primarily
in the gap region $a \lesssim r \lesssim 2a$.  The torque density $dT/dr$ peaks at
$r \simeq 1.45a$, and the region surrounding that peak dominates the integral over all radii.
Moreover, all these torques are positive in net,
but they are locally negative both at small radii ($r \lesssim a$) and at large ($r \gtrsim 1.9a$).
Thus, most of the angular momentum the binary gives the disk is delivered in the gap, where
the gas density is very much lower than in the disk proper.  This point has previously been
emphasized by \cite{Shi11}.  Second, that angular momentum is conveyed to the disk proper
by fluid flows, i.e., Reynolds stresses.  That is why the Reynolds stress is large and positive
from $r \simeq 1.8a$ to $r \simeq 2.5a$.  Outside those regions, Maxwell stress, which always
acts so as to remove angular momentum from the gas and carry it outward, dominates the internal
stresses.  Finally, the net angular momentum change at any given radius is generally positive in
the inner disk because matter continues to pile up between $r \simeq 2a$ and $r \simeq 5a$
throughout the simulation.

\subsubsection{Disk thickness}

    We close this section by commenting on the disk thickness $H/r$ [defined in
Equation~(\ref{scaleheight})], a parameter that will play an important role during the
period when the binary orbit evolves. Our initial data and cooling function were chosen
so as to keep $H$ roughly constant over time at a fixed ratio to the local radius:
$H/r \simeq 0.1$.  However, although the gas temperature stayed very close to the target
entropy at all radii $r > 2a$, and the ratio $H/r$ did stay nearly independent of radius,
its value first rose to $\simeq 0.15$ and then fell slightly (to $\simeq 0.12$ by the end
of the simulation).  The departure from the prediction of simple hydrostatic equilibrium
was proportional to how much the magnetic pressure contributed to support against the vertical
component of gravity.

\subsection{Binary Separation Evolution}\label{sec:binaryevolve}

At $t=40000M$ in \runtwo , we began to evolve the binary orbit, letting it compress
as gravitational radiation removes its orbital energy.  The rate of orbital evolution is
extremely sensitive
to separation: $\dot a/a \propto a^{-4}$ when $a/r_g \gg 1$.  Consequently, even at the
relatively small initial separation assumed here ($a_0 = 20M$), orbital evolution is
comparatively slow at first.  However, it accelerates dramatically after $t \simeq 50000M$.
By the end of \runtwo ($t = 54000 M$), $\dot a/a$ is quite rapid,
and $a \simeq 8M$, small enough to make our PN expansion problematic.

\begin{figure}[htb]
\centerline{
\begin{tabular}{cc}
\includegraphics[width=8cm]{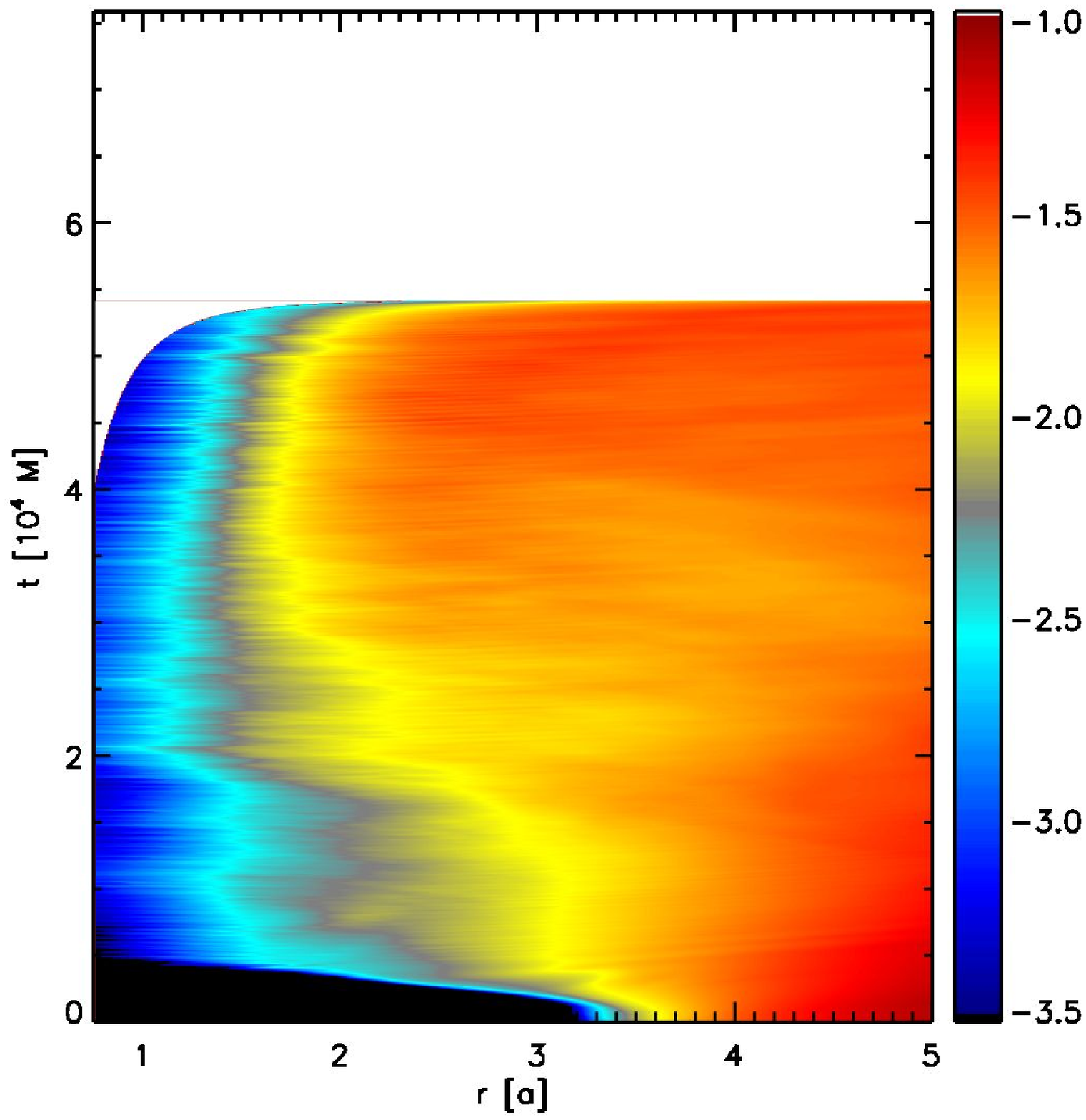}
&\includegraphics[width=8cm]{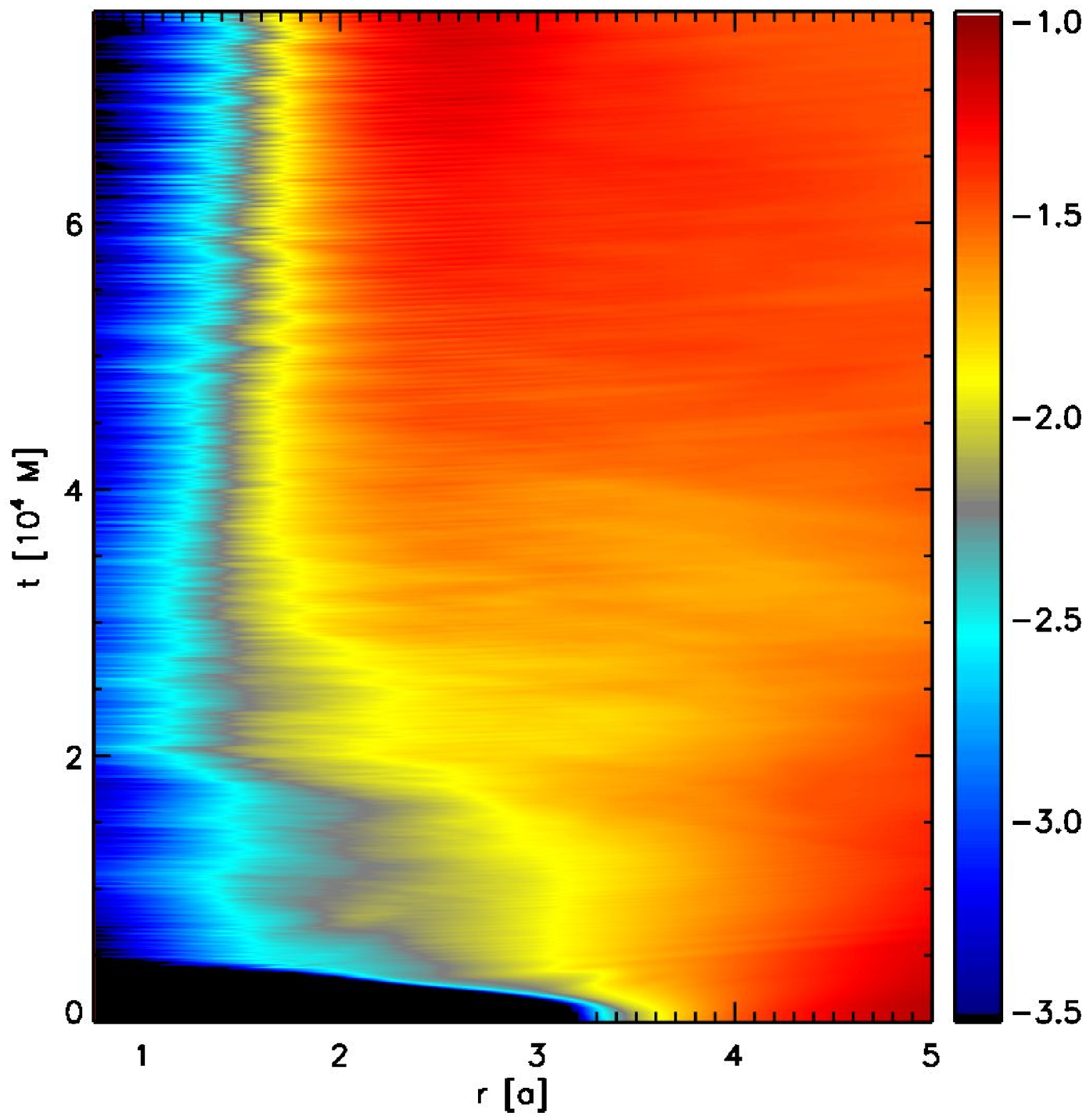}
\end{tabular}
}
\caption{Color contours of $\log \Sigma(r/a(t))$.  The scale
is shown in the color bar. (Left) \runtwo .  (Right) \runthree .}
\label{fig:surfdens_roa_t}
\end{figure}

\begin{figure}[htb]
\centerline{
\begin{tabular}{cc}
\includegraphics[width=8cm]{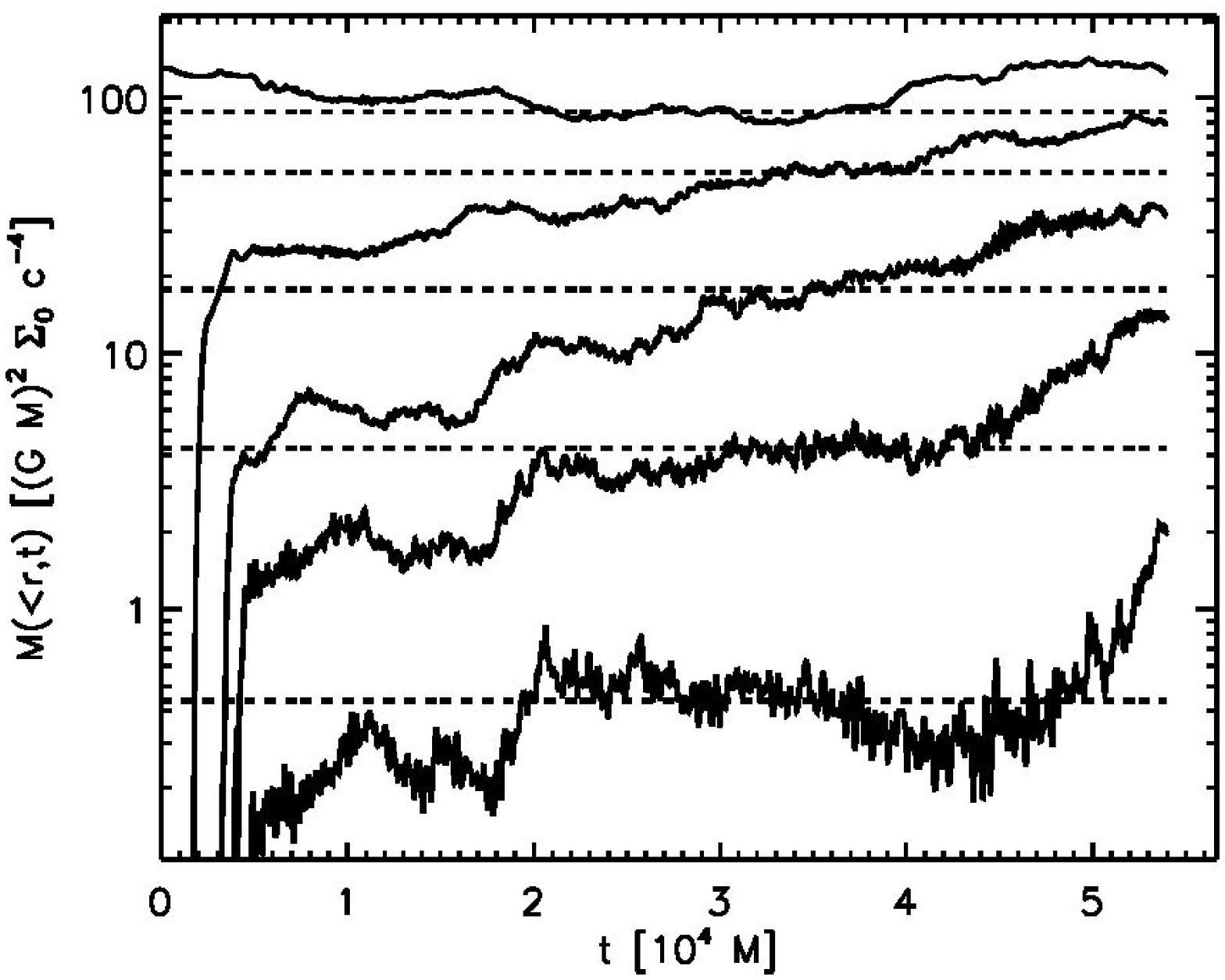}
&\includegraphics[width=8cm]{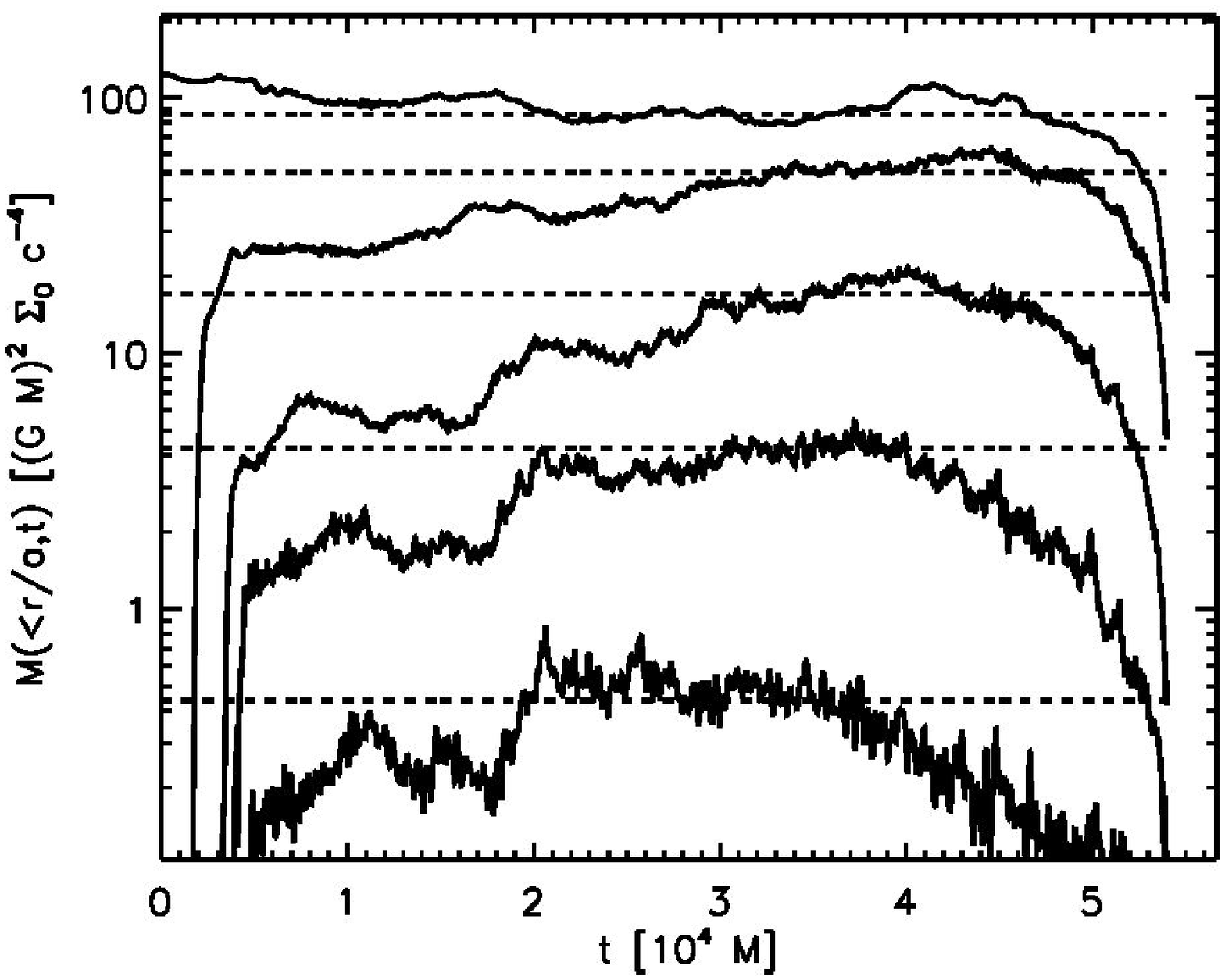}
\end{tabular}
}
\caption{Mass enclosed within several sample radii as functions of time in \runtwo .  From
bottom to top, the radii are $r/a = 1$, 1.5, 2, 3, and 4.  (Left) For fixed $a = a_0$.
(Right) For time-dependent $a(t)$.}
\label{fig:mass_enc}
\end{figure}

While the binary orbit changes relatively slowly, the inner edge of the disk moves inward in
pace with the change in the binary separation, staying close to $\simeq 2a(t)$ (as shown in
Figures~\ref{fig:surfdens_r_t} and \ref{fig:surfdens_roa_t}) until $t \simeq 50000M$.
However, as the orbital evolution becomes more rapid (after $t \simeq 50000M$), although the
inner edge of the disk continues to move inward in terms of absolute distance
(Figure~\ref{fig:surfdens_r_t}), it begins to recede in terms of $r/a(t)$
(Figure~\ref{fig:surfdens_roa_t}).  At the end of the simulation, the disk edge has moved in to
$\simeq 20M$, but that is $\simeq 2.5a$.  Simultaneous with this evolution, the slope of
the inner edge also becomes gentler (Figure~\ref{fig:surfdens_r}).  In other words, the
contrast between the surface density in the disk body and in the gap weakens, particularly
when considering the outer part of the gap.  As shown by the \runthree panel in
Figure~\ref{fig:surfdens_roa_t}, none of this adjustment (in $r/a(t)$ terms) occurs
without binary evolution.

Another view of this process may be seen in Figure~\ref{fig:mass_enc}.  In that figure,
we see the way matter accumulates in the inner disk over time, at first during the
quasi-steady epoch and later during the binary orbital evolution of \runtwo .  The
left-hand panel shows what happens when referred to an absolute radius scale.  When
the binary begins to shrink, the quantity of matter found at small radii grows abruptly,
particularly in the original gap region: the amount of mass inside $r = 40M$ almost
doubles, and the mass inside $r=20M$ increases by a factor of 5 during the
period of binary orbital evolution.  The right-hand panel shows the same events from
a different point of view.  In this figure, we see that the mass enclosed within small
multiples of $a(t)$ declines rapidly as the binary's shrinkage accelerates.  For larger
multiples (e.g., $3a$ and $4a$), the mass enclosed continues to rise for a while after
binary orbital evolution, but eventually drops once the compression becomes rapid.   In
particular, the mass within the gap region (i.e., $r < 2a(t)$) falls by roughly a factor
of 40 during the period of orbital evolution, although this ratio is in fact a bit ill-defined
because $2a(t)$ is almost at the simulation's inner boundary by the end of the simulation.

\begin{figure}[htb]
\centerline{
\begin{tabular}{cc}
\includegraphics[width=8cm]{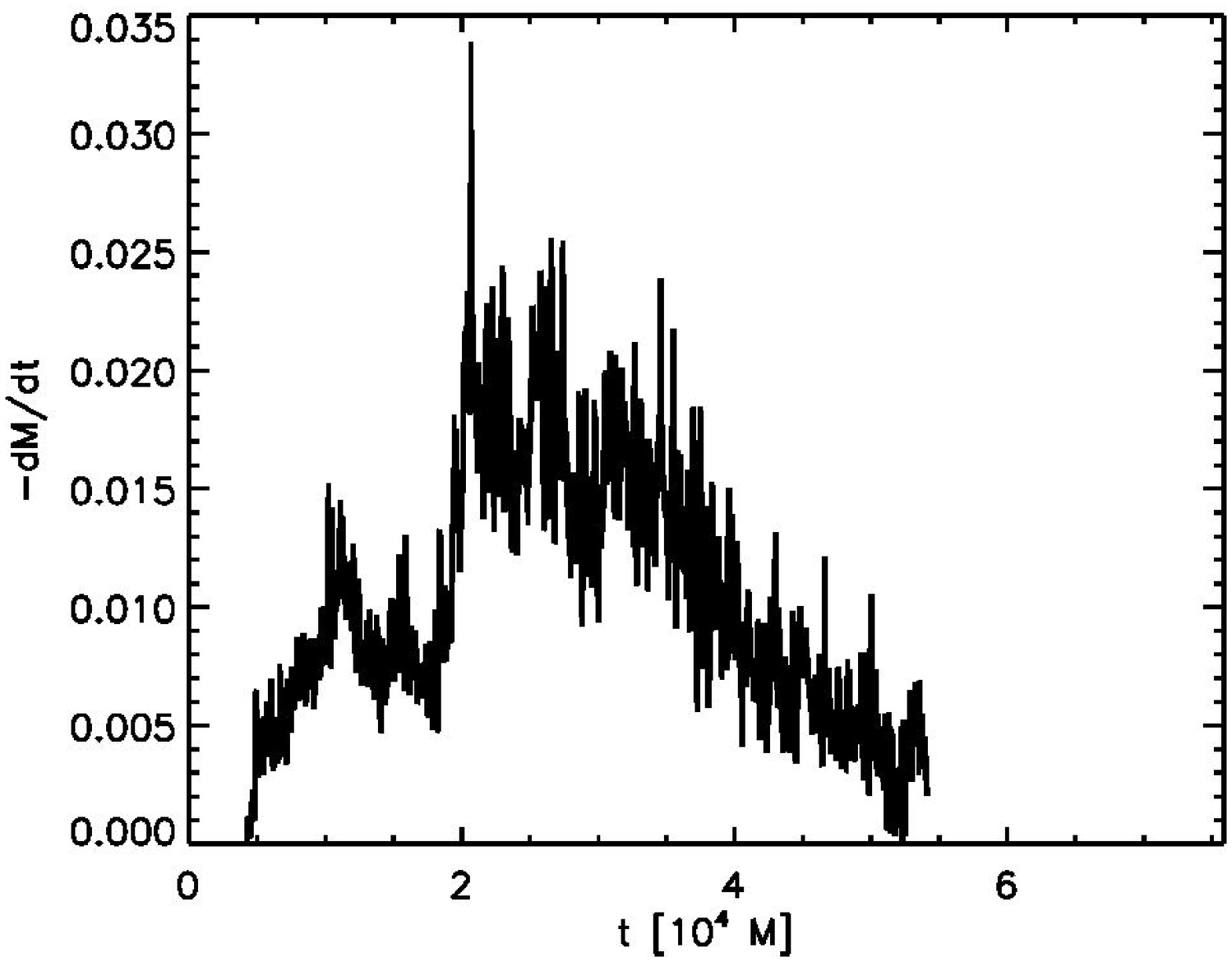}
&\includegraphics[width=8cm]{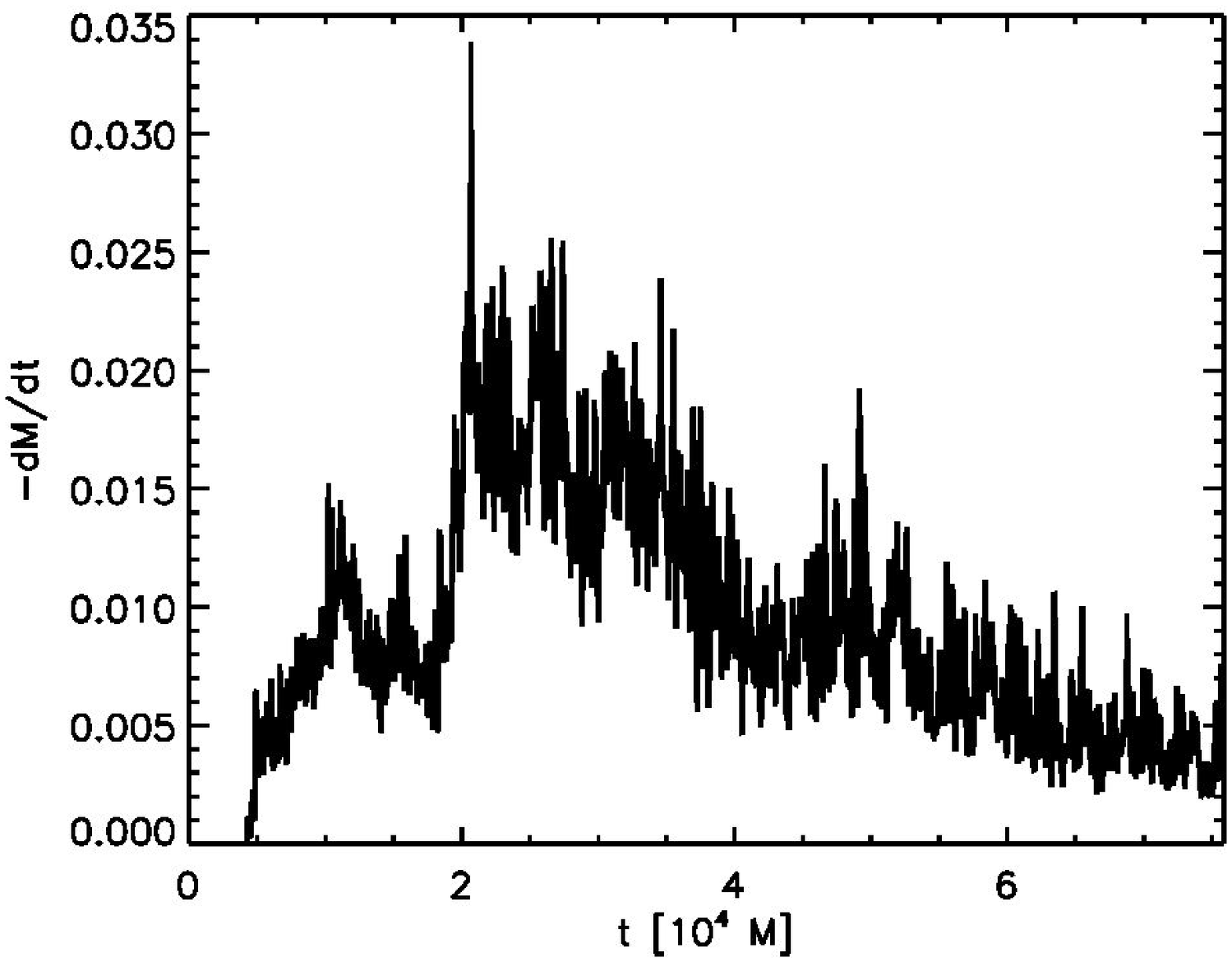}
\end{tabular}
}
\caption{Accretion rate through the inner boundary of the simulation as a function of time.
(Left) \runtwo .  (Right) \runthree .}
\label{fig:accrate_time}
\end{figure}

The accretion rate behaves differently.  It falls (see Figure~\ref{fig:accrate_time}) from
$\simeq 30000M$--$40000M$, even before the binary begins to compress.  Without
binary orbital evolution (\runthree), it levels out from $\simeq 40000M$--$50000M$, before
declining more gradually from $\simeq 50000M$ until the end of \runthree at $\simeq 76000M$.
In \runtwo , the onset of binary evolution at $t=40000M$ leads to a continuing decrease in
the rate at which mass flows through the inner boundary that levels out only after $\simeq
50000M$.  Although the accretion rates in the simulations with and without binary orbital
evolution decline at different times and at different rates, the final accretion rate
in \runtwo, when the binary separation has shrunk to $8M$, is only
20--$30\%$ less than at the same time in \runthree.

\begin{figure}[htb]
\centerline{
\begin{tabular}{cc}
\includegraphics[width=8cm]{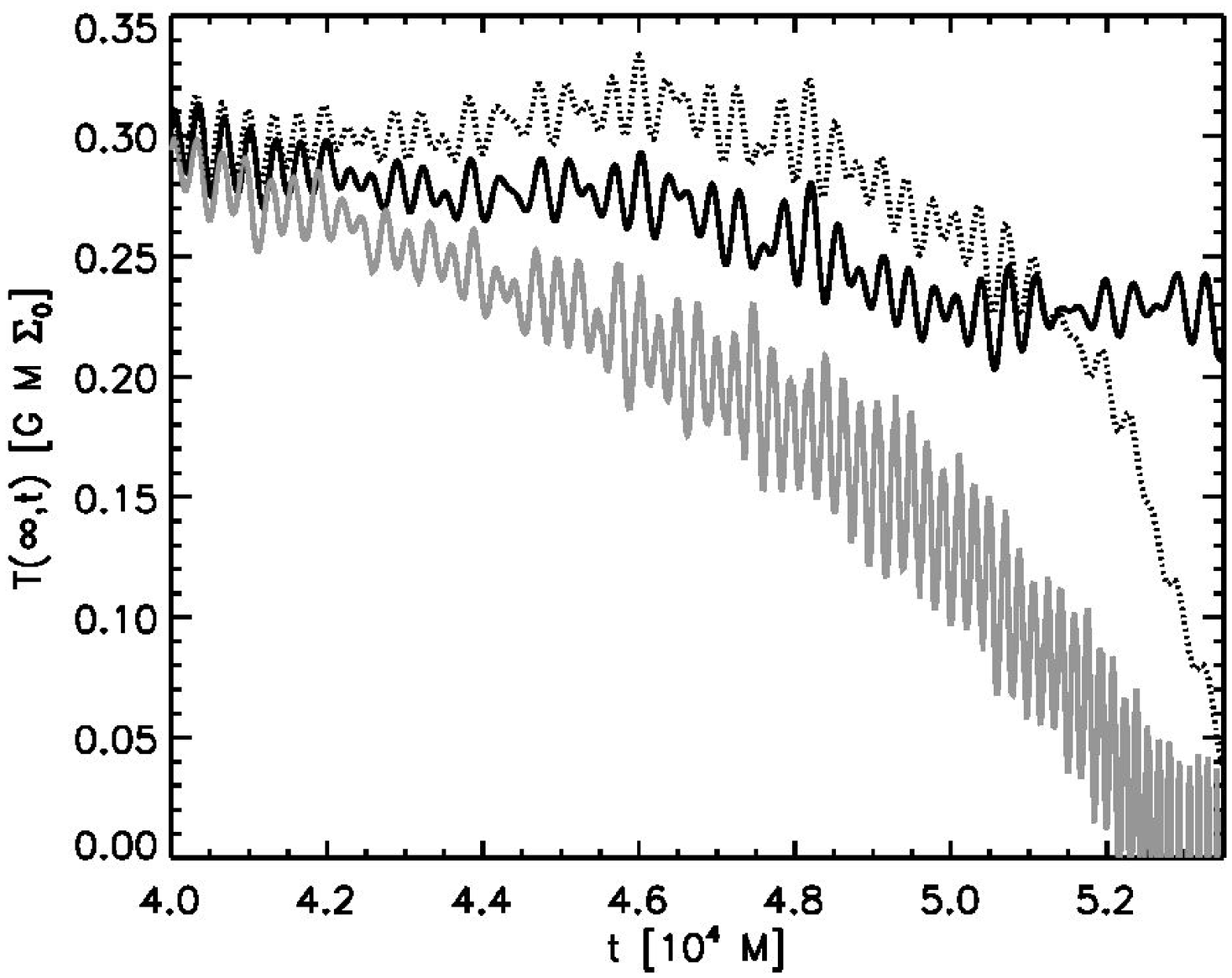}
&\includegraphics[width=8cm]{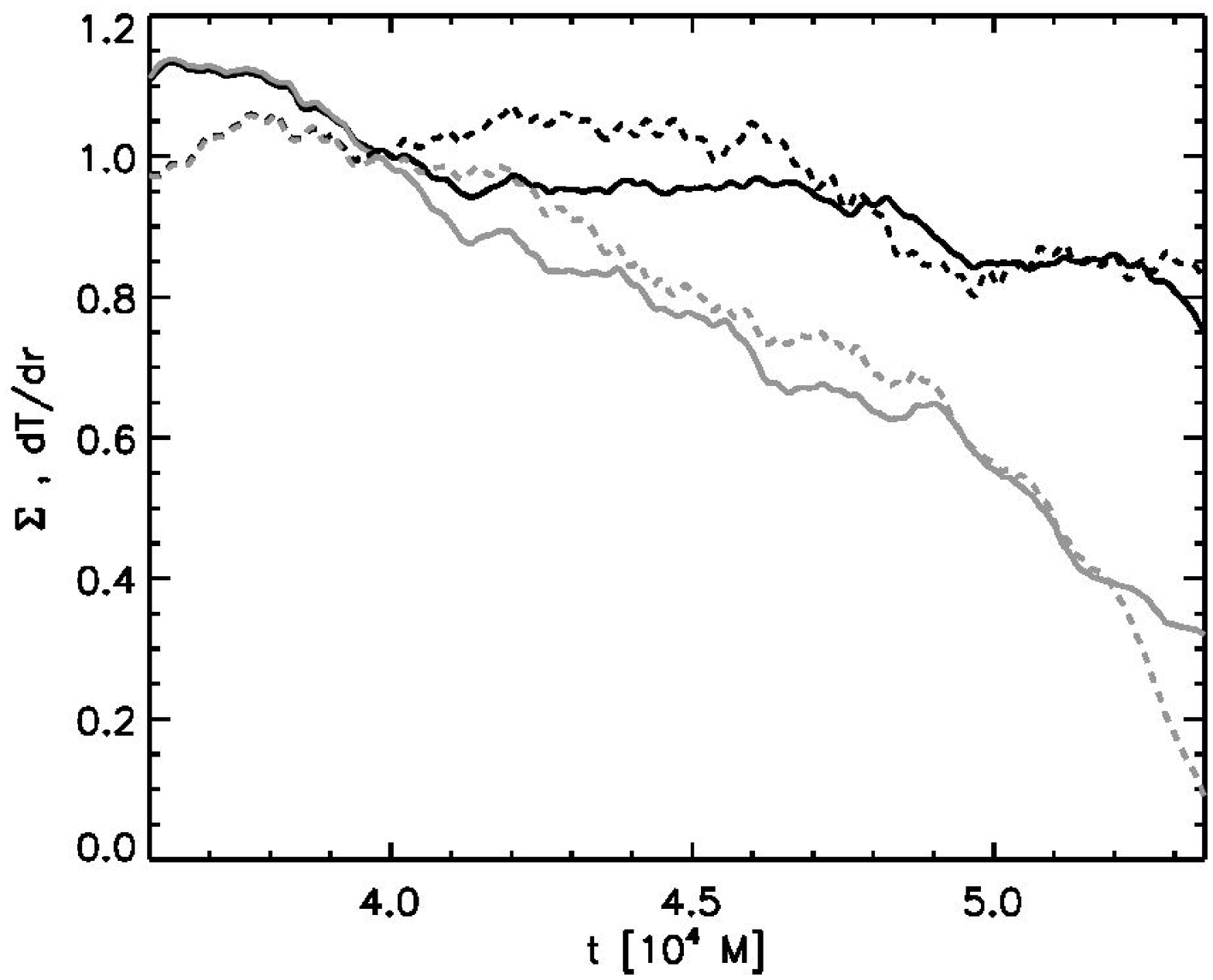}
\end{tabular}
}
\caption{(Left) Integrated torque as a function of time in \runthree (black) and
\runtwo (grey).  Total torque is shown by the solid curves; the dotted curve shows
torque in \runthree including only the radial range $r > \Rmin a_0 / a_{2}(t)$,
where $a_{2}(t)$ is the orbital separation as a function of time in \runtwo.
(Right) Surface density (solid) and torque density at its peak, i.e., at $r=1.45a(t)$
(dashed) in \runthree (black) and \runtwo (gray).}
\label{fig:torque_time}
\end{figure}

Another consequence of the changing relationship between disk material and the binary is
a diminution in the integrated torque when the binary compresses (Figure~\ref{fig:torque_time}).
During the initial slow stages of energy loss due to gravitational wave emission, the
binary continues to exert nearly as much torque on the disk as in \runthree, in which
the binary orbit does not change at all.  However, once the orbital shrinkage begins
to accelerate, the torque plummets; at the end of \runtwo, it has fallen to $\simeq 1/5$
of the value at that time in \runthree .  The greater part of this diminution in torque
is due to the fact that at this stage in the binary's evolution, its separation diminishes
so rapidly that the region between $a$ and $2a$, where most of the torque is expressed, moves
inward faster than the matter can follow.  There is consequently much less matter on which
these torques can be exerted.  The connection between available matter and torque is shown
clearly in the right-hand panel of Figure~\ref{fig:torque_time}, in which one can easily
see that for nearly the entire inspiral the torque density at the location of its maximum
($r=1.45a(t)$) is almost exactly proportional to the surface density there.
However, there is also a smaller part due to an artifact of the simulation.  Its inner boundary
lies at $r_{\rm min} = 15M$.  As soon as $a(t)$ becomes smaller than $15M$, part of the region
in which the binary torque is applied is no longer in the problem volume, so we cannot
calculate any torque occurring there.  As shown in the right-hand panel of
Figure~\ref{fig:torque_time}, this effect becomes significant at $t \simeq 5.2 \times 10^4 M$,
when $a(t) \simeq 13M$.  By the end of \runtwo , $a \simeq 8M$, so that nearly
the entire region where the torque is exerted ($a \lesssim r \lesssim 2a$) has left the
problem volume.  At that point, even if there were significant matter there, our calculation
can neither say what its mass is nor what torque it feels.

\subsection{EM Luminosity: Magnitude, Modulation}\label{sec:emlum}

\begin{figure}[htb]
\centerline{
\includegraphics[width=12cm]{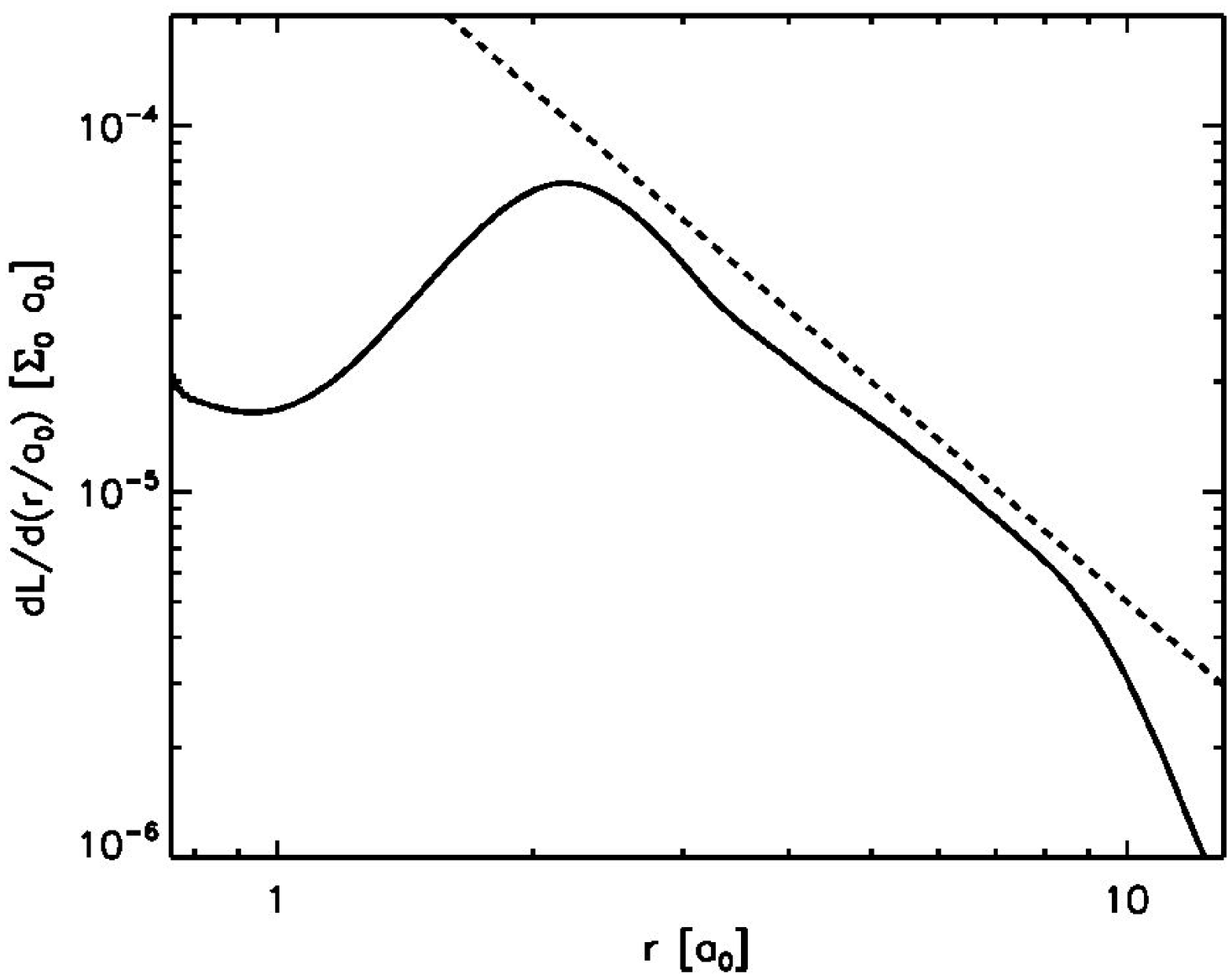}
}
\caption{Luminosity per unit radius averaged over the quasi-steady epoch in \runthree .
The dashed line shows a logarithmic slope of -2.}
\label{fig:dlumdr}
\end{figure}

We define the (coordinate frame) cooling rate per unit radius of the disk by
\begin{equation}
\frac{dL}{dr} = \int \, \sqrt{-g} \, d\theta \, d\phi \mathcal{L}_c u_t .
\end{equation}
During the approximate stationary state, it is best described in terms
of two separate regimes.  As shown in Figure~\ref{fig:dlumdr}, at large radius
($r \gtrsim 2a$), it is very well described by a power-law,
$dL/d(r/a_0) \simeq 5 \times 10^{-4} (r/a_0)^{-2} \Sigma_0 a_0$.
At around $r \simeq 2a$, the cooling rate per unit radius reaches a local maximum
and declines inward.  This distinction neatly corresponds to two different mechanisms for
generating the requisite heat: the dissipation of MHD turbulence associated with mass
accretion (at large radius) and the dissipation of fluid kinetic energy given to the
relatively small amount of gas in the gap by the binary torques (at small radius).
In fact, this identification is confirmed semi-quantitatively.  In time-steady
accretion, the luminosity per unit radius is $(3/2)\dot M c^4/[(r/r_g)^2 GM]$ at radii
where the local orbital angular momentum per unit mass is large compared to the
net angular momentum flux per unit mass.  Our disk is never in inflow equilibrium,
and this expression is not exact when $\dot M$ is a function of radius.  Nonetheless,
taking it as an estimator, it predicts
\begin{equation}
\frac{dL}{dr/a_0} = 4 \times 10^{-4} (\dot M/0.01) (r/a_0)^{-2} \Sigma_0 a_0.
\end{equation}
As Figure~\ref{fig:accrate_radius} shows, the mean accretion rate in code units
at $r=2a$ in \runthree was $\simeq 0.01$, while $\dot M$ at larger radii is
typically similar or perhaps a factor of two greater.  Thus, this prediction of
the luminosity profile on the basis of the time-averaged accretion rate and
expectations derived from time-steady accretion onto a solitary mass quite
accurately matches the actual luminosity profile seen in the simulation.

\begin{figure}[htb]
\centerline{
\includegraphics[width=8cm]{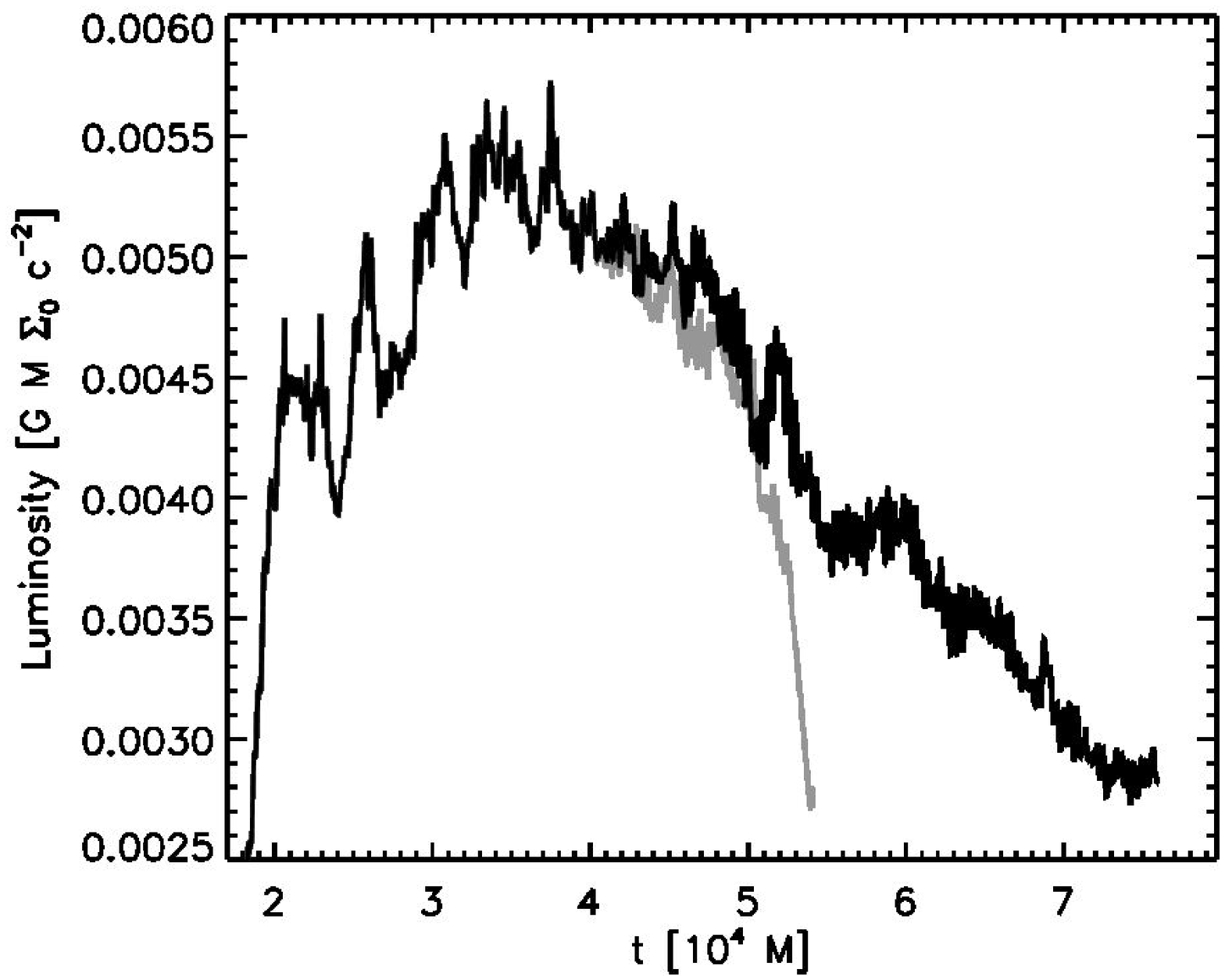}
}
\caption{Luminosity as a function of time.
(Grey) \runtwo .  (Black) \runthree.  We note that the vertical axis' 
range does not include zero in order to accentuate the curves' fluctuations. }
\label{fig:lum_time}
\end{figure}

Integrated over radius, the total luminosity reaches a peak
$\hat L \simeq 5.5 \times 10^{-3}$ at $t \simeq 33000M$ (Figure~\ref{fig:lum_time}),
where $\hat L$ is the integrated luminosity in units of
$GM \Sigma_0 c$.  After reaching this peak, $\hat L$ falls slowly, reaching
$\simeq 3 \times 10^{-3}$ at $t \simeq 76000M$ in \runthree; averaged over the entire
quasi-steady period in this simulation, it is $3.8\times 10^{-3}$.

The light output from \runtwo remains very close to that in \runthree until the binary
orbital evolution becomes rapid at $t\simeq 50000M$.  After that time, it falls more sharply,
so that by the time at which \runtwo stops, $\hat L \simeq 2.7 \times 10^{-3}$; this is,
however, still 2/3 the luminosity in \runthree at the same time.  As the binary shrinks,
the radial distribution of the luminosity changes in parallel, with the peak in surface brightness
moving inward.  We attribute the gradual decline in luminosity to the gradual decline in
accretion rate.  The sharp drop in the final stages of binary orbital shrinkage is due to
the interaction of a boundary effect with genuine dynamics.  As shown by \cite{Shi11}, gas
streams flow inward from the inner edge of a quasi-steady circumbinary disk to
radii $\simeq 1.2a$, where they can be strongly torqued and some of their material flung
back outward toward the disk.  The outward-moving matter shocks against the disk proper
at a radius near that of the surface density peak, and the heat dissipated in these shocks
contributes significantly to the luminosity.  When the binary shrinks,
this mechanism is weakened for two reasons.  The inner boundary of our simulation
($r = 0.8a_0$) eventually becomes larger than $1.2a(t)$; when it does, matter is no longer
thrown outward by binary torques.  At the same time, however, it is possible that the
retreat of the disk's inner edge when measured in terms of $a(t)$ might also lead to
weaker inward streams.

The fact that the energy deposited by binary torques is ultimately radiated in the
disk proper leads to a method of estimating the relative contributions to the total luminosity
coming from accretion and binary torques.  For that reason, and also because the accretion
rate diminishes as the region of the surface density peak is approached from larger radius,
it is a reasonable approximation to suppose that most of the luminosity from the region
of the surface density peak inward has its source in the binary torques.  We can
therefore estimate the work done by the torques by bounding it between $L(r<2a_0)$
and $L(r<3a_0)$.  On this basis, accretion would account for $\simeq 1/2$--$3/4$
of the total (i.e., $\hat L \simeq 1.8$--$2.9 \times 10^{-3}$) and the binary torque for
$\simeq 1/4$--$1/2$ ($\hat L\simeq 0.9$--$2 \times 10^{-3}$).

The rest-mass efficiency of this luminosity is comparable to the rest-mass efficiency
due to accretion that goes all the way to the black hole.  Measured in terms of the
time-dependent luminosity relative to the time-averaged accretion rate through the
inner boundary, the efficiency in \runthree falls from a peak $\simeq 0.06$ achieved
for $20000M \lesssim t \lesssim 45000M$ to $\simeq 0.03$ at the end of this simulation.
There are several reasons that this efficiency is so great even though the potential at
$r=50M$ is an order of magnitude shallower than the potential at the innermost stable
circular orbit (the ``ISCO").  One is that the accretion
rate in the circumbinary disk is roughly twice the accretion rate through the inner
boundary, so the local accretion dissipation in the disk is boosted by that same
factor of two relative to the rate at which mass passes the inner boundary.  Another
is that in a conventional disk around a single black hole the dissipation rate in the
region just outside the ISCO is depressed relative to larger radii because some of
the potential energy released is transported outward by the inter-ring stresses.  In the
Novikov-Thorne model (in which the stresses are assumed to vanish at the ISCO),
almost 40\% of the total luminosity is released outside $r=40M$ when the black
hole has no spin.  This fraction is smaller when the spin is greater, and may be
further reduced to the degree that the net angular momentum flux is smaller \citep{KHH05}.
Lastly, of course, additional energy is deposited in the disk by the work done by
the binary torques.

Translating the peak cooling rate into physical units gives
\begin{equation}
L_{\rm disk} \simeq 2.4 \times 10^{40} (\hat L/10^{-3}) M_6 \tau_0\hbox{~erg/s}.
\end{equation}
Here $\tau_0$ is the Thomson optical depth through a disk of surface density
$\Sigma_0$ and $\hat L$ is the luminosity in code units, i.e., 3--$5\times 10^{-3}$.
In Eddington units, this becomes
$L_{\rm disk}/L_E \simeq 1.7 \times 10^{-4} (\hat L/10^{-3}) \tau_0$.
Thus, for such a system to be readily observable at cosmological distances, it will
be necessary both for the disk to be optically thick to Thomson scattering and for
the mass of the binary to be relatively large.   As a gauge of what might reasonably
be expected, we note that in a steady-state accretion disk around a
solitary black hole, the optical depth of the disk at $r/r_g = 20$ would be
$\sim 2 \times 10^3 (\alpha/0.1)^{-1} (\eta/\dot m)$, where $\eta$ is the usual
rest-mass efficiency and $\dot m$ is the accretion rate in Eddington units.  With this
disk surface density, the luminosity would approach that of a typical AGN when
$M_6$ is at least $\sim 1$.

If this light were radiated thermally, the corresponding effective temperature
would be
\begin{equation}
T_{\rm eff} \simeq 4 \times 10^4 (\hat L/10^{-3})^{1/4} M_6^{-1/4} \tau_0^{1/4}\hbox{~K},
\end{equation}
where we have assumed that the radiating area is $2\pi (2a)^2$.  Thus, it would
emerge primarily in the ultraviolet for fiducial values of black hole mass and
optical depth.

\begin{figure}[htb]
\centerline{
\includegraphics[width=12cm]{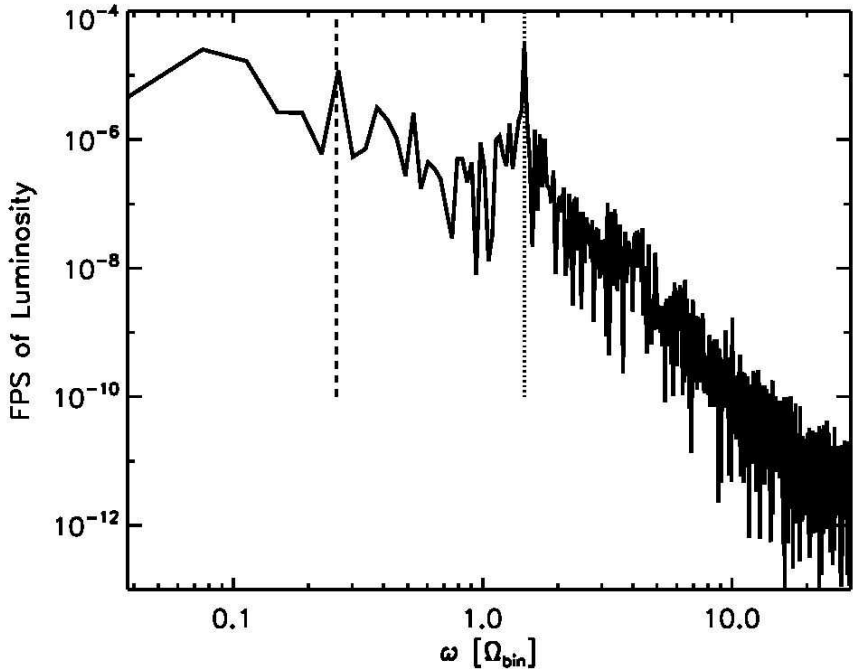}
}
\caption{Fourier power spectrum of the luminosity radiated during the quasi-steady
epoch of \runthree. The vertical lines represent: the orbital frequency at
the surface density maximum (dashes) and the peak in the spectrum (dots).}
\label{fig:lum_pds}
\end{figure}

The luminosity (assumed to be optically thin) exhibits a noticeable modulation as a function
of time, with peak-to-trough contrast of $\simeq 5\%$.  Its Fourier power spectrum shows a strong,
sharp peak at a frequency $1.47\Omega_{\rm bin}$ (see Figure~\ref{fig:lum_pds}) and a weaker
peak at $0.26\Omega_{\rm bin}$.  The latter is the orbital frequency at the radius of the
surface density maximum, $\simeq 2.4a$; because the lump is located at this radius, we
call this frequency $\Omega_{\rm lump}$.  The former we identify with the
rate at which the lump approaches the orbital phase of a member of the binary,
$2(\Omega_{\rm bin} - \Omega_{\rm lump}) = 1.46\Omega_{\rm bin}$.  When the lump draws near
one of the black holes, a new stream forms, falls inward, and is split into two pieces, one of which
gains angular momentum, sweeps back out to the disk, and ultimately shocks against
the disk gas.  It is this process, whose frequency is $1.46\Omega_{\rm bin}$, that modulates
the light curve.  If the binary mass ratio were far from unity, we expect that the modulation
frequency would fall to $\simeq \Omega_{\rm bin} - \Omega_{\rm lump}$.

\section{Discussion}
\label{sec:discussion}

\subsection{Comparison to Newtonian MHD}

   In many respects, the behavior we found in this post-Newtonian regime resembles what was
previously found in the Newtonian limit \citep{Shi11}.  There is very good agreement in the
shapes of their azimuthally-averaged surface density profiles, with any contrasts attributable to
their somewhat different initial conditions.  In both cases, during the quasi-steady epoch the
surface density at the disk's inner edge rises $\propto \exp(3r/a)$, reaches a maximum at
$r \simeq 2.5a$, and then declines to larger radii.

   At early times in both, there is a pair of streams leading from the disk edge to the inner
boundary, which they typically reach at an orbital phase slightly ahead of the nearest member
of the binary.  Both also develop a strong $m=1$ asymmetry (a ``lump") in the surface density at
$r \simeq 2.5a$ at late times.  This asymmetry ultimately causes, in both the Newtonian
and post-Newtonian simulations, a single stream in the gap to become dominant.  Almost the only
contrast in this regard is that the orbit of the lump developed a growing eccentricity in the
Newtonian case, but not in \runthree.

    The level of magnetization is likewise qualitatively similar: the mean plasma $\beta$
in the Newtonian case fell from $\sim 1$ at $\simeq 6a$ to $\simeq 0.3$ at
$r \simeq 2a$, while the value (averaged over the quasi-steady epoch in \runthree) in our
simulations was $\simeq 1.5$ at $r=6a$, grew to $\simeq 2.5$ at the surface density peak,
and then decreased inward.  The magnetic stress-to-pressure ratio $\alpha$ in the disk body
follows the same pattern of close resemblance. It was $\simeq 0.3$ in the disk body in the
Newtonian case, and $\simeq 0.2$ in \runthree .  In the gap, the similarity was more
qualitative than quantitative: in both cases, it rose steeply into the gap, but reached only
$\simeq 0.7$ at $r \simeq a$ in the PN simulation, whereas it climbed to $\simeq 10$ in the
Newtonian one.

     Most strikingly, the luminosity estimated by \cite{Shi11} scales extremely well
to the PN case.  \cite{Shi11} could not directly compute the luminosity because they
assumed an isothermal equation of state.  However, they argued that the work done by
the binary torques would be delivered to the disk and ultimately dissipated there into
heat.  Rewriting in our units their value for the rate at which the torques did work on the
gas gives a luminosity of $0.018 GM\Sigma_p c(a/r_g)^{-1/2}$, where $\Sigma_p$ is the
surface density at the maximum; for our separation ($a=20M)$ and our surface density at
the maximum ($\simeq 0.55$ averaged over the quasi-steady epoch in \runthree),
that becomes $2.2 \times 10^{-3} GM\Sigma_0 c$.  This prediction agrees well with the
upper end of our estimated range for the binary torque share of the luminosity.

\subsection{Comparison to Analytic Estimates of Binary Runaway}

\cite{MP05} predicted that at some point well before the merger, the \bbh
should begin compressing so fast by gravitational radiation that internal stresses
within the disk would not allow it to move inward rapidly enough to stay near the binary.
At the order of magnitude level, this breakaway point would be expected to come when
the gravitational radiation time
\begin{equation}
t_{\rm gr} = \frac{5}{64}\left(\frac{a}{M}\right)^4 \frac{(1+q)^2}{q} M
\end{equation}
becomes shorter than the characteristic disk inflow time
\begin{equation}
t_{\rm in} = \alpha^{-1}(H/r)^{-2}(d\ln\Sigma/d\ln r)^{-1}\Omega^{-1} 
= \alpha^{-1}(H/r)^{-2}(d\ln\Sigma/d\ln r)^{-1}(r/r_g)^{3/2}M.
\end{equation}
In these equations $\Omega$ is the local disk orbital frequency.  The logarithmic
derivative of the surface density enters because spreading of
the inner edge is more rapid when it is especially sharp.

With a typical estimate of the stress level, $\alpha \sim 0.01$, the binary separation
at which $t_{\rm gr}$ and $t_{\rm in}$ would match, and the disk and binary might decouple is
\begin{equation}
a_{\rm dec} = 70 (d\ln\Sigma/d\ln r)^{-2/5}\hoverrscale^{-4/5} M,
\end{equation}
where the fiducial radius at which the inflow time was computed is $r_* = 2a$, and we set $q=1$.
Indeed, it was this sort of estimate that led us to choose the initial conditions for
our simulation.

However, scaling to the actual parameters of our simulation leads to a considerably
smaller predicted value,
\begin{equation}
a_{\rm dec} \simeq 10 [(d\ln\Sigma/d\ln r)/6]^{-2/5}\alphascale^{-2/5}\hoverrscale^{-4/5}M ,
\end{equation}
which is much closer to what is found in \runtwo .  Thus, the substantially
stronger magnetic stresses than predicted by usual $\alpha$-based estimates lead to
decoupling at a much smaller binary separation.  Nonetheless, in the end the inward
motion of the disk is limited by angular momentum transport, so once the magnitude of
those stresses are known, $a_{\rm dec}$ can be estimated quite well by this means.

On the other hand, the meaning of the term ``binary runaway" should also be made more
nuanced.  As we have seen, the accretion rate through the simulation inner boundary decreases
as the binary shrinks, but almost the same decrease in accretion rate occurs when the
binary {\it does not} shrink.  Moreover, the continuing advance of the disk during
the period of orbital evolution brought matter rapidly inward.  The amount of matter
within $30M$ rose by about a factor of 4 while the binary shrank, so that almost as
much matter could be found within that radius as had been within $40M$ at the beginning
of the binary orbital evolution.  Thus, acceleration of binary orbital evolution does
lead to a state in which the surface density at $r < 3a$ is smaller than would
be expected if the orbital evolution were slower, and this diminution in the mass close
to the binary does lead to consequences such as sharply diminished torque (as discussed in
Section~\ref{sec:binaryevolve}) and luminosity (see Section~\ref{sec:emlum}).
On the other hand, neither the torque nor the luminosity falls by as much as an order
of magnitude because the decoupling of binary and disk matter is not complete.  Most notably,
accretion continues at a rate only tens of percent lower than in the absence of inspiral.

Continuing accretion into the binary orbital region has a particularly interesting
consequence.  The material in those streams should, just as happens when
the binary orbital evolution is slower, be captured into orbit around one or the other
of the members of the binary.  It will then settle into two smaller disks, one around
each black hole. In the conditions of our simulation, the inflow time in the individual
black hole disks might be only slightly shorter than the merger time because decoupling
occurs when the binary separation is not a great deal larger than the ISCO, even if both
black holes spin rapidly.  If the circumbinary disk were cooler than in our simulation, so that
$a_{\rm dec} \gg M$, the inflow time in the smaller disks would be shorter than that in the
circumbinary disk by a sizable ratio: $ \sim 3 \times 10^{-4}$ at decoupling if the
scale heights in the inner disks and the outer disk are the same, and even smaller at
later times.  In either case, there could be interesting hydrodynamic interaction between
the two smaller disks as the binary compresses.

\subsection{Different Disk Thermal States}\label{sec:diskthermal}

In terms of an ultimate comparison to observations, a larger question is posed by
the disk's thermal state.  As just shown, the binary separation at
decoupling scales as $(H/r)^{-4/5}$, so the disk's internal pressure ($H \propto c_s \propto
(p/\rho)^{1/2}$) can influence $a_{\rm dec}$.  For reasons of numerical convenience,
we chose parameters yielding a relatively thick disk.  Although the factors controlling
the saturation of magneto-rotational turbulence are still not well understood, a scaling
with disk pressure remains plausible.  If the effective sound speed of the gas
were lower, decoupling might occur at rather larger binary separation, well outside the
domain of relativistic orbits.

Several factors can influence the actual equation of state of the disk.  In ordinary AGN,
local heating due to accretion can make the disk radiation-dominated inside $r \simeq 100M$
when $\dot m \gtrsim 0.3$ and the central mass $M > 10^6 M_{\odot}$ \citep{krolik99}.
Larger central masses lead to radiation dominance even when $\dot m$ is smaller.  When
the disk surrounds a binary, the local heating should be similar at radii $r \gtrsim 2a$,
as shown by Figure~\ref{fig:dlumdr}; the diminished accretion inside $\sim 4a$ is
compensated by dissipation of the work done by the binary torques and delivered to the disk.
In those circumstances, the disk scale height for $r \gtrsim 2a$ is independent
of radius, giving an aspect ratio $H/r \simeq (3/2) (\dot m/\eta) (r/r_g)^{-1}$, where
$\dot m$ is now the {\it local} accretion rate in Eddington units.  Thus, our aspect ratio
of $\simeq 0.15$ would correspond to a nominal $\dot m \simeq 0.4(r/40r_g)$; that is,
this $\dot m(r)$ is the mass accretion rate at $r$ that would, if it reached the black
hole in a flow with $\eta = 0.1$, produce that fraction of an Eddington luminosity.

Given the relatively large leakage fraction through the inner edge of the circumbinary
disk, the accretion rate onto the two black holes will in general be smaller than the
accretion rate in the disk by only a factor of a few.  They might therefore generate
a sizable luminosity with a spectrum not too different from that of a generic AGN.
Because the density in the gap is considerably smaller than in the disk, this
luminosity may irradiate the disk, particularly if it is relatively thick.  The inner
edge of the disk could be heated by this means, as well.

Thus, there remains considerable uncertainty in the thickness profile of a circumbinary
disk surrounding a relatively compact binary black hole.  The particular thickness we
have simulated lies within that range of uncertainty, but may not be generic.

\subsection{Distinctive EM Signals}

Given a sufficient external mass supply rate, the luminosity from the circumbinary
disk alone could be great enough to be detected, even from a cosmological distance
(Section~\ref{sec:emlum}).  However, until the binary separation becomes as small as
that considered here, the luminosity from matter accreting onto the two individual
black holes would dominate the circumbinary luminosity by a large ratio.  In many
respects, a binary black hole with $a \gg 10M$ should strongly resemble a conventional
AGN.  However, when the separation is as small as the $\sim 20M$ scale studied in our simulations,
contrasts with ordinary AGN continua might occur due both to the additional heating
near the inner disk and the gap from $\sim 2.5a$ down to $\sim a/3$ in the range of
radii in which a thermal disk exists.  The supplementary heating due to the binary
torques will increase the luminosity of the portion of the disk near $r \simeq 2a$,
but the temperature in this region is smaller than the hottest part of the accretion
flow by a ratio $\sim (1.5r_{\rm ISCO}/2a)^{3/4}$, where the factor 1.5 multiplying
the ISCO radius is meant to account approximately for the displacement of the temperature
maximum from the ISCO.  Consequently, the additional luminosity will appear at rather
longer wavelengths than the peak of the thermal continuum.   On the other hand, radiation
from the gap region will be very different from what might be expected from a conventional
disk in those radii.  The dissipation rate is much smaller because the motions are
laminar, not turbulent; moreover, whatever light does issue from that region is unlikely
to be effectively thermalized, and would therefore emerge at considerably shorter
wavelengths.  Thus, the luminosity at wavelengths intermediate between those
characteristic of the innermost part of the disk and those characteristic of
$r \simeq 2a$ would be significantly suppressed.

The periodic modulation in the heating rate of the circumbinary disk that we have found might
make its emission easier to isolate.  The key question governing that ``might" is how
effectively optical depth in the disk blurs the modulation.  Our cooling function, which
operates at a characteristic rate $\sim \Omega(r)$ filters out variations on timescales
$\ll \Omega^{-1}$, but optical depth would impose a rather more severe upper bound on
the maximum effective frequency of variation.  According to \cite{KK87},
a constant-density sphere of radius $R$ and optical depth $\tau$ suppresses the amplitude of
a periodic signal of angular frequency $\omega$ injected at its center by a factor
$\simeq 3 c/(R\tau\omega)$ when $\omega R \tau/c \gg 1$.  To apply this estimator,
we suppose that a stratified disk segment with scale height $H$ can be approximated by
a homogeneous sphere of radius $R=H$.  For binary separation $a$, the relevant orbital radius
is $\simeq 2.5a$ and the signal frequency $\omega = 1.46\Omega_{\rm bin} \simeq 1.5(GM/a^3)^{1/2}$
for total binary mass $M$.  As shown in Figure~\ref{fig:surfdens_equatorial}, the surface
density in the lump grows to be $\simeq \Sigma_0$, so we take $\tau = \tau_0$.  The suppression
factor is then $\simeq 0.024 (\tau_0/1000)^{-1} (a/20r_g)^{1/2} (H/0.15r)^{-1}$.  In other
words, when the relevant region of the disk is optically thick, the luminosity in the modulated
component is independent of surface density, so that its fractional modulation decreases as
the luminosity increases.

As we have discussed in section~\ref{sec:diskthermal}, there is also considerable
uncertainty in $H/r$, even given a disk surface density.  If the disk thickness and optical
depth were determined by the considerations of conventional time-steady accretion flows
around single black holes, the characteristic cooling time $\tau H/c$ can be identified
with $(\alpha\Omega)^{-1}$.   The fluctuation suppression factor could then be estimated
by $\simeq 3 \alpha \Omega/\omega$.  Here the relevant orbital frequency is
$\Omega(2.4a) = 0.26\Omega_{\rm bin}$, while
$0.74\Omega_{\rm bin} \leq \omega \leq 1.46\Omega_{\rm bin}$ (the upper limit applies in
the equal-mass case, the lower limit when the masses are very unequal).  Thus, the
suppression factor estimated in this way is $\simeq 0.1(\alpha/0.2)$ for equal black
hole masses, rising to double that in the limit of very unequal masses.
However, this estimate is made uncertain by the fact that the disk in the vicinity of
the surface density peak is certainly {\it not} in a state of inflow equilibrium.
In addition, just as for the other estimate, the association of the periodic modulation
with the lump means that any estimate based on assumptions of axisymmetry likely
underestimates the local optical depth.

Both estimates suggest that the modulation will be suppressed by at least a factor of
several, but both are also subject to considerable uncertainty, making the actual outcome
unclear.  It is worth pointing out that in the event the modulation is detectable, the
period of the modulation would allow an estimate
of the binary orbital period.  When the binary mass ratio is unity, the binary orbital
frequency is 0.68 times the frequency of the modulation; as the mass ratio departs
from unity, the binary orbital frequency should rise toward $\simeq 1.36$ times
the modulation frequency.

Finally, we remark that our predictions of EM signals from circumbinary disks around
merging black holes are complementary to those previously made \citep{Bode12} in two ways.
In the previous work, the period of EM emission {\it began} when the binary
separation shrank to $8M$; that is when our calculation {\it ends}.  In addition,
that effort expressly excluded the disk proper, which they defined as $r \geq 16M$;
in our work, that is the location of the overwhelming majority of the emission.
Our effort also differs from previous work in this area in that we explicitly
include radiation losses in the gas's energy equation and also tie the rate of radiation
directly to the instantaneous local thermodynamic state of the gas (albeit in only
a formal way).  In addition, our discussion of the observability of periodic modulation
in the lightcurve takes into account possible suppression of the variation due to optical
depth in the source.

\section{Summary}\label{sec:conclusions}

By describing the binary black hole spacetime at separations of tens of gravitational radii
through the PN approximation, we have been able to simulate many orbits of
fluid motion around such a system in fully relativistic MHD.  In so doing, we have
demonstrated that the qualitative properties of circumbinary disks in such a regime
are well described by an extrapolation from their properties in the Newtonian limit:
Matter piles up at $\simeq 2.5a$, while smaller radii are largely cleared of mass;
nonetheless, accretion continues through the inner gap, albeit reduced by a factor
of a few from the rate at which it is supplied at larger radii.

At the same time, however, we have also investigated the initial stages of strongly
relativistic behavior in the form of the disk's response to binary orbital evolution by
gravitational wave emission.  By carrying the disk's evolution through the transition
from the epoch in which its characteristic inflow time is short compared to the binary
evolution timescale all the way to the epoch in which the binary evolves much faster
than the disk, we have established the time at which the binary ``runs away" from the
disk, and more importantly, the degree to which it does so.  This decoupling
causes a drop in both the torque the binary exerts on the disk and in the disk luminosity.
However, a sizable fraction of the accretion rate at large radius continues to makes
its way to the binary throughout this period.

The binary separation at which this decoupling occurs is rather smaller than
commonly estimated, largely because the internal stresses produced by MHD turbulence
in the disk are considerably greater than typical applications of the $\alpha$-model
had guessed.  The actual value of $a_{\rm dec}$ is sensitive to the disk's thermodynamics
to the degree that the absolute level of the internal stresses are proportional to the
disk's internal pressure.  Because accretion continues, luminosity released when the
accreting gas reaches the black holes may illuminate the disk and heat it.  This
sort of feedback has the potential to keep the disk's inflow rate high, self-consistently
sustaining the accretion rate.

Given the sort of accretion rates associated with AGN, the inner regions of circumbinary
disks around binary black holes with separations of tens of gravitational radii can
be almost as bright as AGN, although there may be identifying alterations in the
shapes of their optical/UV continua.

We have also shown that the work done on streams passing from the inner edge of
the circumbinary disk through the evacuated gap around the binary is carried back
to the disk and dissipated there.  Because the disk generically develops a
non-axisymmetric density distribution at $\simeq 2.5$ binary separations, the
dissipation rate is modulated periodically with $\sim 5\%$ fractional amplitude.
In the right circumstances, this modulation might be detectable, although optical
depth in the disk is likely to diminish its fractional amplitude, particularly when
the accretion rate is high enough to make the system luminous.  If this modulation
can be detected, its period would provide an estimator of the binary's orbital
frequency with factor of 2 accuracy.

A number of our results may be sensitive to the particular parameters chosen, most
importantly equal masses in the binary, spinless black holes, and perfect alignment
between the orientation of the binary's orbital angular momentum and the disk's
angular momentum.  Moreover, these choices can interact: for example, black hole
spins oblique to the gas orbital plane can induce changes in that plane.  Future work
exploring a variety of choices for these parameters may reveal additional effects.

\acknowledgments

This work was supported by several NSF grants. 
S.C.N, M.C, and Y.Z. received support from AST-1028087 
and J.H.K. from AST-1028111.
M.C., B.C.M, H.N, Y.Z. also acknowledge partial support 
from PHY-0929114, PHY-0969855, 
PHY-0903782, OCI-0832606 and DRL-1136221.
NY acknowledges support from NSF grant PHY-1114374 and NASA grant NNX11AI49G, under sub-award 00001944. 
We thank C. Lousto (RIT), John Hawley (UVa), K. Sorathia (JHU), J. Schnittman (Goddard) 
and Jiming Shi (UC Berkeley) for valuable discussions on the manuscript. 
Computational resources were provided by the Ranger system 
at the Texas Advance Computing Center (Teragrid allocation TG-PHY060027N), 
which is supported in part by the NSF, and by NewHorizons at Rochester
Institute of Technology, which was supported by 
NSF grant No. PHY-0722703, DMS-0820923 and AST-1028087.

\appendix

\section{Hydrostationary Torus Solutions in General Axisymmetric Spacetimes}
\label{sec:hydr-torus-solut}

Here, we describe a method for calculating axisymmetric non-magnetized gas distributions 
supported by pressure gradients and rotation within an axisymmetric spacetime. 
We will assume a general spacetime with Killing vectors 
$\left( \partial / \partial \phi \right)^a$ and 
$\left( \partial / \partial t    \right)^a$
that can be expressed in the simple form (in coordinates similar to spherical Boyer-Lindquist coordinates):
\beq{
g_{\mu \nu} =  \left[ \begin{array}{cccc}
g_{t t} & 0 &0 &g_{t \phi} \\[0.25cm]
0 & g_{r r} & 0 & 0 \\[0.25cm]
0 & 0 & g_{\theta \theta} & 0 \\[0.25cm]
g_{t \phi} & 0 & 0 & g_{\phi \phi} \\[0.25cm]
\end{array} \right] \quad ,
\label{axi-sym-metric}
}
which means that the inverse metric is 
\beq{
g^{\mu \nu} =  \left[ \begin{array}{cccc}
-\frac{g_{\phi \phi}}{A} & 0 &0 & \frac{g_{t \phi}}{A} \\[0.25cm]
0 & \frac{1}{g_{r r}} & 0 & 0 \\[0.25cm]
0 & 0 & \frac{1}{g_{\theta \theta}} & 0 \\[0.25cm]
\frac{g_{t \phi}}{A}  & 0 & 0 & -\frac{g_{t t}}{A} \\[0.25cm]
\end{array} \right] \,.
\label{axi-sym-inv-metric}
}
where $A = g_{t \phi}^2 - g_{t t} \, g_{\phi \phi}$.  We have verified that the $\phi$-average of our 
PN spacetime, $\hat{g}_{\mu \nu}$, has this form to within the accuracy of our PN procedure.

The initial state of the simulation consists of matter in axisymmetric 
hydrostatic equilibrium with a specific angular momentum profile, $\ell$.
We start from the discussion of \cite{dVHK03}, which is based
on \cite{1985ApJ...288....1C} and other citations
mentioned therein.   
The disk is centered about the equator of the black hole's spin; we will eventually 
assume that it is initially isentropic.  
The time-independent and axisymmetric Euler-Lagrange equations 
reduce, essentially, to 
\beq{
\frac{\partial_i h}{h} + \frac{1}{2} u^2_t \partial_i u_t^{-2} - \frac{\Omega}{1 - \ell \Omega} \partial_i \ell
= 0  
\,,
\label{enthalpy-diff-eq}
}
where the angular frequency---$\Omega = u^\phi/u^t$---is not a simple function of the 
specific angular momentum---$\ell = -u_\phi/u_t$. 
The $4$-velocity, $u^\mu$, in our symmetry has zero components: $u^r = u^\theta = u_r = u_\theta = 0$.  
One can show, from the normalization condition $u_\mu u^\mu = -1$, that 
\beq{
u_t = - \left[ -g^{tt} + 2 \ell \,g^{t\phi} - \ell^2 g^{\phi\phi}  \right]^{-1/2}
\,,
\label{ut-eq}
}
and
\beq{
\Omega = \frac{ g^{t\phi} - \ell \,g^{\phi\phi} }{ g^{tt} - \ell \,g^{t\phi} }
\,.
\label{Omega-l-relation}
}
The solutions assume that 
\beq{
\Omega = \eta \,\lambda^{-q}
\,,
\label{Omega-constraint1}
}
where $\eta$ and $q$ are yet to be determined parameters,
and $\lambda$ is defined by 
\beqa{
\lambda^2 &=& \frac{\ell}{\Omega} 
\nonumber \\ 
&=& \ell \,\frac{g^{tt} - \ell \,g^{t\phi}}{g^{t\phi} - \ell \,g^{\phi\phi}}
\,.
\label{lambda-def}
}
We can eliminate $\Omega$ from this system by combining equations~(\ref{Omega-constraint1})
and (\ref{lambda-def}) to yield a non-linear algebraic equation for $\ell=\ell(r,\theta)$
in terms of the metric: 
\beq{
R(\ell) = g^{t\phi} \left[ \ell^2 +  \lambda^2(\ell) \right] - g^{tt} \ell - g^{\phi\phi} \ell \,\lambda^2(\ell)
= 0
\,,
\label{nonlinear-l-eq2}
}
where  
\beqa{
\ell &=& \Omega \,\lambda^2 
\nonumber \\
&=& \eta \,\lambda^{2-q}  
\,,
\label{l-lambda-eq}
}
or
\beq{
\lambda = \left( \frac{\ell}{\eta} \right)^{1/\left(2-q\right)}  
\,. 
\label{lambda-l-eq}
}
Also, we can show that 
\beqa{
\Omega &=& \eta^{-2/\left(q-2\right)} \ \ell^{q/\left(q-2\right)}
\nonumber \\ 
& \equiv & k \,\ell^\zeta
\,, \label{Omega-l-new-eq}
}
where $k = \eta^{-2/\left(q-2\right)}$ and $\zeta = q/\left(q-2\right)$. 

Typically, one ``solves'' equation~(\ref{nonlinear-l-eq2}) by approximating $\lambda^2$ with its Schwarzschild value: 
$\lambda^2 \simeq -g^{tt}/g^{\phi\phi}$ \citep{dVHK03,Noble09,Farris11}.  For our $\phi$-averaged spacetimes, the Schwarzschild approximation
is not so good\footnote{When using the approximate method, we found the disk to undergo a low frequency breathing mode that dominated 
the early evolution of the disk.}. 
Therefore, we need a better solution. We can solve this equation to roundoff precision
by using a Newton-Raphson scheme. To do so, we will need to know $\partial R(\ell) / \partial \ell$: 
\beq{
\pderiv{R}{\ell} = 
g^{t\phi} \left[ 2 \,\ell +  \pderiv{\lambda^2}{\ell} \right]  - g^{tt}  
- g^{\phi\phi} \left[  \lambda^2  + \ell \,\pderiv{\lambda^2}{\ell} \right]
\,,
\label{dR-dl}
}
where $\partial \lambda^2/\partial \ell = 
2 \lambda^2/[(2-q)\ell]$. 
Let us come back to equation~(\ref{enthalpy-diff-eq}). A solution to this equation yields our disk solution. 
The solution process involves integrating it from the inner disk's edge
--- located at $r_\mathrm{in}$ --- to a point within the disk:
\beq{
\int^h_{h_\mathrm{in}} \frac{d h}{h}
= - \frac{1}{2} \int^{u_t}_{u_{t \mathrm{in}}}  \frac{ d (u_t)^{-2} }{ (u_t)^{-2} }
+ \int^{\ell}_{\ell_\mathrm{in}} \frac{k \,\ell^\zeta }{ 1 - k \,\ell^{\zeta + 1}} d\ell \,. 
\label{enthalpy-int-eq}
}
With the boundary condition, $h_\mathrm{in} = h(r_\mathrm{in},\theta=\pi/2) = 1$,
one can solve this integral equation for $h=h(r,\theta)$:
\beq{
h = \frac{  u_{t \mathrm{in}} f(\ell_\mathrm{in})  }{ u_{t}(r,\theta)  f(\ell(r,\theta)) } 
\,,
\label{enthalpy-solution}
}
where $u_t$ is given by equation~(\ref{ut-eq}), $\ell$ is found using Newton-Raphson on equation~(\ref{nonlinear-l-eq2}), 
$\ell_\mathrm{in} = \ell(r_\mathrm{in},\pi/2)$ is a boundary value, and 
$f(\ell) \equiv \left|  1 - k \, \ell^{\zeta+1}  \right|^{1/\left(\zeta+1\right)}$. 

We want a distribution that resembles a torus, which has a 
pressure maximum at some radius $\rpmax$, and has finite extent.
The parameters $\left\{\ell_\mathrm{in}, q, \eta\right\}$ determine whether we get such a solution. 
We would like to replace one of the degrees of freedom with $\rpmax$, however, there is not a closed-form solution for $\rpmax$
in terms of any of the original parameters.  We know that the fluid attains the Keplerian angular momentum 
at the pressure maximum ($\rpmax$) as the pressure
gradient must be zero there.  It means that $\ell_\mathrm{in}$ should be super-Keplerian
at the inner edge ($r_\mathrm{in}$), and $\ell_\mathrm{out}$ should be 
sub-Keplerian at the outer edge ($r_\mathrm{out}$). 
We therefore know that $\ell_\mathrm{in} > \ell_K(r_\mathrm{in})$,  
$\ell_\mathrm{out} < \ell_K(r_\mathrm{out})$, and $\ell_p = \ell_K(\rpmax)$,
where $\ell_K$ is the Keplerian specific angular momentum (see
the next section below).  However, we only have two free parameters, and now have three constraints
(if we specify all $\ell_\mathrm{in}, \ell_\mathrm{out}$ and $\ell_p$). 
It may be possible to change equation~(\ref{l-lambda-eq}) to look like:
\beq{
\ell = \eta \left( \lambda - \lambda_0 \right)^{2-q}  
\,, \label{general-l-lambda-eq}
}
and then find $\lambda_0$ with this third constraint.  Using these three constraints, however, does not yield a closed-form solution
for $q$, $\eta$, and $\lambda_0$. Therefore, another Newton-Raphson procedure would be required.
Hence, we relax the constraint on $\ell_\mathrm{out} < \ell_K(r_\mathrm{out})$, and let $\rout$ be a result of our procedure. 
Using $\lambda_0=0$, we find that 
\beqa{
q &=& 2 - \frac{ \log{\left(\ell_\mathrm{in}/\ell_p\right)} }
{ \log{\left(  \lambda_\mathrm{in} / \lambda_p \right)} } 
\,,
\label{q-soln}
\\
\eta &=& \frac{ \ell_p }{ \lambda_p^{2-q} } \label{eta-soln}
\,,
}
where $\lambda_p = \lambda(\ell_p, \rpmax)$ and
$\lambda_\mathrm{in} = \lambda(\ell_\mathrm{in}, r_\mathrm{in})$ given by equation~(\ref{lambda-def}). 

We follow the solution process based on one described in~\cite{1985ApJ...288....1C}, and is the following:
\begin{enumerate}
\item Chose values of $(r_\mathrm{in}, \rpmax, \ell_\mathrm{in})$, and derive $q$ and $\eta$ using
equations~(\ref{q-soln}) and (\ref{eta-soln}); 
\item Calculate $\lambda$ (and then $\ell$) at a new $(r,\theta)$ via 
Newton-Raphson (using equations~(\ref{nonlinear-l-eq2}) and (\ref{lambda-l-eq}))
close to the previous location, so that the old location's value can be used as a seed to the Newton iteration to 
successfully find a solution;  
\item Calculate $u_t$ via equation~(\ref{ut-eq}) then $h$ at $(r,\theta)$ via equation~(\ref{enthalpy-solution});
\end{enumerate}

The solution process progresses throughout $(r,\theta)$ space until the boundary of the disk is found, where
$h(r,\theta)=1$.
The path we take starts at $(r,\theta)=(r_\mathrm{in},\pi/2)$, 
moves in increasing $r$ along the $\theta=\pi/2$ line, and we test to see if $h$ is increasing.  If $h$ is not increasing at first, 
then we stop and try a different set of parameters.  If $h$ is increasing at first, we then proceed until the outer edge
of the disk is reached;  this radius is denoted as $\rout$.   Then, 
$\forall r \in \left[\rin, \rout\right]$, we start from the $\theta=\pi/2$ solution and proceed backward and 
forward in $\theta$ along constant $r$ to find $h(r,\theta)$.  

The initial torus solution used herein is parameterized by 
$\ell_\mathrm{in}=8.743$, $\eta = 1.961$, $q = 1.642$, $\rin = 3 a_0 = 60M$, $\rpmax = 5 a_0 = 100M$, $\rout = 11.75 a_0 = 235M$. 

\begin{figure}[htb]
\centerline{
\includegraphics[width=12cm]{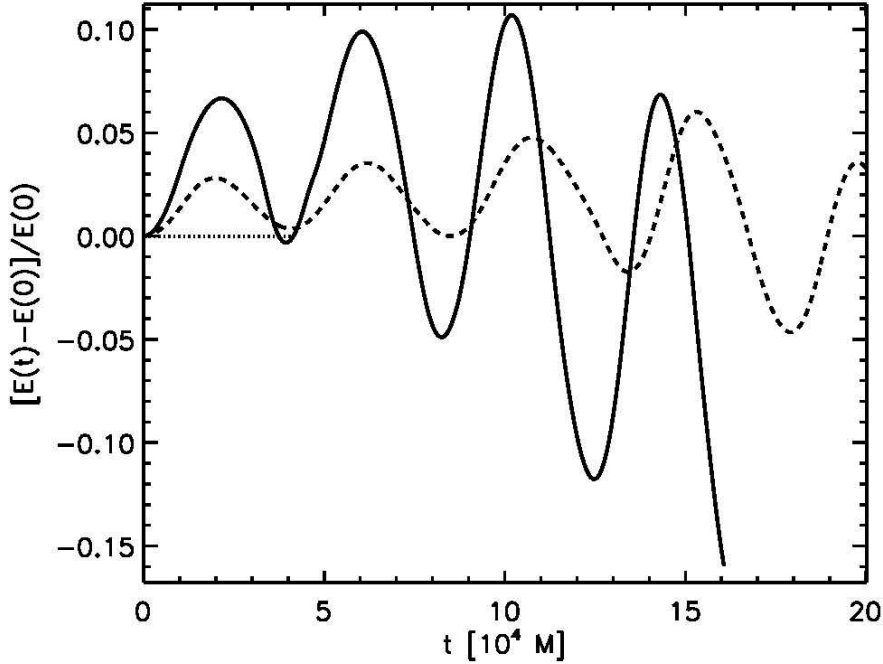}
}
\caption{Change in total energy relative from its initial value versus time of non-magnetized tori from different runs.  
The first run used the so-called ``approximate method'' and was evolved in a PN BBH spacetime (solid);  the second run used 
our new and more accurate procedure, but used the same spacetime as the first (dashes); 
the last used the same disk from the second run, but was evolved in the $\phi$-average of the other runs' spacetime (dots).  The 
curve for the last run (dots) oscillates with amplitude $\sim 10^{-8}$, which is why it appears consistent with zero in this figure. 
All the disks share the same parameters: $\rin = 2 a_0$, $\rpmax = 4 a_0$, $a_0 = 30 M$. 
\label{fig:tori-energy-evolution}}
\end{figure}

\subsection{General Keplerian Velocity}
\label{sec:gener-kepl-veloc}

The stationary torus solution described in Section~\ref{sec:hydr-torus-solut} requires the Keplerian, or 
circular equatorial geodesic orbits, of the spacetime.  Since we do not calculate $\hat{g}_{\mu \nu}$ in closed form, 
we require equations for these orbits based on a generalized metric of the form Equation~(\ref{axi-sym-metric}).  Here, 
we state the equations governing the Keplerian orbits. 

Keplerian orbits in our spacetime have $4$-velocity, 
$u^\mu = [u^t, 0, 0, u^\phi] = u^t [1, 0, 0, \Omega_K]$.  
$\Omega_K$ is found from the $r$-component of the geodesic equation,
$du^r/d\tau  = - {\Gamma^r}_{\mu \nu} u^\mu u^\nu  = 0$,
which ultimately yields
\beqa{
\Omega_{K \, \pm} &=& \frac{1}{ {\Gamma^r}_{\phi \phi} }   \left[ 
- {\Gamma^r}_{t \phi}
\pm 
\sqrt{
\left( {\Gamma^r}_{t \phi} \right)^2 
- {\Gamma^r}_{\phi \phi} {\Gamma^r}_{t t}
}
\right]
\nonumber \\
&=& - \frac{1}{ \partial_r g_{\phi\phi} }   \left[ 
 \partial_r g_{t \phi} 
\pm
\sqrt{
\left( \partial_r g_{t \phi} \right)^2 
- \left(\partial_r g_{\phi\phi}\right) \left(\partial_r g_{t t}\right)
}
\right]
\,.
\label{Omega-K-soln2}
}
$\Omega_{K \, -}$ and $\Omega_{K \, +}$
are the prograde angular velocity and retrograde angular velocity, respectively.  
The $4$-velocity components are found by the normalization condition: 
\beqa{
u^t &=& [ g_{t t} + 2 \,\Omega \,g_{t \phi} + \Omega^2 g_{\phi \phi}  ]^{-1/2}
\,,
\label{ut-Keplerian}
\\
u^\phi &=& \Omega \,u^t
\,. \label{uphi-Keplerian}
}
The Keplerian specific angular momentum, $\ell_K$, is found by the relation between $\ell$ and $\Omega$:
\beq{
\ell = - \frac{ g_{t \phi} + \Omega \,g_{\phi \phi} }{  g_{t t} + \Omega \,g_{t \phi} } 
\,. \label{l-Omega-eq}
}

\section{Resolution Requirements}
\label{sec:resolution-requirements}
Our grid resolution was chosen to adequately resolve the MRI, to resolve 
the spiral density waves generated by the binary's potential, and to involve
cells that are nearly cubical.  We discuss each choice in turn. 

Many recent studies have  explored the resolution dependence of global MHD accretion disk simulations
\citep{hgk11,2011arXiv1106.4019S,2012ApJ...744..187S}.   \cite{hgk11} found that many 
global properties of the disk nearly asymptote with increasing resolution once the following criteria are 
satisfied:
\beq{
  \NN{z} \gtrsim 16 \left(\frac{\beta}{100}\right)^{1/2} \left(\frac{\beta_z}{\beta}\right)^{1/2}  
\left(\frac{\QQ{z}}{10}\right)
\,,
\label{Nz-per-scaleheight}
}
\beq{
\NN{3} \gtrsim 790 \left(\frac{0.1}{H/r}\right) \left(\frac{\beta}{10}\right)^{1/2}  \left(\frac{\QQ{3}}{25}\right)
\,,
\label{nphi-condition}
}
where $\NN{z}$ is the number of cells per scale height, $H$, $\QQ{z} > 10$ is the recommended quality factor
of the simulation, $\beta_z \equiv \langle p \rangle / \langle \left|\sqrt{g_{zz}} B^z\right|^2 \rangle$.  
We use spherical coordinates, so $\NN{\theta}$ is needed instead:
\beq{
\NN{z} = \frac{H}{\Delta z} = \frac{H}{r \Delta \theta} = \frac{H/r}{\Delta \theta}  \equiv N_{H/r}
\quad ,
\label{Nz-Nth-relation}
}
where  $N_{H/r}$ is the number of cells in the poloidal direction per scale height. 
We see from prior simulations (e.g., \cite{Noble10}) that $\beta \simeq 10$ and $\beta_z/\beta \simeq 50$ 
are reasonable for a disk in its asymptotic steady 
state, suggesting that $N_{H/r} > 36$.  The initial condition values of $\beta \simeq 100$ 
and $\beta_z/\beta \simeq 1$, however, yield a weaker constraint ($N_{H/r} > 16$) on the resolution.  
Thus, we setup a grid such that $N_{H/r} \simeq 36$  with $H/r=0.1$, our simulation's scale height.
This is satisfied by the $\xtwo$ discretization described in Section~\ref{sec:grid-bound-cond}.  We also note
that our condition satisfies the recommendation of $N_{H/r} > 32$ by \cite{2011arXiv1106.4019S}. 

The more severe constraint is on the azimuthal symmetry.  Both \cite{hgk11} and \cite{2011arXiv1106.4019S} 
suggest that past simulations under-resolved 
the azimuthal direction and that one should cover the full azimuthal range $\phi \in [0,2\pi]$ instead of 
assuming quarter- or half-circle symmetry.   Since $\Delta \phi$ 
limits the time step size, we were only able to afford $\NN{3} = 400$ as anything larger was 
impractical given our computational resources at the time.  We were optimistic with this 
resolution, however, since the thinnest run of 
\cite{Noble10} failed to satisfy  Equation~(\ref{nphi-condition}) yet still resolved the MRI with $\QQ{3} > 25$
throughout most of the disk's body.  

\begin{figure}[htb]
\centerline{
\begin{tabular}{c}
\includegraphics[width=12cm]{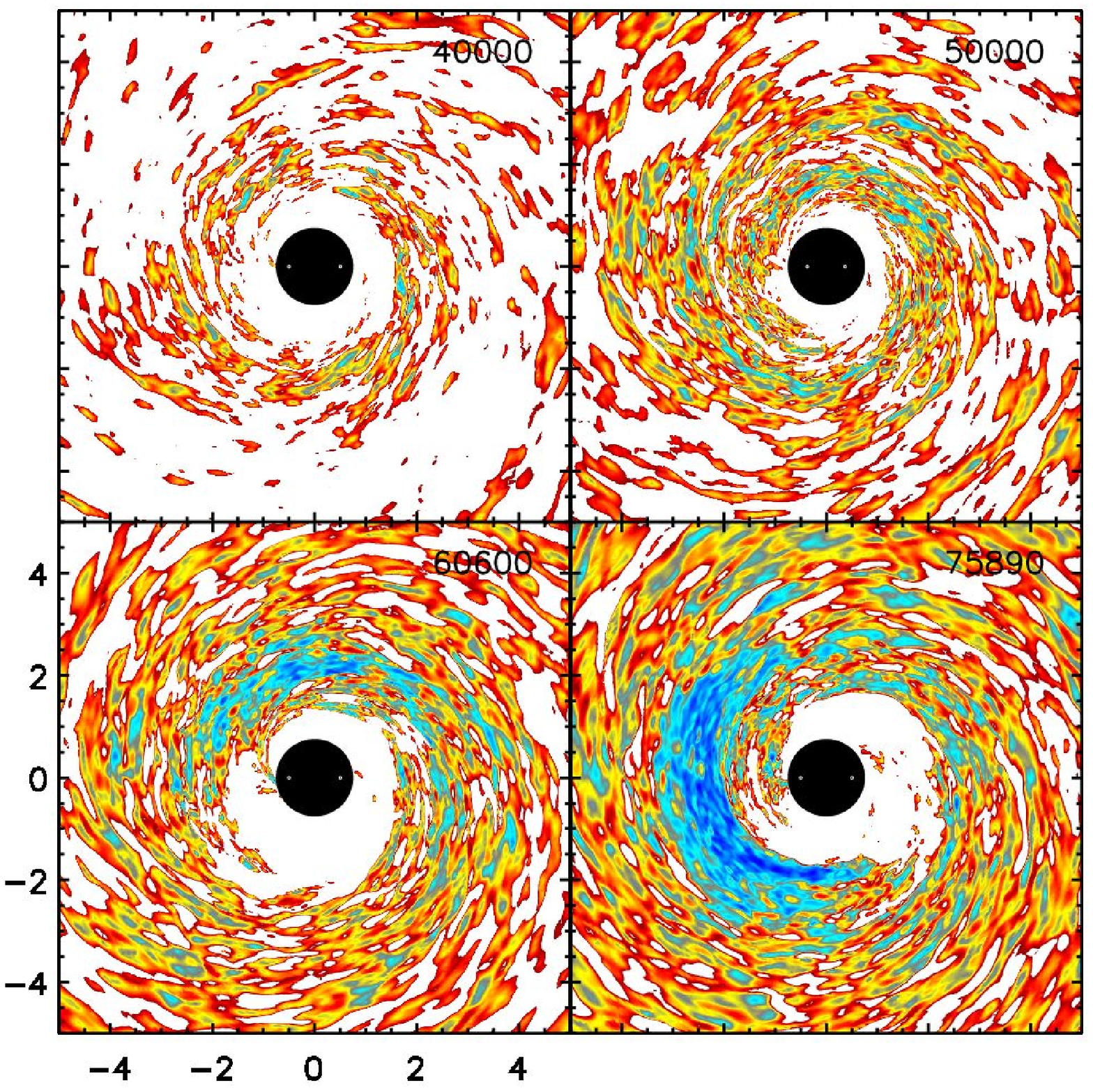}\\
\includegraphics[width=11cm]{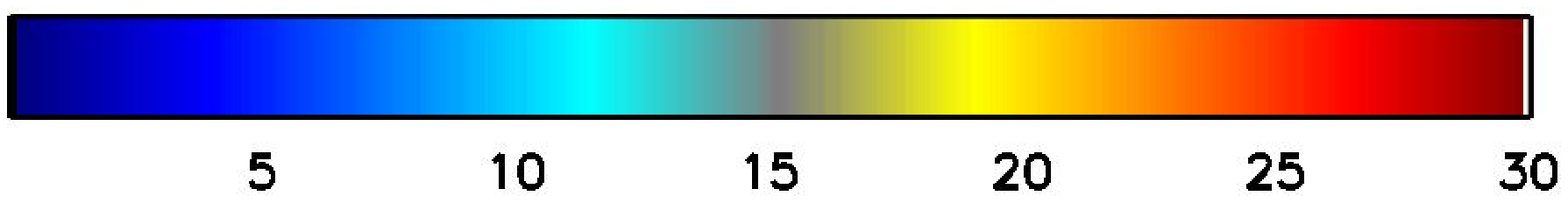}
\end{tabular}
}
\caption{
$\langle \QQ{2} \rangle_{\rho}$ from \runtwo (top row) and \runthree (second row) at late times in each simulation. 
The times of each snapshot are specified in the upper-right corner of each frame in units of $M$.  The vertical 
and horizontal axes are in units of $a_0=20M$. 
We note that the $t=40000M$ snapshot is shared by \runtwo and \runthree. The color map used to make the snapshots is 
given in the bottom row. 
\label{fig:mri-q2-run23}}
\end{figure}

\begin{figure}[htb]
\centerline{
\begin{tabular}{c}
\includegraphics[width=12cm]{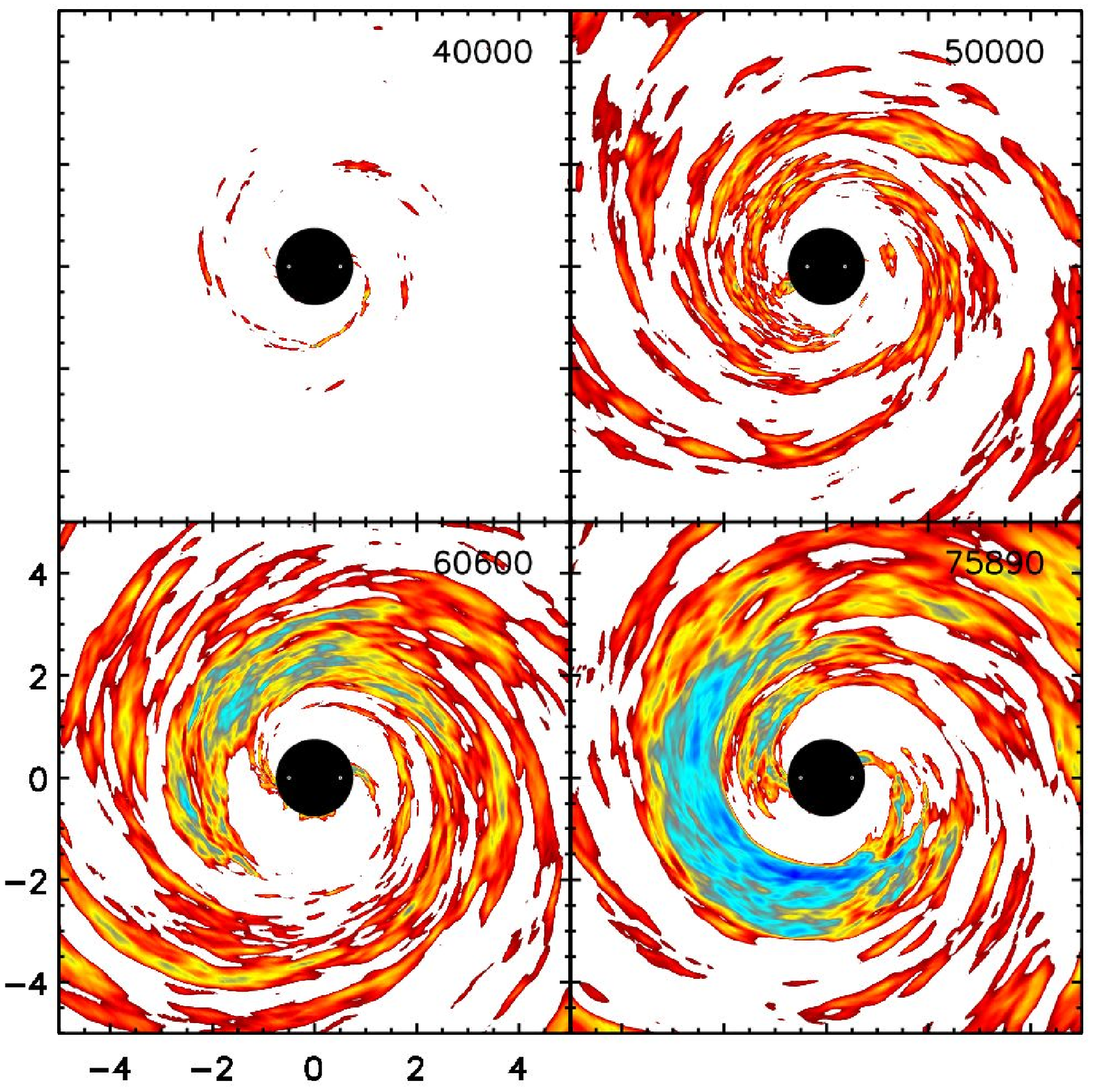}\\
\includegraphics[width=11cm]{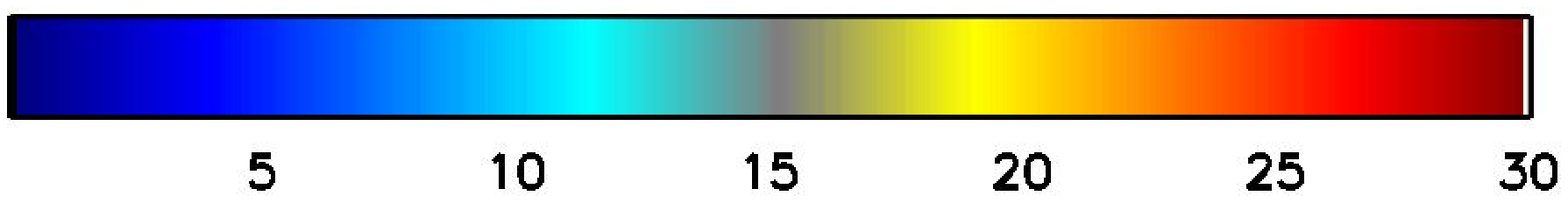}
\end{tabular}
}
\caption{
Same as in Figure~\ref{fig:mri-q2-run23}, but for $\langle \QQ{3} \rangle_{\rho}$.
\label{fig:mri-q3-run23}}
\end{figure}

We demonstrate how well \runtwo and \runthree resolve the MRI in Figures~\ref{fig:mri-q2-run23}~-~\ref{fig:mri-q3-run23}, where
we show mass-weighted averages of the $\QQ{2}$ and $\QQ{3}$ MRI quality factors: 
\beq{
\QQ{i} = \frac{2 \pi \left|b^i\right|}{ \Delta \xx{i} \, \Omega_K(r) \, \sqrt{\rho h + 2 p_{m} } }
\quad . \label{mri-quality-factors}
}
The averages were made over $\xtwo$ in the following way: 
\beq{
\langle \QQ{i} \rangle_{\rho} \equiv \frac{\int_0^{1} \QQ{i} \rho \, \sqrt{-g} \, d\xtwo}{\int_0^{1} \rho \, \sqrt{-g} \, d\xtwo}
\quad . \label{Qi-avg-x2}
}
A mass-weighting is used to calculate $\langle Q^i \rangle_{\rho}$ in order to bias 
the integral over the turbulent portion of the disk (the disk's bulk) rather than 
the laminar regions (e.g., corona, funnel).   We find that the $\QQ{z}$ constraint, 
i.e. $\langle \QQ{2} \rangle_{\rho} > 10$, 
is satisfied for all times and regions in either \runtwo or \runthree except 
for the densest parts of the lump at late times in \runthree.   
Similarly, the $\QQ{3}$ constraint, i.e. $\langle \QQ{3} \rangle_{\rho} > 25$,  
is satisfied for all times and regions in either \runtwo or \runthree except 
for in the lump at late times in \runthree.   We further note that
$\langle B^{r} B_{\phi} \rangle/\langle p_m \rangle \simeq 0.3-0.35$  when averaged over the 
quasi-steady period of \runtwo and \runthree;  this level is consistent with the 
asymptotic value found in resolution studies about point masses 
\citep{2008NewA...13..244B,2009ApJ...694.1010G,2010ApJ...713...52D,2011ApJ...730...94S,hgk11,2011arXiv1106.4019S,2012ApJ...744..187S}.

We also aim to resolve the spiral density waves generated by the binary's time-dependent quadrupolar potential.  
This means that we need about $\sim10$ cells per wavelength of the sound wave generated by the binary, 
$\lambda_d = 2 \,\pi \,c_s/\Omegabin$, where $c_s =  \left( H/r \right) r \, \Omega_K$ is 
the speed of sound.  We use the Newtonian approximates $\Omegabin \simeq  a^{-3/2}$ and $\Omega_K \simeq r^{-3/2}$
to simplify $\lambda_d$:  $\lambda_d  \simeq  2 \pi \left(H/r\right) r \left(a/r\right)^{3/2}$. 
We want to resolve $\lambda_d$ in $\phi$ and $r$ out to $\rpmax$, which means that we want to satisfy another
quality condition: $\lambda_d / \Delta r  , \lambda_d / r \Delta \phi  \simeq Q_d$,  where  $Q_d$ will be the target
number of cells per spiral density wavelength.  Using the grid specifications described in Section~\ref{sec:grid-bound-cond}, 
it is easy to show that the resolution constraints become: 
\beq{
\None \simeq 305 \left(\frac{\rpmax/a_0}{5}\right)^{3/2} 
\left(\frac{0.1}{H/r}\right) 
\left(\frac{\ln \left(\Rmax/\Rmin\right)}{\ln \left(13/0.75\right)}\right) 
\left(\frac{Q_d}{6}\right) 
\quad , 
}
\beq{
\Nthree \simeq 671  \left(\frac{0.1}{H/r}\right)  \left(\frac{\rpmax/a_0}{5} \right)^{3/2} \left(\frac{Q_d}{6}\right)
\quad . 
\label{Nphi-olr-constraint-2}
}
We find that the spiral density wave criterion is stricter than the MRI criterion when 
\beq{
Q_d \left(\frac{r}{a}\right)^{3/2} > \beta^{1/2} \QQ{3}
\,.
\label{olr-over-mri-phi}
}
Again, we did not satisfy the constraint on azimuthal resolution with our choice of $\Nthree = 400$.  In this case, 
we were reassured by evidence found by \cite{Shi11} that found that the spiral density waves were extended over
a large azimuthal extent and required far fewer cells than expected in that direction to resolve.   In practice, 
we find that spiral density waves are short-lived as they propagate through the turbulent, shear flow of the disk; 
they are hardly ever seen in snapshots of intrinsic quantities past $r\simeq 3a_0$. 

Since gridscale dissipation scales with the ratio of the cell extent
to characteristic length scales of the physical quantities, cells that are oblate
may effectively lead to anisotropic dissipation.  Because physical dissipation mechanisms 
are isotropic, this effect could lead to unphysical artifacts. 
For this reason, we attempt to make the cells within the bulk of the disk---where most of the 
dissipation occurs---as isotropic as possible.   
Our runs use grids with $\Delta r : r \Delta \theta : r \Delta \phi : \simeq  3 : 1 : 3$ as measured in
the $\theta = \pi / 2$ plane.  \cite{2011arXiv1106.4019S} suggested that  $\Delta r = r \Delta \phi$ 
and $\Delta \phi / \Delta \theta \le 2$;  we satisfy the former, and violate the latter by a slim margin: 
$\Delta \phi$ should be just $2/3$ times the size we use. 
We note that the poloidal extent increases off the equator, so that the cells become more cubical at larger 
$\left|\theta-\pi/2\right|$.

\section{Mass, Energy, and Angular Momentum Budgets}
\label{sec:conservation-laws}
Space and time gradients of the accretion flow's extensive quantities, mass ($M$), energy ($E$) and angular momentum ($J$) 
are fundamental for understanding the disk's evolution and structure.  In stationary spacetimes, $M$, $E$, and $J$ are 
all conserved.   In our $\left(t,\phi\right)$-dependent  spacetime, $E$ and $J$ are no longer strictly conserved. 
We describe here how several functions used throughout the paper are derived from the evolution equations of mass, energy 
and angular momentum. 

\subsection{Angular Momentum}
\label{sec:angular-momentum}

We begin this section with the angular momentum equation because of its import to accretion physics. 
We follow the notation and derivation procedure outlined in~\cite{Farris11}.

An extensive quantity, $J$, is the integral over the spatial volume of the time component of the its associated current, $j^\mu$: 
$J  =  \int j^t  \, \sqrt{-g} \, dV$, 
where $dV$ is the spatial volume component in the spacelike hypersurface (e.g., $dr d\theta d\phi$).
We are interested in the azimuthal component of the momentum as that is
the dominant component of the gas' and the binary's momenta.  We therefore recognize that 
$j^\mu = {T^\mu}_\nu \phi^\nu$, and 
$\phi^\nu = \left(\partial_\phi\right)^\nu = \partial x^\nu/\partial \phi = [0,0,0,1]$ in spherical 
coordinates, which is what we use.  We wish to calculate $d^2 J/dt dr$ . 
If $J$ is locally conserved perfectly,  $\nabla_\mu j^\mu = 0$.  
In our case it will not be conserved exactly, and 
exploring the radial gradient of its volume integral will help us understand how MHD
stresses and the binary's gravitational torque compete over the run of the flow.
This quantity is: 
\beqa{
\deriv{}{r} \int \left( \nabla_\mu j^\mu \right) \, \sqrt{-g} \, dr d\theta d\phi  
& = &  \partial_r \int \left( \partial_\mu \sqrt{-g} \, j^\mu \right) \,  dr d\theta d\phi  
\nonumber \\
& = &  \partial_r \partial_t  J + \partial_r \partial_i  \int j^i \, \sqrt{-g}\,  dr d\theta d\phi
\nonumber \\
& = &  \partial_r \partial_t  J
+      \partial_r       \int  {T^r}_\phi   \,\sqrt{-g}  \, d\theta d\phi  \,, 
\label{radial-ang-mom-nonconvergence8}
}
where the last equality results from the fact that $j^\theta$ is zero on the axis, and 
$j^\phi(\phi=0) = j^\phi(\phi=2\pi)$. 
On the other hand, we know from 
the stress-energy EOM---$\nabla_\mu {T^\mu}_\nu = -\mathcal{F}_\nu$---that 
\beqa{
\deriv{}{r} \int \left( \nabla_\mu j^\mu \right) \, \sqrt{-g} \, dr d\theta d\phi  
& = &   \int \left( \nabla_\mu j^\mu \right) \, \sqrt{-g} \, d\theta d\phi 
\nonumber \\
& = &  \deriv{T}{r}  - \int \mathcal{F}_\phi \, \sqrt{-g} \, d\theta d\phi  \,,
\label{radial-ang-mom-nonconvergence-2nd-6} 
}
where the torque density, $dT/dr$, can be expressed as 
\beq{
\deriv{T}{r}  =  \int {T^\mu}_\nu {\Gamma^\nu}_{\mu \phi} \,  \sqrt{-g} \, d\theta d\phi
\ = \ \frac{1}{2} \int T^{\mu \nu} \partial_\phi g_{\mu \nu} \,  \sqrt{-g} \, d\theta d\phi
\,. \label{final-torque-eq}
}
We remind the reader that $\mathcal{F}_\nu$ is the radiative cooling flux (see Section~\ref{sec:simulation-details} for details).

Therefore, equating the two equations~(\ref{radial-ang-mom-nonconvergence8}) and~(\ref{radial-ang-mom-nonconvergence-2nd-6}),
we have
\beq{
\begin{array}{ccccccccccc}
\partial_r \partial_t  J  & = &\deriv{T}{r} & - & \left\{ \mathcal{F}_\phi \right\}
& - & \partial_r  \left\{ {T^r}_\phi \right\} &&&&
\\
& =  &\deriv{T}{r}  &- &\left\{ \mathcal{F}_\phi \right\}
& - &\partial_r  \left\{   {M^r}_\phi \right\}
& - &\partial_r  \left\{   {R^r}_\phi \right\}
& - &\partial_r  \left\{   {A^r}_\phi \right\}
\,,
\end{array}
\label{ang-mom-profile-4}
}
where we have used here the shorthand
\beq{
\left\{ X \right\} \equiv \int \sqrt{-g} \, X  d\theta \, d\phi \ = \langle X \rangle  \int \sqrt{-g} \, d\theta \, d\phi 
\,. \label{shell-integral}
}
Also, ${M^r}_\phi$, ${R^r}_\phi$, and ${A^r}_\phi$ are---respectively---the Maxwell (MHD) stress, Reynolds stress, and 
advected flux of angular momentum.  We note that ${M^\mu}_\nu = 2p_m u^\mu u_\nu + p_m {\delta^\mu}_\nu - b^\mu b_\nu$ is the EM part of ${T^\mu}_\nu$, while 
$\left( {R^\mu}_\nu + {A^\mu}_\nu \right) = {{T_H}^\mu}_\nu = \rho h u^\mu u_\nu + p {\delta^\mu}_\nu$ is the hydrodynamic part.  
The Reynolds stress alone is more complicated to calculate 
as we have to find the perturbation from the 
mean flow: 
\beq{
{R^r}_\phi = \rho h \, \delta u^r \,  \delta u_\phi 
\,, \label{reynolds-stress}
}
where 
\beq{
\delta u^\mu \equiv u^\mu - \left\{ \rho u^\mu \right\} / \left\{ \rho  \right\}
\,. \label{velocity-perturbation}
}
We note that we include the enthalpy as it technically contributes to the 
stress; its contribution is insignificant, however, for our relatively cool flow.  
The quantities $\left\{ {R^\mu}_{\nu} \right\}$ and $\left\{ {A^\mu}_{\nu} \right\}$ are not 
calculated during the simulation, but found approximately from other shell-integrated 
quantities we do calculate;  $\left\{ {M^\mu}_{\nu} \right\}$,  $\left\{ \mathcal{F}_\mu \right\}$, and 
$dT/dr$  \emph{are} calculated as stated above during the run. 
One can easily show from equations~(\ref{reynolds-stress}) and~(\ref{velocity-perturbation}) that 
\beq{
\left\{ {R^r}_\phi \right\} 
\ = \ \left\{ \rho h \, \delta u^r \, \delta u_\phi  \right\}
\ \simeq \  \left\{ {{T_H}^r}_\phi \right\} - \left\{ {A^r}_\phi \right\} 
\label{reynolds-derivation-11}
}
where ${{T_H}^r}_\phi$ is the 
hydrodynamic part of ${T^r}_\phi$, and $\left\{ {A^r}_\phi \right\}$ is calculated approximately 
as  
\beq{ 
\left\{ {A^r}_\phi  \right\} \simeq \frac{\left\{ \rho \ell \right\}  \left\{ \rho h u^r \right\} }{\left\{ \rho \right\}}
\,. \label{advected-flux}
}
Here, $\ell = - u_\phi / u_t$ as its defined in Appendix~\ref{sec:hydr-torus-solut}. 
The approximations used to find equations~(\ref{reynolds-derivation-11}-\ref{advected-flux}) 
include: 1) $h \simeq 1$, and 2) $u_t \simeq -1$. 
We have demonstrated that these assumptions are valid to the few percent level in the bound portion of the 
flow for our simulations described in this paper.

\subsection{Energy}
\label{sec:energy}

Torques and stresses do work on the gas, transporting angular momentum.
This work can be dissipated in the disk, changing its internal energy,
which is eventually radiated away in part.
Here we calculate the partitions in which the energy can move into; 
this calculation is nearly identical to that for $d^2 J/dtdr$ in Appendix~\ref{sec:angular-momentum}
The current associated with $E$ is $e^\mu = {T^\mu}_\nu t^\nu$, where $t^\mu$ is the $4$-vector along time coordinate, 
$t^\mu = \left[1,0,0,0\right]$.  They are related by $E = \int e^t \sqrt{-g} \,dV$.  
Just as with $j^\mu$, the divergence of $e^\mu$ is not exactly zero,
because of the time-dependent spacetime. Using a similar analysis as before, we get 
\beq{
\partial_r \partial_t  E \ =  \ dW/dr  \ - \ \left\{ \mathcal{F}_t \right\}
\ - \ \partial_r  \left\{ {T^r}_t \right\}
\nonumber 
}
where $dW/dr = \left\{\frac{1}{2} T^{\mu \nu} \partial_t g_{\mu \nu} \right\}$ is the work done by the spacetime on 
the matter. 

\subsection{Mass Accretion Rate}
\label{sec:acc-rate}
The current $j^\mu = \rho u^\mu$ is associated with the conserved quantity $M$, 
so we have $M = \int \rho u^t  \sqrt{-g} \,dV$, and 
\beq{
\deriv{M}{t}  
              \ = \ - \int \rho u^r  \sqrt{-g} \,d\theta d\phi \,.  \label{dMdt-5}
}
by using a similar technique to obtain Equation~(\ref{radial-ang-mom-nonconvergence8}).

\bibliography{bhm_references}

\end{document}